\documentclass[useAMS,usenatbib]{mn2e}

\usepackage{pdflscape}
\usepackage{graphics}
\usepackage{epsfig}
\usepackage{latexsym}
\usepackage{amsfonts}
\usepackage{amsmath}
\usepackage{amssymb}
\usepackage[english]{babel}
\usepackage{graphicx}
\usepackage{url}
\usepackage{textcomp}

\usepackage{amssymb}
\usepackage{graphics}
\usepackage{graphicx}
\usepackage{epstopdf}

\usepackage[usenames]{color}

\def\red#1{\textcolor{red}{ #1}}  % flag for editing
\def\blue#1{\textcolor{blue}{ #1}}  % flag for editing
\newcommand{\be}{\begin{equation}}
\newcommand{\ee}{\end{equation}}
\newcommand{\ba}{\begin{eqnarray}}
\newcommand{\ea}{\end{eqnarray}}
\newcommand{\bc}{\begin{center}}
\newcommand{\ec}{\end{center}}

\usepackage{ifpdf}
\ifpdf
  \usepackage[pdftex]{hyperref}
\else
  \usepackage[hypertex]{hyperref}
\fi

\title[SSC and equipartition in PWNe]{The effects of
magnetic field, age, and intrinsic luminosity 
on Crab-like pulsar wind nebulae
}
\author[Torres, Mart\'in, de O\~na Wilhelmi, \& Cillis]{D. F. Torres$^{1,2}$, J. Mart\'in$^{1}$, E. de O\~na Wilhelmi$^{1}$, \& Analia Cillis$^3$\\
$^1$Institute of Space Sciences (IEEC-CSIC), Campus UAB,  Torre C5, 2a planta, 08193 Barcelona, Spain \\
$^2$Instituci\'o Catalana de Recerca i Estudis Avan\c{c}ats (ICREA) Barcelona, Spain\\
$^3$Instituto de Astronom\'ia y F\'isica del Espacio,  Casilla de Correo 67 - Suc. 28 (C1428ZAA), Buenos Aires, Argentina \\
}

\begin{document}

\date{}

%\pubyear{2013}

\maketitle

\label{firstpage}

\begin{abstract}

We investigate the time-dependent behavior of Crab-like pulsar wind nebulae (PWNe) generating
a set of models using 4 different initial spin-down luminosities ($L_0 =\{1,0.1,0.01,0.001\} \times L_{0, {\rm Crab}}$), 8 values of magnetic fraction ($\eta =$ 0.001, 0.01, 0.03, 0.1, 0.5, 0.9, 0.99, and 0.999, i.e., from fully particle dominated to fully magnetically dominated nebulae), and 3 distinctive ages: 940, 3000, and 9000 years.
We find that the self-synchrotron Compton (SSC) contribution is irrelevant for $L_{SD}$=0.1, 1, and 10\% of the Crab power, disregarding the age and the magnetic fraction. SSC only becomes relevant for highly energetic ($\sim 70\%$ of the Crab),  particle dominated nebulae at low ages (of less than a few kyr), located in a FIR background with  relatively low energy density. Since no pulsar other than Crab is known to have these features, 
these results clarify why the Crab Nebula, and only it, is SSC dominated.
No young PWN would be detectable at TeV energies if the pulsar's spin-down power is 0.1\% Crab or lower. For 1\% of the Crab spin-down, only particle dominated nebulae can be detected by H.E.S.S.-like telescopes when young enough (with details depending on the precise injection and environmental parameters). Above 10\% of the Crab's power, all PWNe are detectable by H.E.S.S.-like telescopes if they are particle dominated, no matter the age.
The impact of the magnetic fraction on the final SED is varied and important, generating order of magnitude variations in the luminosity output for systems that are otherwise the same (equal $P$, $\dot P$, injection, and environment).

\end{abstract}


%%%%%%%%%%%%%%%%%%%%%%%%%%%%%%%%%%%%%%%%%%%%%%%%%%
\section{Introduction}
%%%%%%%%%%%%%%%%%%%%%%%%%%%%%%%%%%%%%%%%%%%%%%%%%%

Since the observation of the first unidentified TeV gamma-ray source (Aharonian et al. 2002, Albert et al. 2008),
more than twenty pulsar wind nebula (PWNe) have been identified at very-high-energies (VHE; $E>100$ GeV) by the current generation of Cherenkov telescopes. PWNe are thus the most numerous population of VHE Galactic sources. 
These PWNe are associated with young ($\tau<$10$^5$ years, {here, only the very young are considered}) and energetic pulsars ($\dot{E}>10^{33}$ erg s$^{-1}$), and usually display extended emission up to a few tens of parsecs (Rieger et al. 2013). 
The majority of PWNe were observed by the H.E.S.S. experiment during the Survey of the Galactic plane performed since 2004 (see Gast et al. 2012 for the current status).
Up to that time, only the Crab Nebula has been detected having 
a steady gamma-ray flux about 1 TeV  (Weekes et al. 1989).  
However, in the next few years, the number of PWNe expected to be detected with the forthcoming Cherenkov Telescope Array (CTA, Actis et al 2011) will rise up to 300--500 (de O\~na Wilhelmi et al. 2013), providing an unprecedented database to study the fraction of the pulsar energy that is transferred to the particles, or the magnetic field in the nebula, or what rules the injection power in the surroundings of the pulsar. 

The main features of the Crab Nebula non-thermal emission, extending over 21 decades of frequencies, has been satisfactorily described by the formation of a PWN based, to a large extent, on a simple magneto-hydrodynamic (MHD) model for the interaction of a cold ultra-relativistic electron-positron wind with the interstellar medium (Kennel \& Coroniti 1984). 
Recent, more detailed two-dimensional MHD simulations have further into such a concept (Bogovalov et al. 2005, Volpi et al. 2008). { Here, we use a free expansion model for the nebulae (see Appendix) and we do not consider systems in the reverberation phase and beyond (see e.g., Gelfand et al. 2009). }

The $\gamma$-ray luminosity detected, which is believed to be the result of Comptonization of soft photon fields by relativistic electrons injected by the pulsar during its lifetime, has prompted the development of time-dependent models (e.g., Tanaka \& Takahara 2011, Bucciantini et al. 2011, Mart\'in et al. 2012). The latter have been used to study some of the members of the PWNe population but a systematic study is still lacking. However,  a few basic questions still remain:
Why is Crab the only PWN that is self-synchrotron (SSC)-dominated?
Why are the PWNe that we see at TeV energies particle dominated? Is there any observational bias behind this fact? At which sensitivity do we expect to map the whole phase space between particle and magnetic dominated nebula? What defines TeV observability of PWNe?
This work addresses these questions by making a phase-space exploration of Crab-like PWNe-models.

%%%%%%%%%%%%%%%%%%%%%%%%%%%%%%%%%%%%%%%%%%%%%%%%%%
\section{Method}
%%%%%%%%%%%%%%%%%%%%%%%%%%%%%%%%%%%%%%%%%%%%%%%%%%

\begin{figure}
\begin{center}
\includegraphics[scale=0.495]{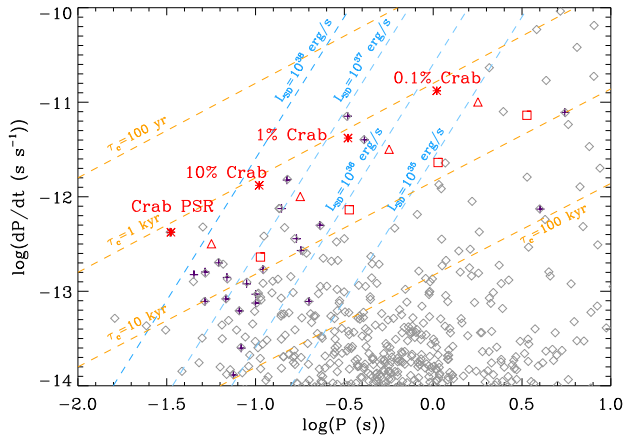} 
\end{center}
\caption{$P\dot P$ diagram of ATNF pulsars (in grey), together with the TeV detected PWNe (in violet),
and the 4 fake pulsars adopted for this study (in red). The latter are shown at three different ages. See the text for a discussion. }
\label{PPdot}
\end{figure}

To investigate the behavior of a Crab-like PWNe with different parameters, we generated a set of fake { (i.e. synthetic)} PWNe-models using the Crab Nebula as starting scaling. The Crab Nebula model we adopt was presented in Mart\'in et al. (2012) 
and Torres et al. (2013). At the Crab's age (taken as 940 years, to account for the non-simultaneity of the data points often used in its SED, which are actually obtained along 40 years), the set of parameters given in Table \ref{TableAppendix} yields to a perfect fit to the observational data of its nebula, from radio to gamma-rays.

We consider 4 different intrinsic luminosities with respect to the Crab one ($L_0 =\{1,0.1,0.01,0.001\} \times L_{0, {\rm Crab}}$). Additionally, we require $\tau_0$ and $\tau_c$ to be the same as those of Crab, as well as we take the same moment of inertia and braking index. All pulsars which have a braking index measurement show an $n$-value smaller than 3 (Espinoza et al. 2011), like Crab.
These requirements
sets the properties of the fake pulsars, as well as define that all of them are young. We use Eq. (\ref{tau0}) and the definition of $\tau_c$ (see Appendix) to derive, e.g., $\dot P$ as a function of $P$,
and Eq. (\ref{Ldet}) to derive $P$ as a function of the chosen $L_0$. The definition of $n$ can then be used to define $\ddot P$.
Using this approach, and an initial spin-down power equaling that of Crab, 10\%, 1\%, and 0.1\% of the latter, we have defined the properties of  4 fake pulsars, which we show in Table \ref{TableAppendix2} at an age of 940 years (the first row in that Table are Crab's observational data). Their position in the $P\dot P$-diagram is shown in Fig. \ref{PPdot}.
{ These 4 simulated pulsars cover a wide range of young systems putatively powering a nebula}, from the powerful Crab, to a magnetar-like case with 0.1\% of its power (i.e. like PSR J1550-5418). The two intermediate cases, with luminosities of 10\% and 1\% of Crab are similar to, e.g., PSR J1124-5916 or J1930+1852, and J1119-6127, respectively.

\begin{table}
\centering
\scriptsize
\caption{Physical magnitudes for the Crab Nebula model. }
\vspace{0.2cm}
\begin{tabular}{llll}
\hline
Magnitude & Symbol & Value\\
\hline
Moment of inertia (g cm$^2$) & $I$ & $10^{45}$\\
Breaking index & $n$ & 2.509\\
Initial spin-down age (yr) & $\tau_0$ & 730\\
Age (yr) & $t_{age}$ & 940 \\
Initial spin-down luminosity (erg/s) & $L_0$ & $3.1 \times 10^{39}$ \\
Spin-down luminosity at $t_{age}$ (erg/s) & $L(t_{age})$ & $4.5 \times 10^{38}$ \\
Distance (kpc) & $D$ & 2\\
Ejected mass ($M_\odot$) & $M_{ej}$ & 9.5\\
SN explosion energy (erg) & $E_0$ & $10^{51}$\\
PWN radius today (pc) & $R_{PWN}(t_{age})$ & 2.1 \\
\hline
%CMB temperature (K) & $T_{CMB}$ & 2.73\\
%CMB energy density (eV/cm$^3$) & $w_{CMB}$ & 0.25\\
FIR temperature (K) & $T_{FIR}$ & 70\\
FIR energy density (eV/cm$^3$) & $w_{FIR}$ & 0.5\\
NIR temperature (K) & $T_{NIR}$ & 5000\\
NIR energy density (eV/cm$^3$) & $w_{NIR}$ & 1\\
%ISM density (cm$^{-3}$) & $n_H$ & 1\\
\hline
%Minimum energy at injection & $\gamma_{min}$ & $1$\\
Break energy & $\gamma_b$ & $7 \times 10^5$\\
Low energy index & $\alpha_1$ & 1.5\\
High energy index & $\alpha_2$ & 2.5\\
Max. energy at injection at $t_{age}$ &  $\gamma_{max}(t_{age})$ & $1.2 \times 10^{10}$  \\
Shock radius fraction & $\varepsilon$ & $1/3$\\
Magnetic field ($\mu$G) & $B(t_{age})$ & 84.2 \\
Magnetic fraction  & $\eta$ & 0.03 \\ 
\hline
\hline
\label{TableAppendix}
\end{tabular}
\end{table}

\begin{table}
\centering
\scriptsize
\caption{Properties of the fake pulsars considered for the study, at an age of 940 years.}
\vspace{0.2cm}
\begin{tabular}{lll}
\hline
$L_0$ & P & $\dot P$ \\ % & $\ddot P$ \\
  ($L_{0,Crab}$)          & (s) & (s s$^{-1}$) \\ % & s s$^{-2}$ \\
\hline
1& 0.0334  & $4.2 \times 10^{-13}$ \\ % & 1.33e-23  \\
$ 0.1$& 0.1048  &$1.3 \times 10^{-12}$  \\ % & 4.18e-23  \\
$ 0.01$ & 0.3314  & $4.2\times 10^{-12}$  \\ % & 1.32e-22  \\
$ 0.001$ & 1.0479  & $1.3\times 10^{-11}$ \\ % & 4.18e-22  \\
\hline
\hline
\label{TableAppendix2}
\end{tabular}
\end{table}

The spectral energy distribution (SED) of the nebulae (consisting of synchrotron and inverse Compton (IC) radiation) is determined by several factors, including the magnetic field strength, the age of the system, and the background photon fields. The Appendix  briefly discusses the underlying time-dependent  model.
To account for the PWNe phase space, we considered 8 values of magnetic fraction $\eta$ (0.001, 0.01, 0.03, 0.1, 0.5, 0.9, 0.99, and 0.999, from fully particle dominated to fully magnetically dominated nebulae), and 3 distinctive ages: 940, 3000, and 9000 years, in addition of the 4 values of $L_0$. Therefore, the explored phase space of PWNe models is constructed by $8\times 4 \times 3=96$ cases.
The supernova (SN) explosion energy is fixed in our models, as is the ejected mass, the injection parameters, and the environmental variables. 
It should be noted that these assumptions (particularly, to assume the same injection or environmental parameters than those in the Crab Nebula) will not necessarily reflect the reality of a particular PWN (below we present a discussion on how the photon field and injection spectrum would affect the results). Here, we are not looking for fits of the multi-wavelength emission of a particular source, but rather searching for common trends in the phase-space of PWN models. For the study we are doing, maintaining these parameters fixed is essential to shed light on the behavior of the generated luminosities and SEDs as a function of the initial spin-down power and the magnetic fraction.

For instance, the contribution of the IC yield on the far infra-red (FIR) background ($T \sim 70$ K) would increase with respect to the cosmic microwave background (CMB) if we consider a steeper spectrum of electrons than the one in the Crab Nebula. 
In such a case, more electrons  (with energies of a few TeV)  are able to generate TeV photons via interacting with the FIR background, increasing its contributions relatively to the one from the CMB (Aharonian et al. 1997). 
On the contrary, the optical / near infra-red (NIR) background ($T \sim 5000$ K) hardly plays any role. Assuming the Thompson limit, the IC emissivity is $q(E) \propto w T^{(\alpha-3)/2} E^{-(\alpha+1)/2}$, where $w$ is the energy density and $T$ is the temperature of the photon background, and $\alpha$ is the slope of the electron distribution (Blumenthal \& Gould 1970).
It is possible to compare the two contributions by estimating the ratio 
$q_1 / q_2 = (w_1 / w_2) (T_1/T_2)^{(\alpha-3)/2}$. 
{ Supposing $\alpha \sim 1.5$,  then the ratio between the IC contribution 
of NIR ($T \sim$ 5000 K) to FIR ($T \sim$ 80 K) is about ~0.045 (for equal energy densities; and as per the quoted formula).} 
In addition, the Klein-Nishina effect, operative for VHE
production with NIR photons, would reduce the IC-NIR yield significantly. If we use the ratio above to compare the IC-FIR with
the IC-CMB yield, the contribution of dust with $T \sim 70$ K interacting with electrons distributed with a slope of $\alpha =2.5$
will be similar to that of the CMB when $w_{FIR} \sim 2.2 w_{CMB} \sim 0.5$ eV cm$^{-3}$. 
The fixed values of the photon backgrounds on our Crab-like models are close to the Galactic averages (see Porter et al. 2006) and should be enough to gather general trends, which is the aim of this exercise. 
Finally, the distance to Crab is taken as fiducial. Given the relatively short distance to the Crab Nebula ($D=2$ kpc), the conclusions reached on the lack of detectable TeV emission for some configurations will hold for pulsars located  farther away. In any case, we provide both, luminosities and fluxes, when showing SEDs.

%%%%%%%%%%%%%%%%%%%%%%%%%%%%%%%%%%%%%%%%%%%%%%%%%%
\section{Results}
%%%%%%%%%%%%%%%%%%%%%%%%%%%%%%%%%%%%%%%%%%%%%%%%%%

\begin{table*}
\centering
\scriptsize
\caption{Physical magnitudes for the fake PWNe sets. The symbol ``\ldots\ " 
stands for the same value
shown in the column to the left.}
\vspace{0.2cm}
\begin{tabular}{l l l l l l l l }
\hline
\hline
 {\red {$L_0=L_{0,Crab}$=$3.1 \times 10^{39}$ erg s$^{-1}$}} \\
\hline
\hline
Magnitude & Symbol & {\blue {$\eta$=0.001}} & {\blue { $\eta$=0.03}} & {\blue {$\eta$=0.1}} & {\blue {$\eta$=0.5}} & {\blue {$\eta$=0.9}} & {\blue {$\eta$=0.999}}  \\
\hline
Age (yr) & $t_{age}$ & 940\\
Spin-down luminosity at $t_{age}$ (erg/s) & $L(t_{age})$ & $4.5 \times 10^{38}$ & \ldots & \ldots & \ldots & \ldots & \ldots\\
Maximum energy at injection at $t_{age}$ &  $\gamma_{max}(t_{age})$ & $2.3 \times 10^9$ & $1.2 \times 10^{10}$ & $2.3 \times 10^{10}$ & $5.1 \times 10^{10}$ & $6.8 \times 10^{10}$ & $7.2 \times 10^{10}$\\
Magnetic field ($\mu$G) & $B(t_{age})$ & 15.4 & 84.2 & 153.8 & 343.8 & 461.3 & 486.0\\
PWN radius today (pc) & $R_{PWN}(t_{age})$ & 2.1 & \ldots & \ldots & \ldots & \ldots & \ldots\\
\hline
Age (yr) & $t_{age}$ & 3000\\
%Initial spin-down luminosity (erg/s) & $L_0$ & $3.1 \times 10^{38}$ & \ldots & \ldots & \ldots & \ldots & \ldots\\
Spin-down luminosity at $t_{age}$ (erg/s) & $L(t_{age})$ & $7.0 \times 10^{37}$ & \ldots & \ldots & \ldots & \ldots & \ldots\\
Maximum energy at injection at $t_{age}$ &  $\gamma_{max}(t_{age})$ & $9.0 \times 10^8$ & $4.9 \times 10^9$ & $9.0 \times 10^9$ & $2.0 \times 10^{10}$ & $2.7 \times 10^{10}$ & $2.8 \times 10^{10}$\\
Magnetic field ($\mu$G) & $B(t_{age})$ & 1.7 & 9.3 & 17.0 & 38.0 & 50.9 & 53.7\\
PWN radius today (pc) & $R_{PWN}(t_{age})$ & 8.5 & \ldots & \ldots & \ldots & \ldots & \ldots\\
\hline
Age (yr) & $t_{age}$ & 9000\\
%Initial spin-down luminosity (erg/s) & $L_0$ & $3.1 \times 10^{38}$ & \ldots & \ldots & \ldots & \ldots & \ldots\\
Spin-down luminosity at $t_{age}$ (erg/s) & $L(t_{age})$ & $7.5 \times 10^{36}$ & \ldots & \ldots & \ldots & \ldots & \ldots\\
Maximum energy at injection at $t_{age}$ &  $\gamma_{max}(t_{age})$ & $2.9 \times 10^8$ & $1.6 \times 10^9$ & $2.9 \times 10^9$ & $6.6 \times 10^9$ & $8.8 \times 10^9$ & $9.3 \times 10^9$\\
Magnetic field ($\mu$G) & $B(t_{age})$ & 0.2 & 0.9 & 1.7 & 3.8 & 5.1 & 5.4\\
PWN radius today (pc) & $R_{PWN}(t_{age})$ & 31.6 & \ldots & \ldots & \ldots & \ldots & \ldots\\
\hline
\hline
 {\red {$L_0= 0.1 \, L_{0,Crab}$=$3.1 \times 10^{38}$ erg s$^{-1}$ }} \\
\hline
\hline
Age (yr) & $t_{age}$ & 940\\
Spin-down luminosity at $t_{age}$ (erg/s) & $L(t_{age})$ & $4.5 \times 10^{37}$ & \ldots & \ldots & \ldots & \ldots & \ldots\\
Maximum energy at injection at $t_{age}$ &  $\gamma_{max}(t_{age})$ & $7.2 \times 10^8$ & $3.9 \times 10^9$ & $7.2 \times 10^9$ & $1.6 \times 10^{10}$ & $2.2 \times 10^{10}$ & $2.3 \times 10^{10}$\\
Magnetic field ($\mu$G) & $B(t_{age})$ & 9.7 & 53.1 & 97.0 & 216.9 & 291.0 & 306.6\\
PWN radius today (pc) & $R_{PWN}(t_{age})$ & 1.3 & \ldots & \ldots & \ldots & \ldots & \ldots\\
\hline
Age (yr) & $t_{age}$ & 3000\\
%Initial spin-down luminosity (erg/s) & $L_0$ & $3.1 \times 10^{38}$ & \ldots & \ldots & \ldots & \ldots & \ldots\\
Spin-down luminosity at $t_{age}$ (erg/s) & $L(t_{age})$ & $7.0 \times 10^{36}$ & \ldots & \ldots & \ldots & \ldots & \ldots\\
Maximum energy at injection at $t_{age}$ &  $\gamma_{max}(t_{age})$ & $2.8 \times 10^8$ & $1.6 \times 10^9$ & $2.8 \times 10^9$ & $6.3 \times 10^9$ & $8.5 \times 10^9$ & $9.0 \times 10^9$\\
Magnetic field ($\mu$G) & $B(t_{age})$ & 1.1 & 5.9 & 10.7 & 23.9 & 32.1 & 33.9\\
PWN radius today (pc) & $R_{PWN}(t_{age})$ & 5.3 & \ldots & \ldots & \ldots & \ldots & \ldots\\
\hline
Age (yr) & $t_{age}$ & 9000\\
%Initial spin-down luminosity (erg/s) & $L_0$ & $3.1 \times 10^{38}$ & \ldots & \ldots & \ldots & \ldots & \ldots\\
Spin-down luminosity at $t_{age}$ (erg/s) & $L(t_{age})$ & $7.5 \times 10^{35}$ & \ldots & \ldots & \ldots & \ldots & \ldots\\
Maximum energy at injection at $t_{age}$ &  $\gamma_{max}(t_{age})$ & $9.3 \times 10^7$ & $5.1 \times 10^8$ & $9.3 \times 10^8$ & $2.1 \times 10^9$ & $2.8 \times 10^9$ & $2.9 \times 10^9$\\
Magnetic field ($\mu$G) & $B(t_{age})$ & 0.1 & 0.6 & 1.1 & 2.4 & 3.2 & 3.4\\
PWN radius today (pc) & $R_{PWN}(t_{age})$ & 19.9 & \ldots & \ldots & \ldots & \ldots & \ldots\\
\hline
\hline
 {\red {$L_0= 0.01 \, L_{0,Crab}$=$3.1 \times 10^{37}$ erg s$^{-1}$}} \\
\hline
\hline
Age (yr) & $t_{age}$ & 940\\
Spin-down luminosity at $t_{age}$ (erg/s) & $L(t_{age})$ & $4.5 \times 10^{36}$ & \ldots & \ldots & \ldots & \ldots & \ldots\\
Maximum energy at injection at $t_{age}$ &  $\gamma_{max}(t_{age})$ & $2.3 \times 10^8$ & $1.2 \times 10^9$ & $2.3 \times 10^9$ & $5.1 \times 10^9$ & $6.8 \times 10^9$ & $7.2 \times 10^9$\\
Magnetic field ($\mu$G) & $B(t_{age})$ & 6.1 & 33.5 & 61.2 & 136.9 & 183.6 & 193.5\\
PWN radius today (pc) & $R_{PWN}(t_{age})$ & 0.8 & \ldots & \ldots & \ldots\\
\hline
Age (yr) & $t_{age}$ & 3000\\
%Initial spin-down luminosity (erg/s) & $L_0$ & $3.1 \times 10^{38}$ & \ldots & \ldots & \ldots & \ldots & \ldots\\
Spin-down luminosity at $t_{age}$ (erg/s) & $L(t_{age})$ & $7.0 \times 10^{35}$ & \ldots & \ldots & \ldots & \ldots & \ldots\\
Maximum energy at injection at $t_{age}$ &  $\gamma_{max}(t_{age})$ & $9.0 \times 10^7$ & $4.9 \times 10^8$ & $9.0 \times 10^8$ & $2.0 \times 10^9$ & $2.7 \times 10^9$ & $2.8 \times 10^9$\\
Magnetic field ($\mu$G) & $B(t_{age})$ & 0.7 & 3.7 & 6.8 & 15.1 & 20.3 & 21.4\\
PWN radius today (pc) & $R_{PWN}(t_{age})$ & 3.4 & \ldots & \ldots & \ldots & \ldots & \ldots\\
\hline
Age (yr) & $t_{age}$ & 9000\\
%Initial spin-down luminosity (erg/s) & $L_0$ & $3.1 \times 10^{38}$ & \ldots & \ldots & \ldots & \ldots & \ldots\\
Spin-down luminosity at $t_{age}$ (erg/s) & $L(t_{age})$ & $7.5 \times 10^{34}$ & \ldots & \ldots & \ldots & \ldots & \ldots\\
Maximum energy at injection at $t_{age}$ &  $\gamma_{max}(t_{age})$ & $2.9 \times 10^7$ & $1.6 \times 10^8$ & $2.9 \times 10^8$ & $6.7 \times 10^8$ & $8.8 \times 10^8$ & $9.3 \times 10^8$\\
Magnetic field ($\mu$G) & $B(t_{age})$ & 0.07 & 0.4 & 0.7 & 1.5 & 2.0 & 2.1\\
PWN radius today (pc) & $R_{PWN}(t_{age})$ & 12.6 & \ldots & \ldots & \ldots & \ldots & \ldots\\
\hline
\hline
 {\red {$L_0= 0.001 \, L_{0,Crab}$=$3.1 \times 10^{36}$ erg s$^{-1}$}} \\
\hline
\hline
Age (yr) & $t_{age}$ & 940\\
Spin-down luminosity at $t_{age}$ (erg/s) & $L(t_{age})$ & $4.5 \times 10^{35}$ & \ldots & \ldots & \ldots & \ldots & \ldots\\
Maximum energy at injection at $t_{age}$ &  $\gamma_{max}(t_{age})$ & $7.2 \times 10^7$ & $3.9 \times 10^8$ & $7.2 \times 10^8$ & $1.6 \times 10^9$ & $2.2 \times 10^9$ & $2.3 \times 10^9$\\
Magnetic field ($\mu$G) & $B(t_{age})$ & 3.8 & 21.2 & 38.6 & 86.4 & 115.9 & 122.1\\
PWN radius today (pc) & $R_{PWN}(t_{age})$ & 0.5 & \ldots & \ldots & \ldots & \ldots & \ldots\\
\hline
Age (yr) & $t_{age}$ & 3000\\
%Initial spin-down luminosity (erg/s) & $L_0$ & $3.1 \times 10^{38}$ & \ldots & \ldots & \ldots & \ldots\\
Spin-down luminosity at $t_{age}$ (erg/s) & $L(t_{age})$ & $7.0 \times 10^{34}$ & \ldots & \ldots & \ldots & \ldots & \ldots\\
Maximum energy at injection at $t_{age}$ &  $\gamma_{max}(t_{age})$ & $2.8 \times 10^7$ & $1.6 \times 10^8$ & $2.8 \times 10^8$ & $6.3 \times 10^8$ & $8.5 \times 10^8$ & $9.0 \times 10^8$\\
Magnetic field ($\mu$G) & $B(t_{age})$ & 0.4 & 2.3 & 4.3 & 9.5 & 12.8 & 13.5\\
PWN radius today (pc) & $R_{PWN}(t_{age})$ & 2.1 & \ldots & \ldots & \ldots & \ldots & \ldots\\
\hline
Age (yr) & $t_{age}$ & 9000\\
%Initial spin-down luminosity (erg/s) & $L_0$ & $3.1 \times 10^{38}$ & \ldots & \ldots & \ldots & \ldots\\
Spin-down luminosity at $t_{age}$ (erg/s) & $L(t_{age})$ & $7.5 \times 10^{33}$ & \ldots & \ldots & \ldots & \ldots & \ldots\\
Maximum energy at injection at $t_{age}$ &  $\gamma_{max}(t_{age})$ & $9.3 \times 10^6$ & $5.1 \times 10^7$ & $9.3 \times 10^7$ & $2.0 \times 10^8$ & $2.8 \times 10^8$ & $2.9 \times 10^8$\\
Magnetic field ($\mu$G) & $B(t_{age})$ & 0.04 & 0.2 & 0.4 & 1.0 & 1.3 & 1.4\\
PWN radius today (pc) & $R_{PWN}(t_{age})$ & 7.9 & \ldots & \ldots & \ldots & \ldots & \ldots\\
\hline
\hline
\end{tabular}
\label{param}
\end{table*}

Table \ref{param} shows the results for the scaled models for the different parameters considered. The results varying the total luminosity from the largest to the smallest are listed from top to bottom, each one considering three evolutionary ages (940, 3000 and 9000 years), and different magnetic fractions, increasing from left to right. Table \ref{param} also quotes the intrinsic sizes of the simulated nebulae, magnetic fields, and maximum energy at the selected age, for the different models.
A systematic comparison among the different results will be done in the following.

%%%%%%%%%%%%%%%%%%%%%%%%%%%%%%%%%%%%%%%%%%%%%%%%%%
\subsection{IC contributions for different age and pulsar spin-down power}
%%%%%%%%%%%%%%%%%%%%%%%%%%%%%%%%%%%%%%%%%%%%%%%%%%

\begin{landscape}
\begin{figure}
\begin{center}
\includegraphics[scale=0.3245]{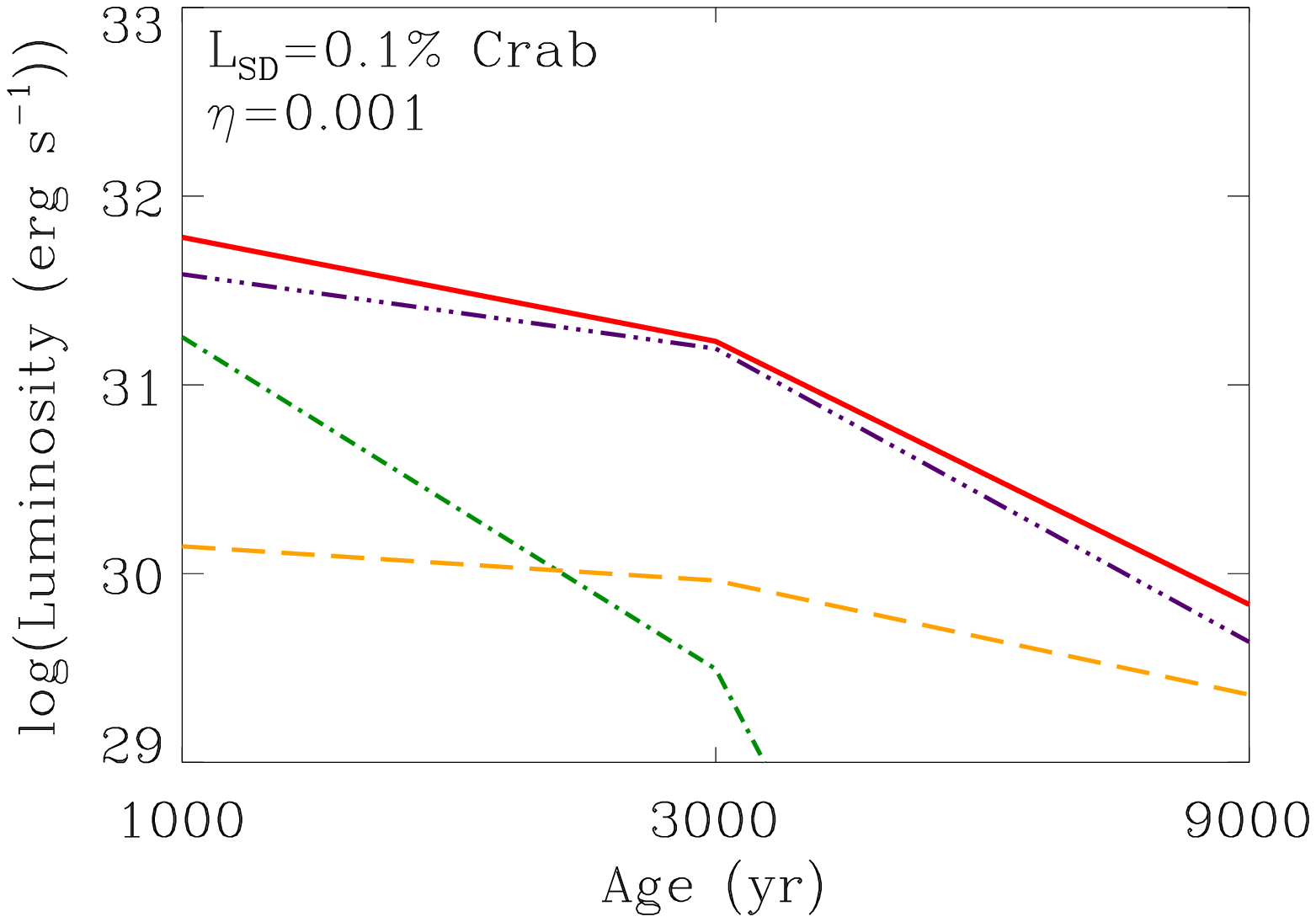}  \includegraphics[scale=0.3245]{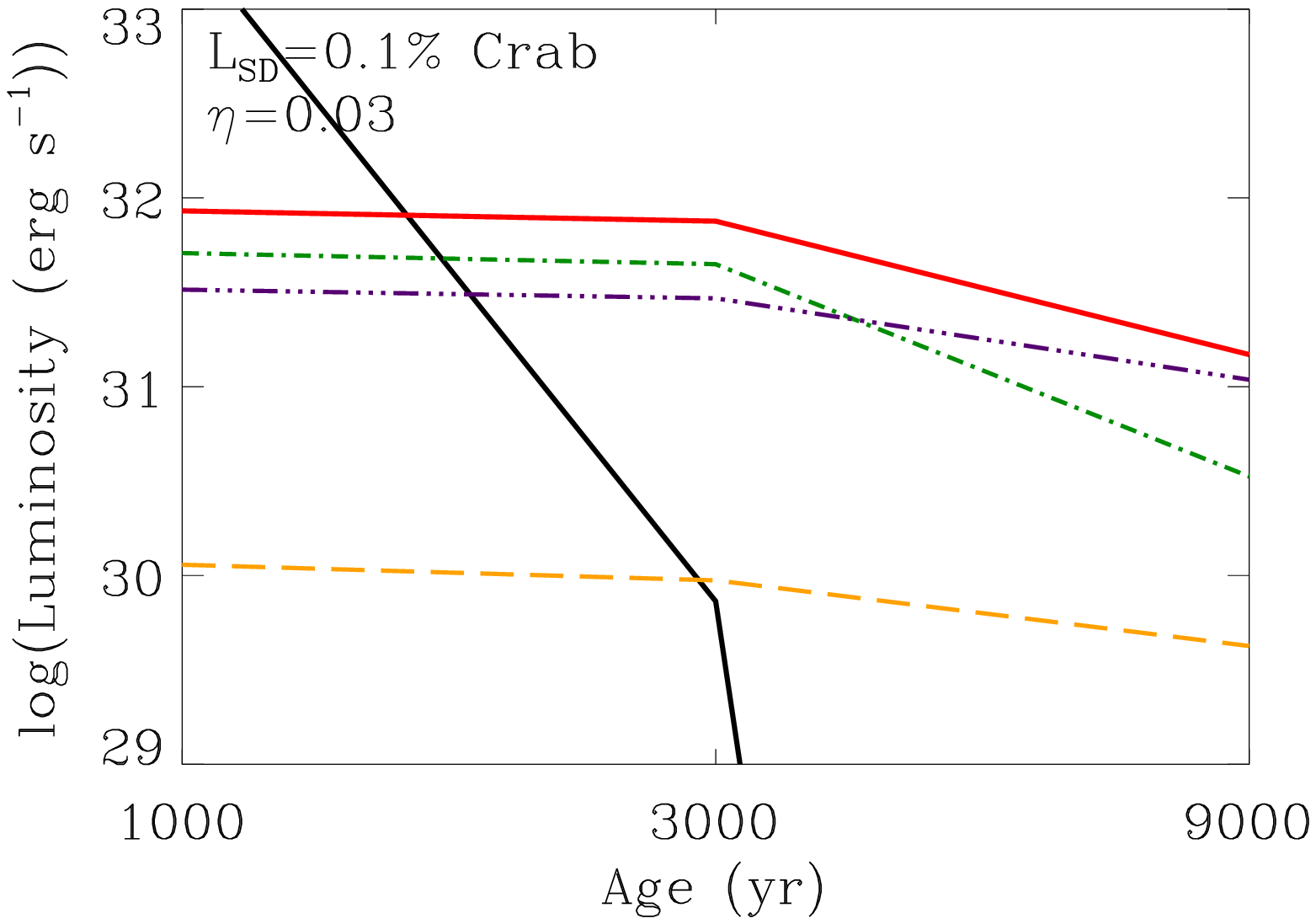} 
\includegraphics[scale=0.3245]{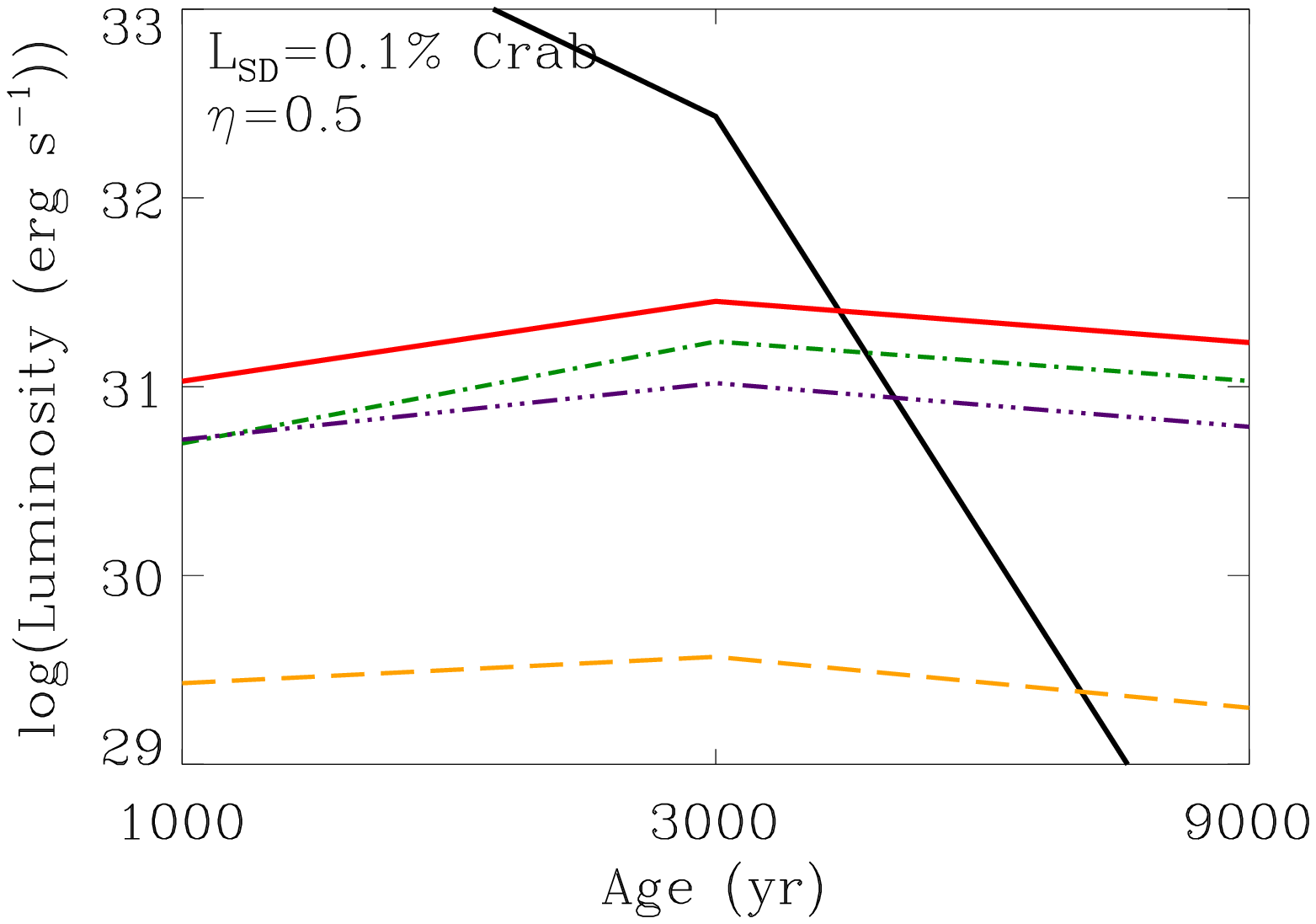}  \includegraphics[scale=0.3245]{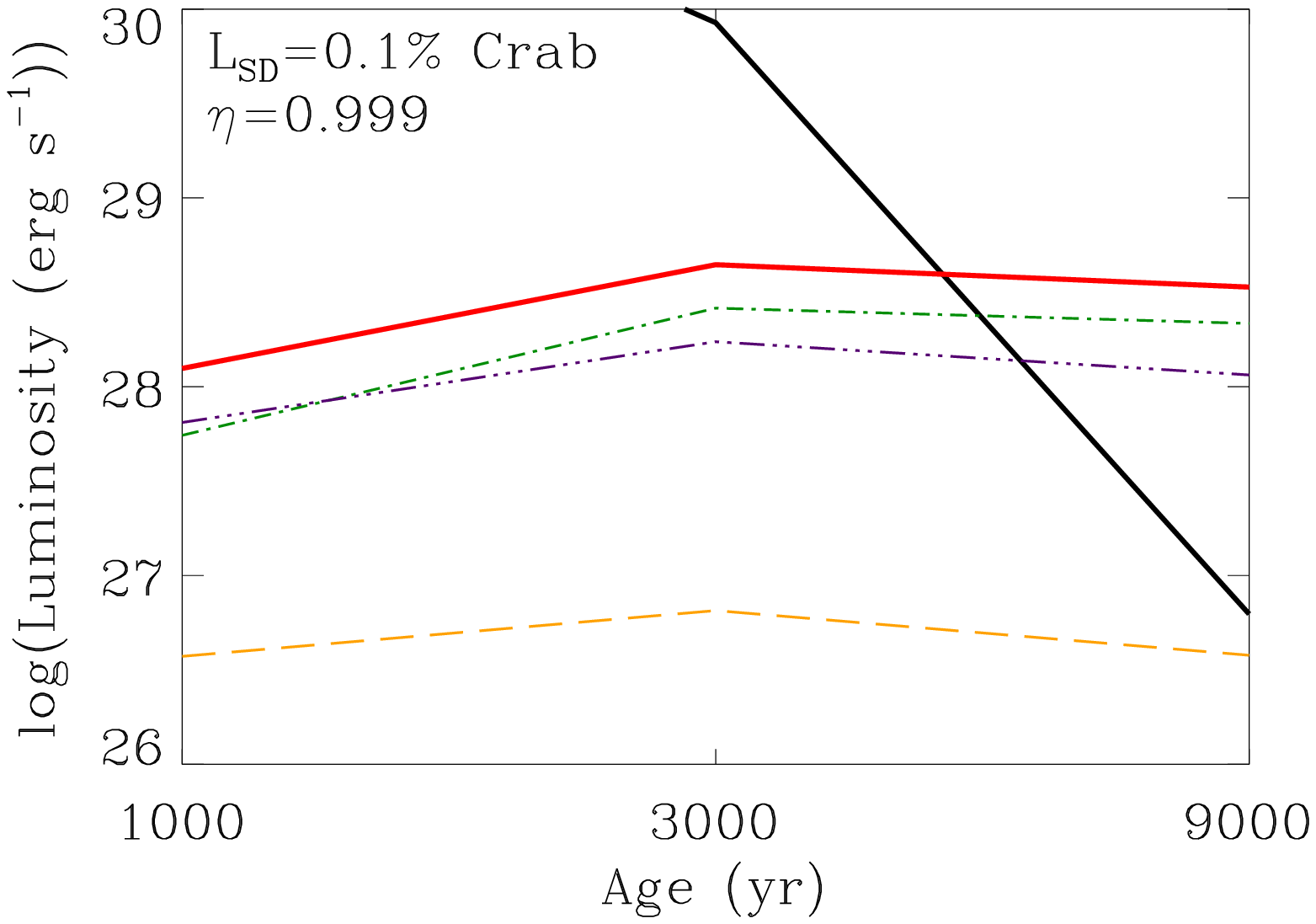} \\
\includegraphics[scale=0.3245]{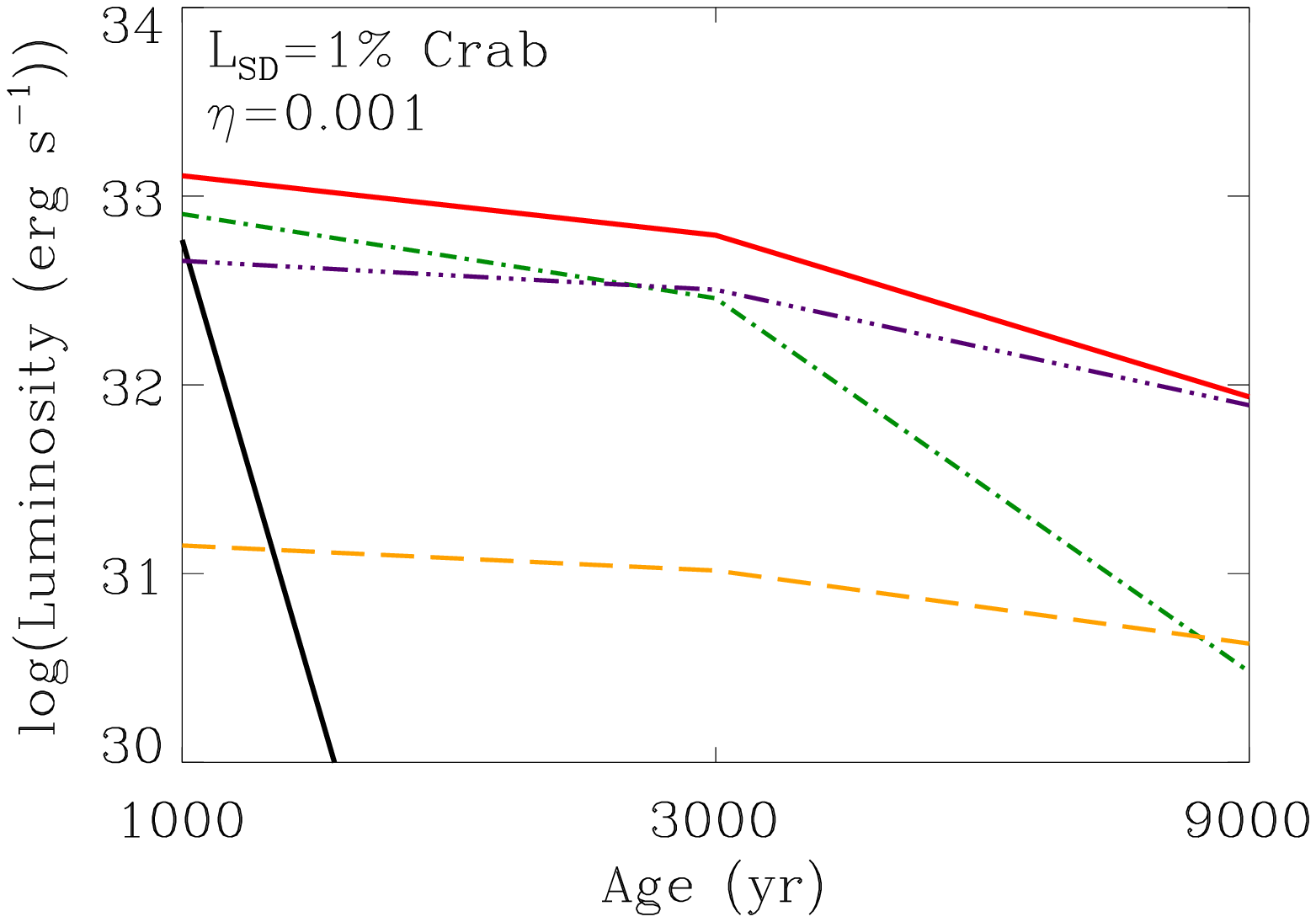}  \includegraphics[scale=0.3245]{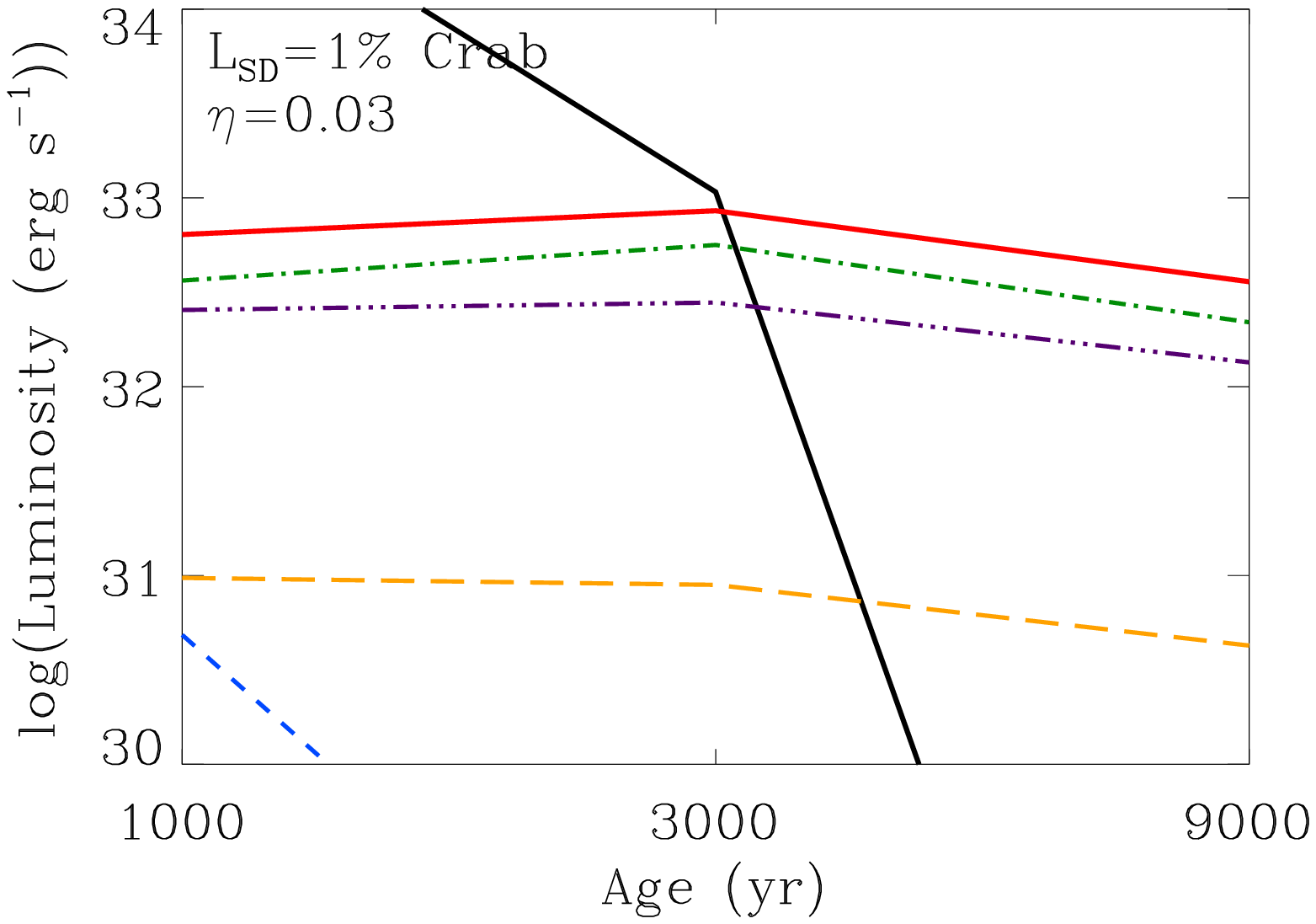} 
\includegraphics[scale=0.3245]{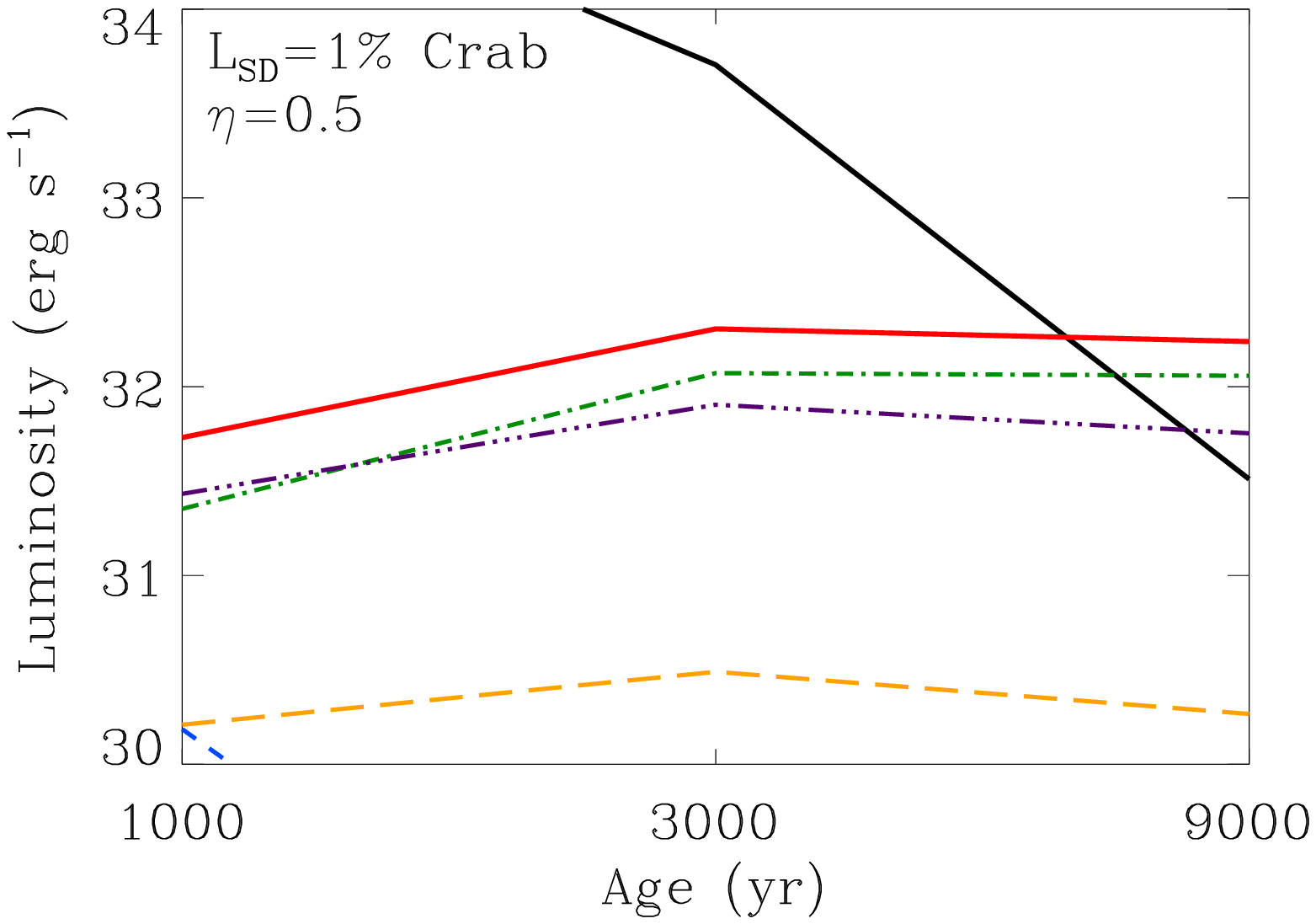}  \includegraphics[scale=0.3245]{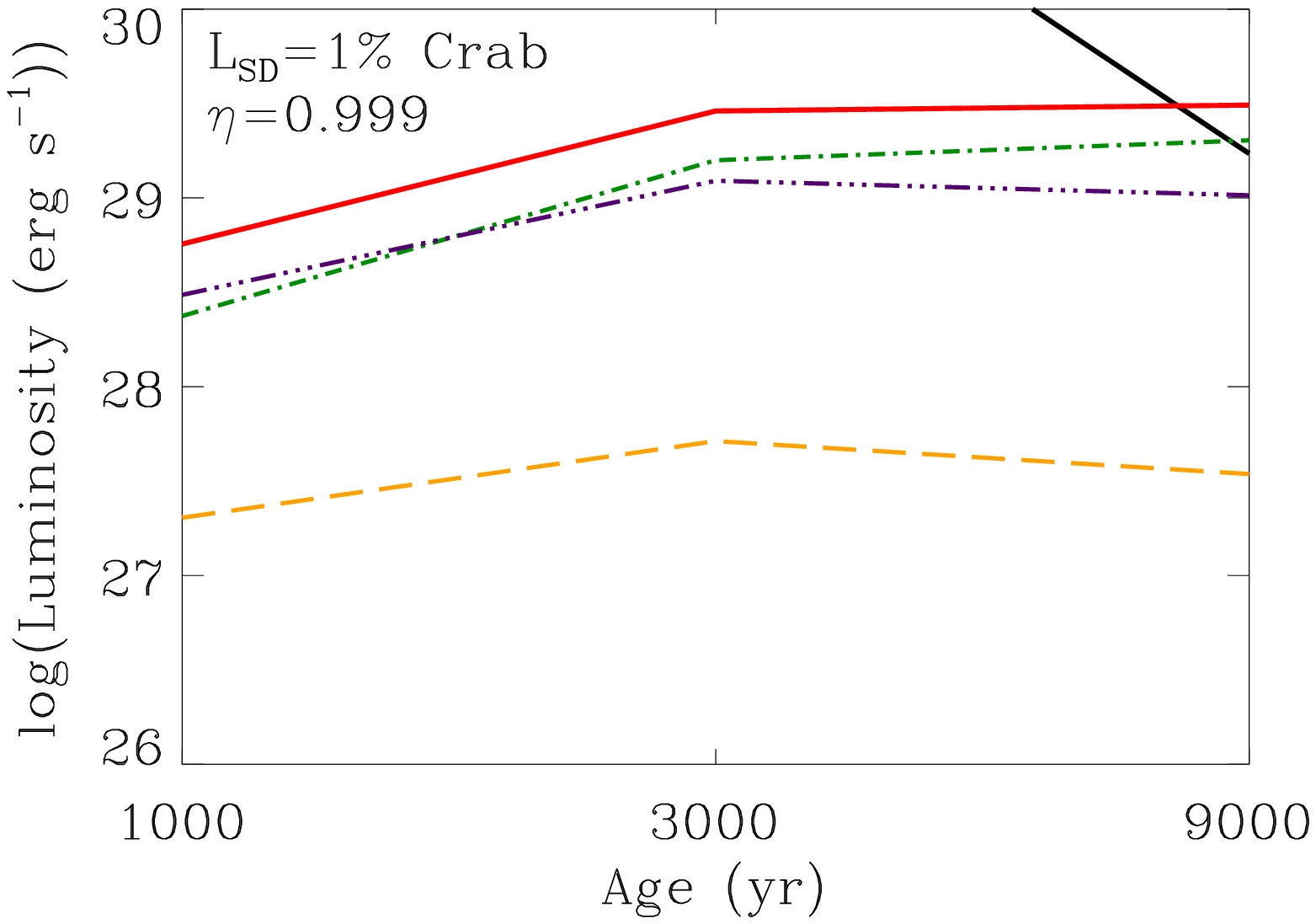}\\
\includegraphics[scale=0.3245]{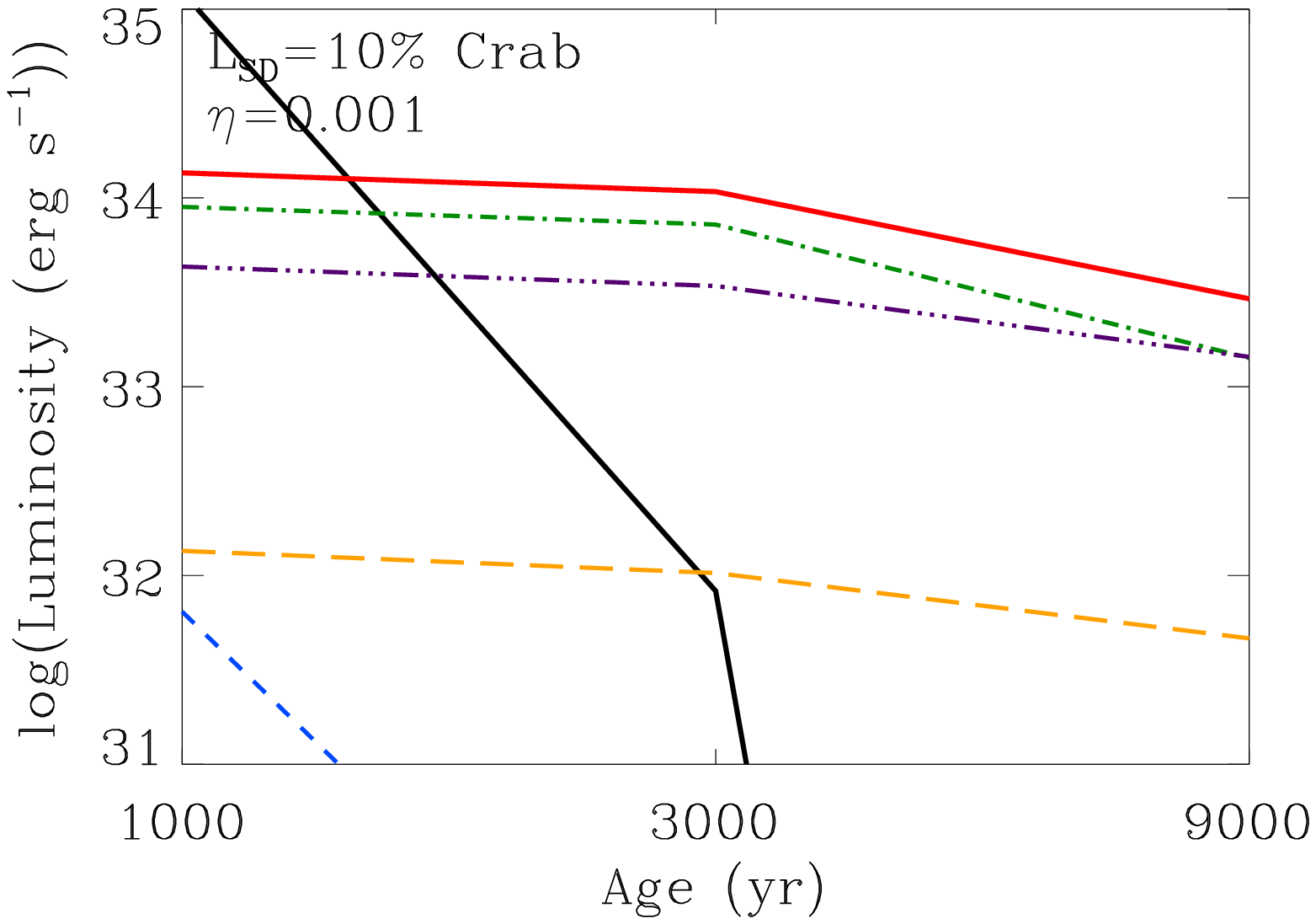}  \includegraphics[scale=0.3245]{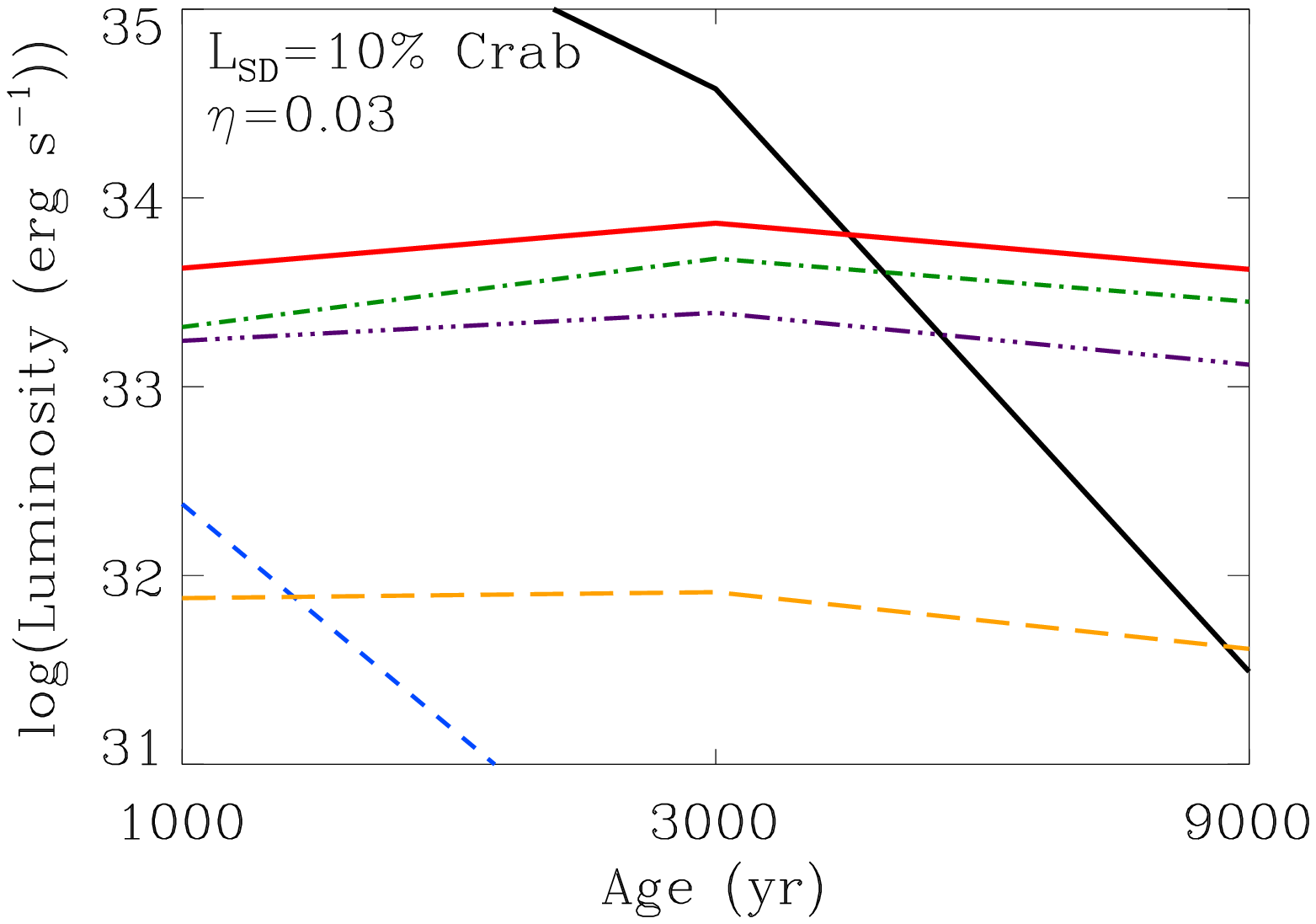} 
\includegraphics[scale=0.3245]{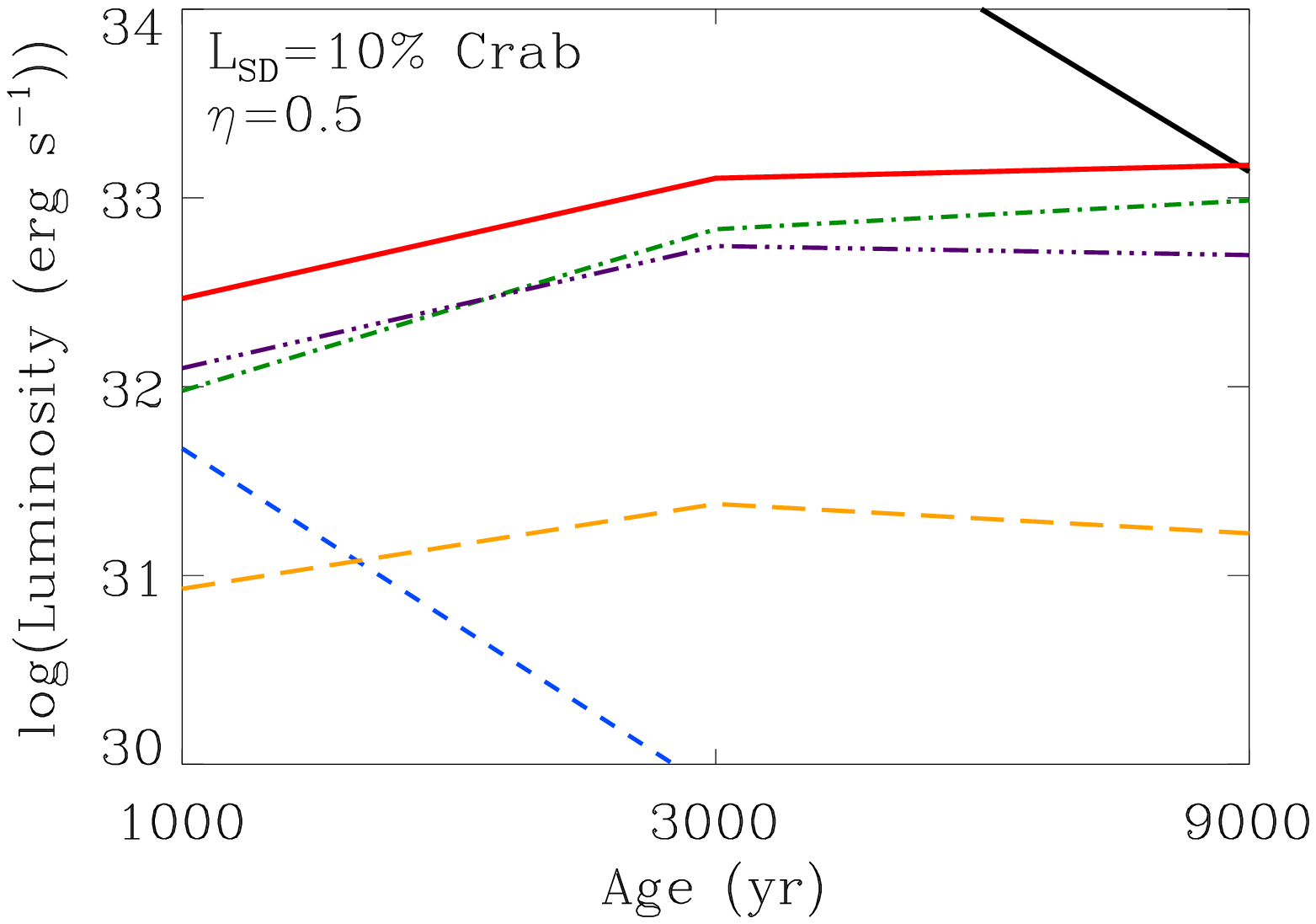}  \includegraphics[scale=0.3245]{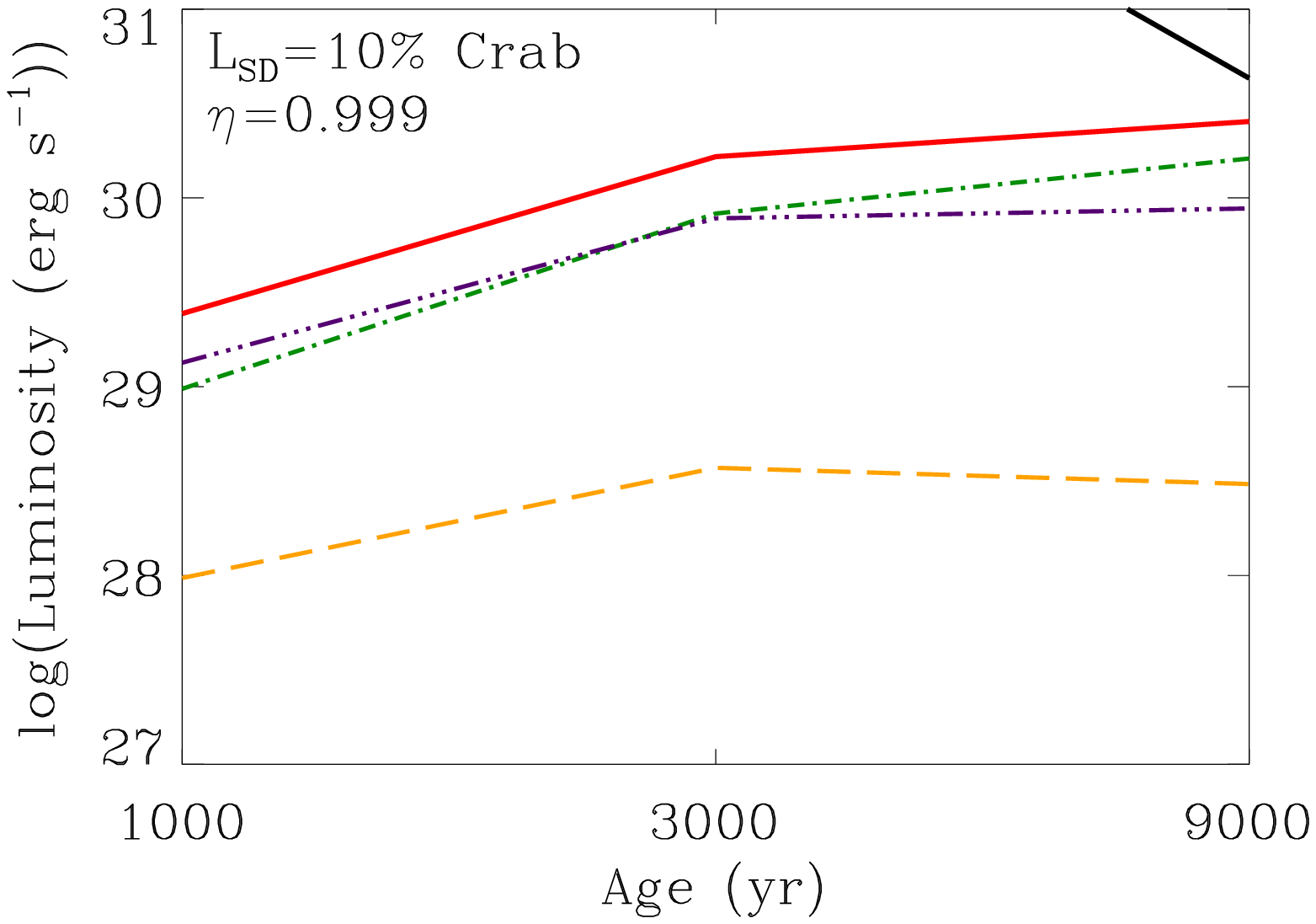}\\
\includegraphics[scale=0.3245]{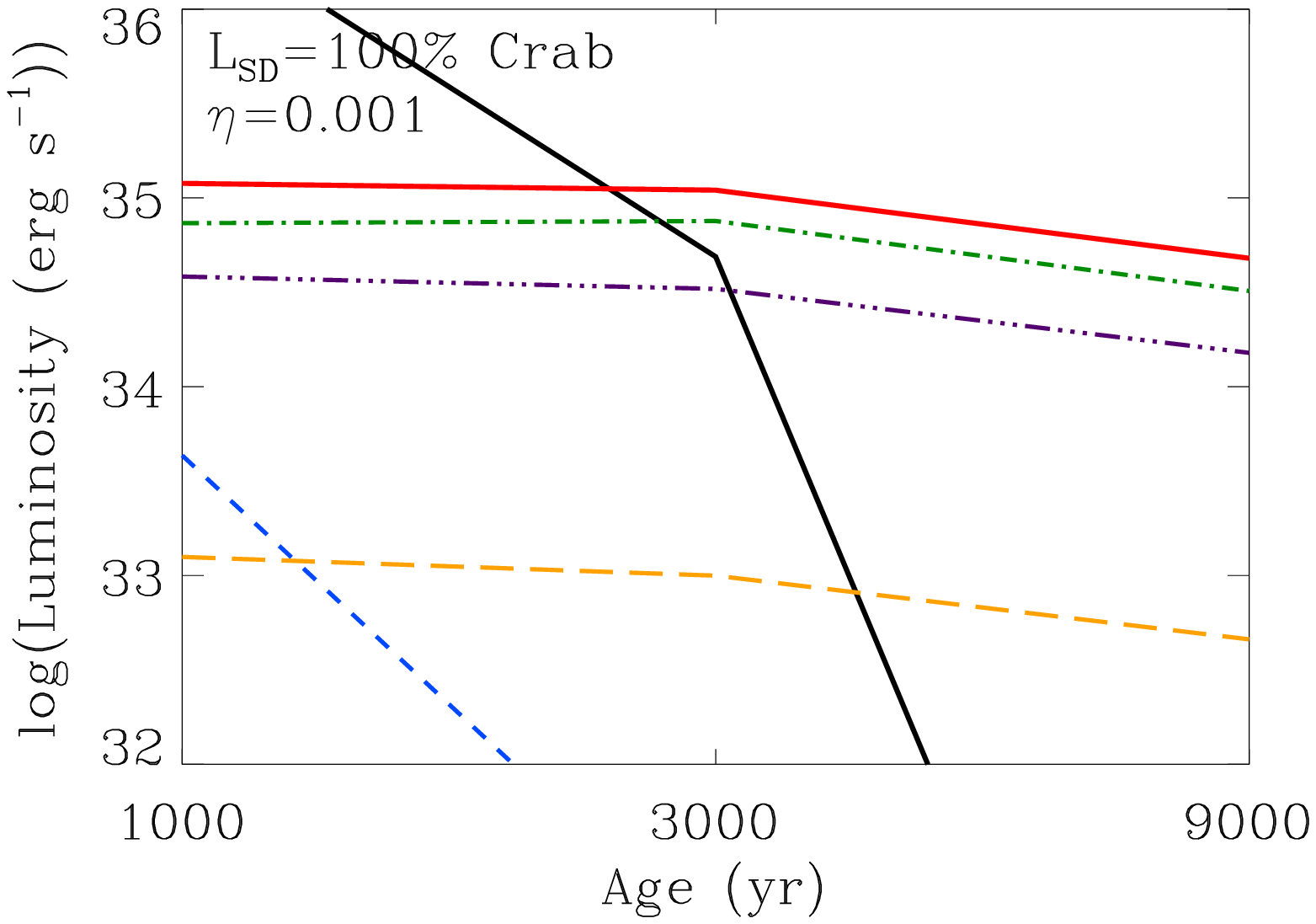}  \includegraphics[scale=0.3245]{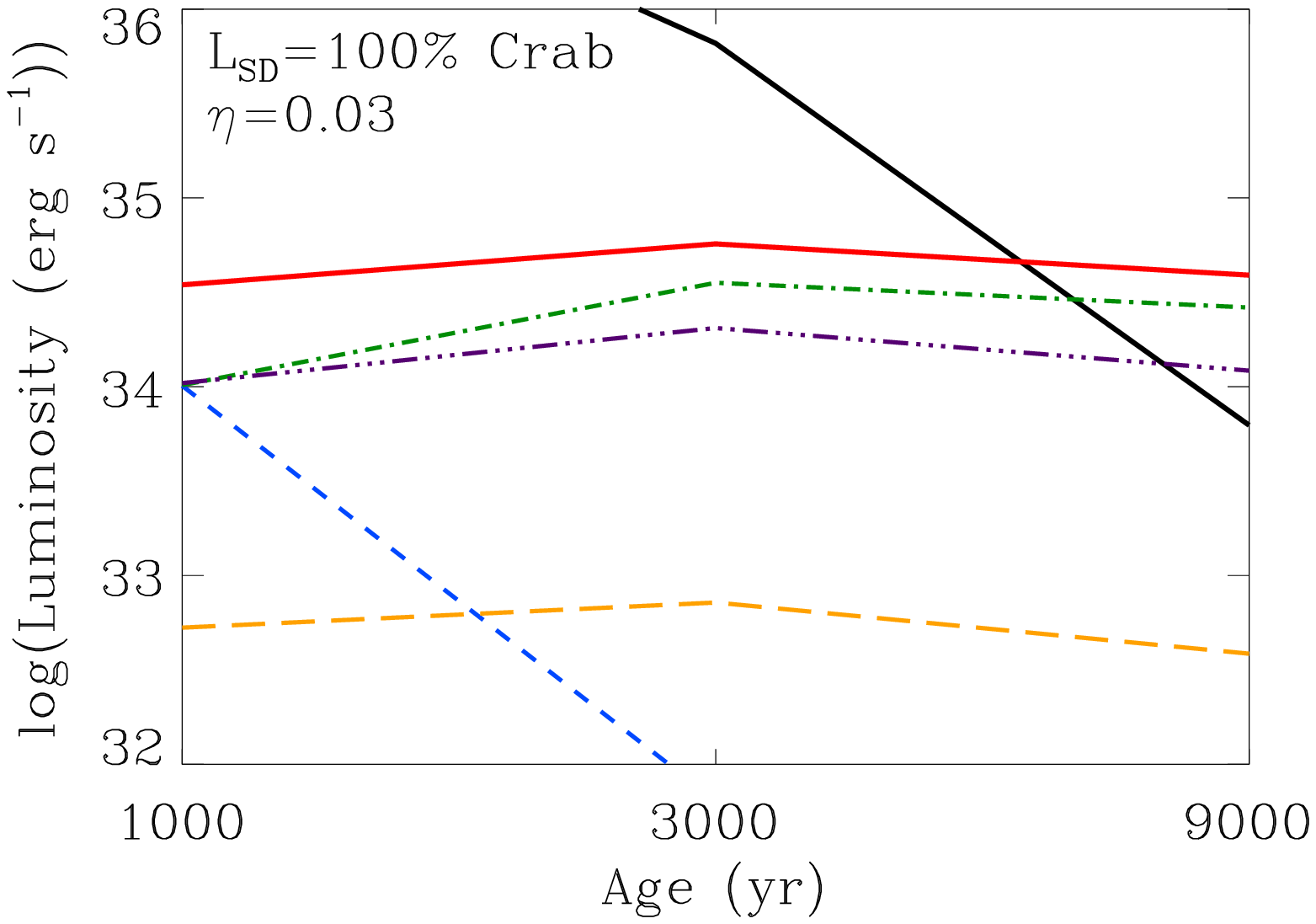} 
\includegraphics[scale=0.3245]{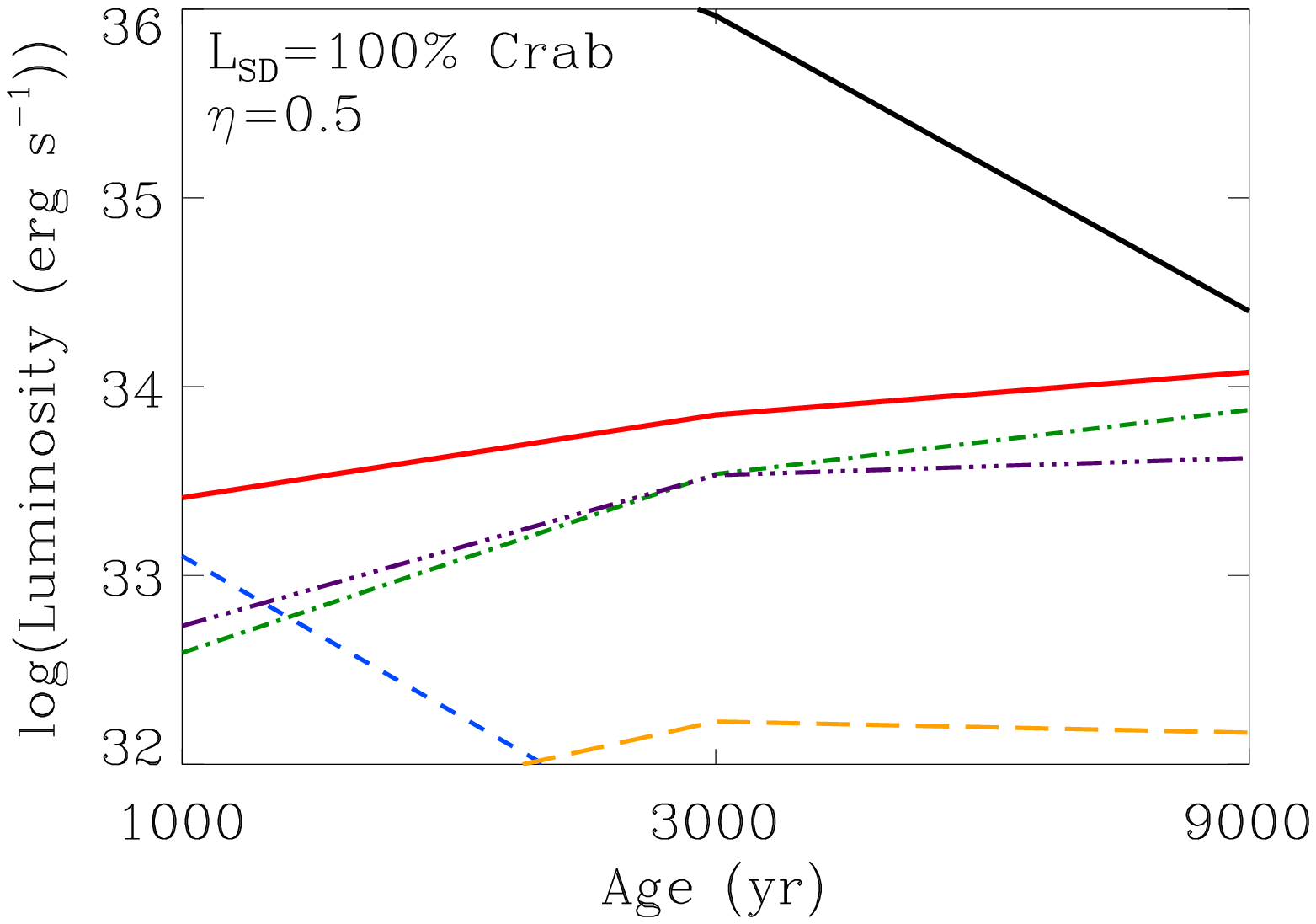}  \includegraphics[scale=0.3245]{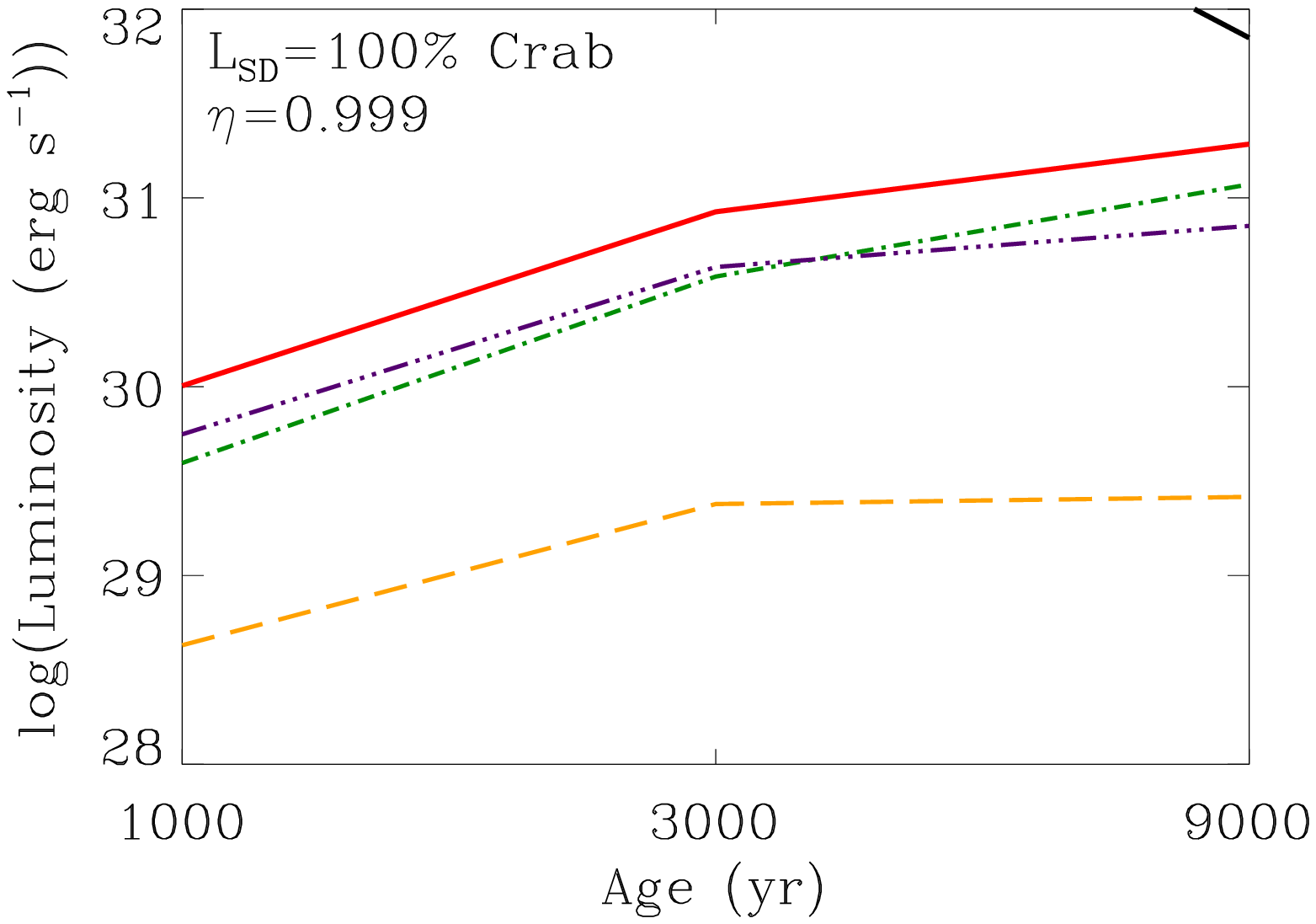}
\end{center}
\caption{Luminosities between 1 and 10 TeV of the IC contributions of the spectrum as a function of age. We fix $L_{SD}$=0.1, 1, 10, and 100\% of the 
Crab Nebula (from top to bottom) and a magnetic fraction of 0.001, 0.03, 0.5 \& 0.999 (from left to right). 
The black solid line is the synchrotron luminosity calculated between 1 and 
10 keV. The other components are: total IC (red solid line), IC-CMB (green dot-dashed line), IC-FIR (purple triple dotted-dashed line), 
IC-NIR (orange dashed line) and SSC (blue-dashed line).}
\label{lum-age}
\end{figure}
\end{landscape}

\begin{landscape}
\begin{figure}
\begin{center}
\includegraphics[scale=0.3245]{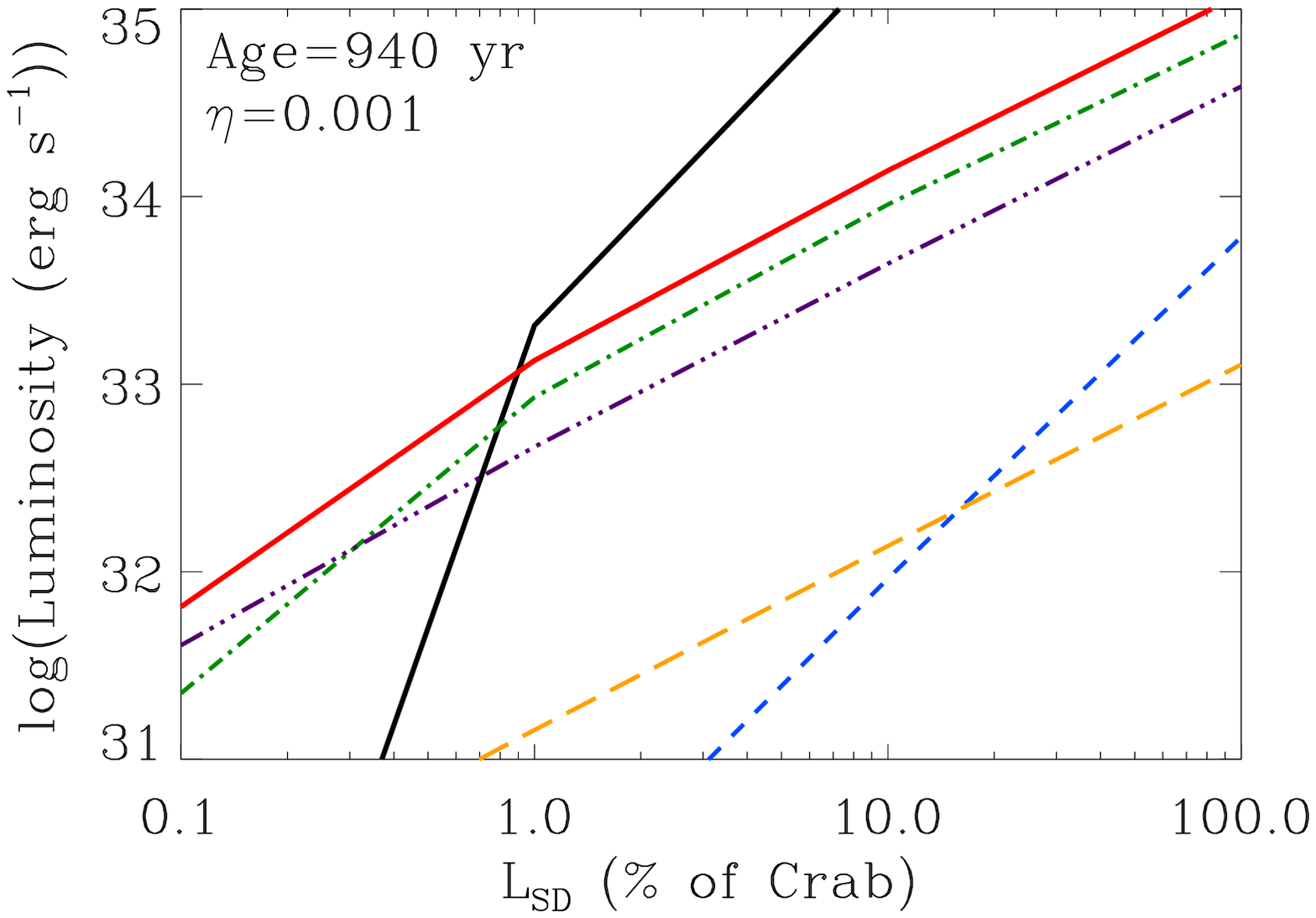}  \includegraphics[scale=0.3245]{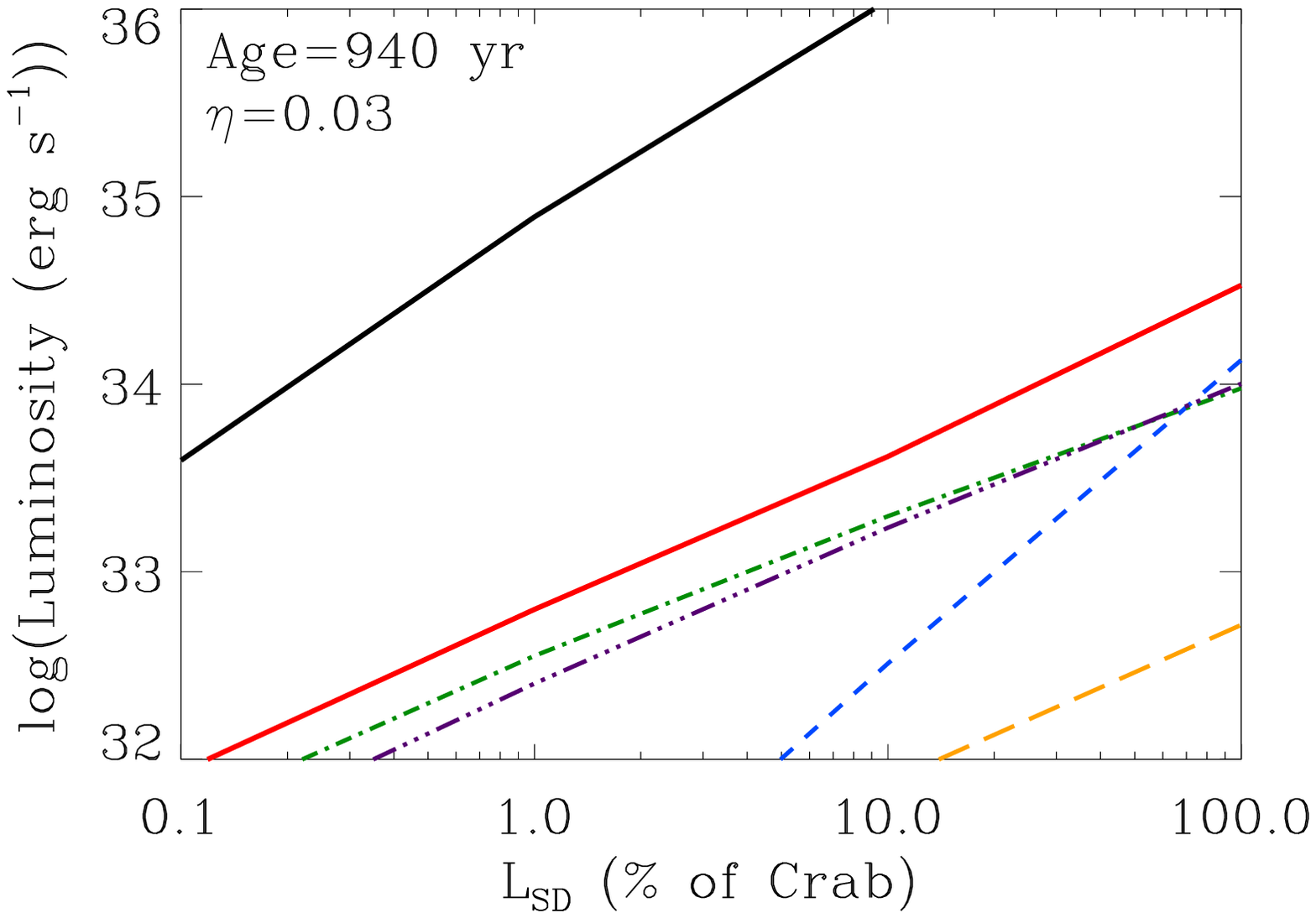} 
\includegraphics[scale=0.3245]{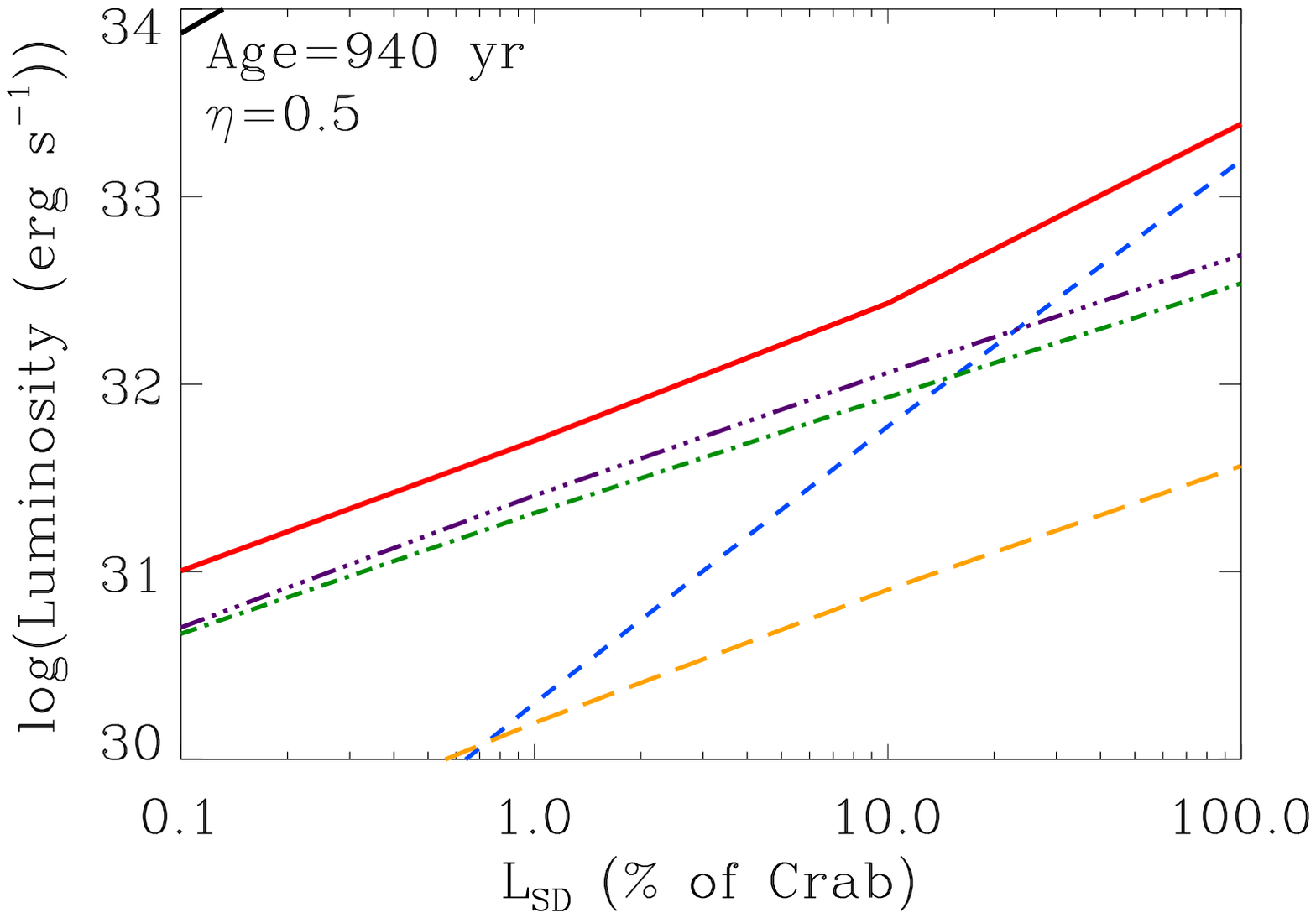}  \includegraphics[scale=0.3245]{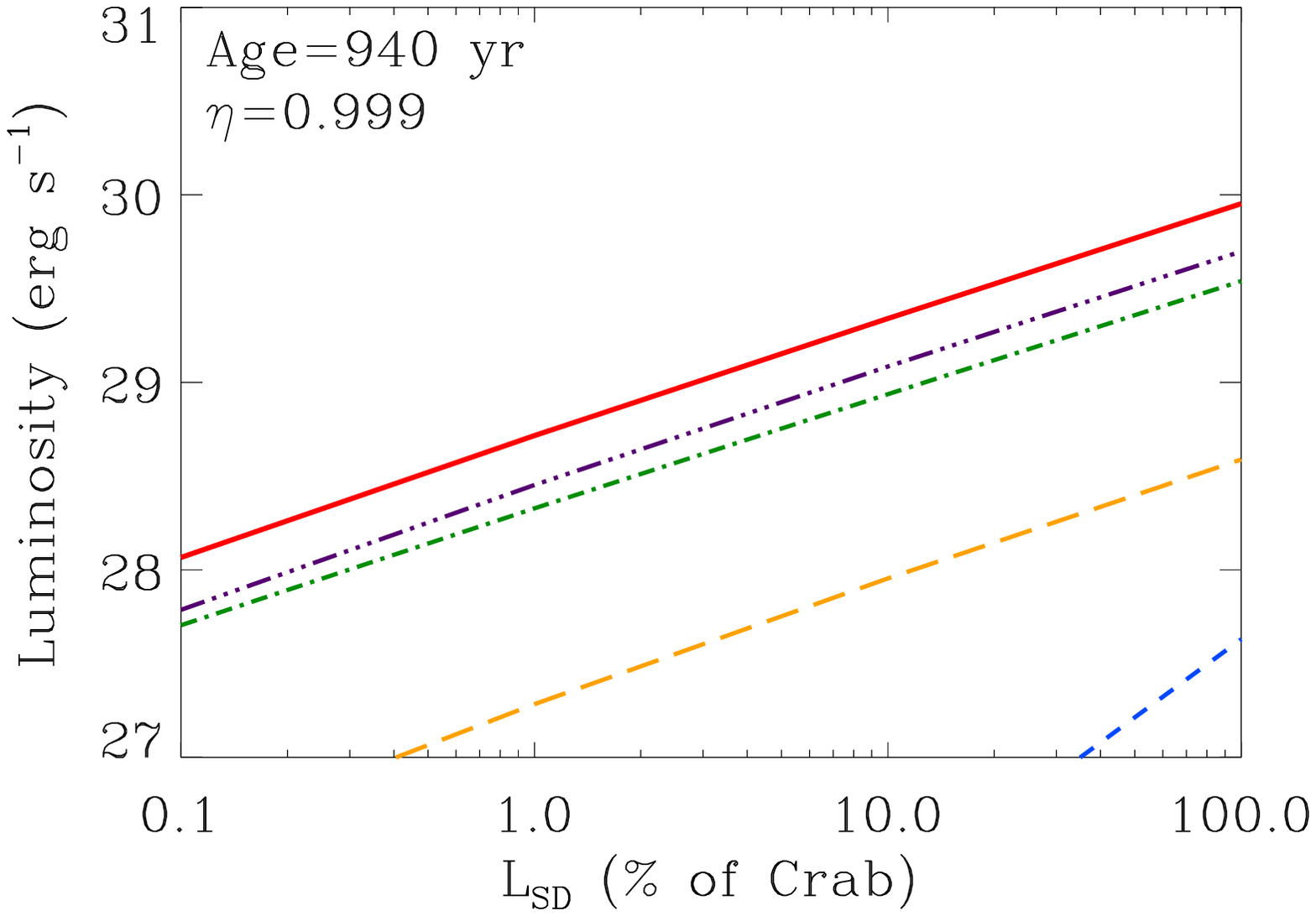}\\
\includegraphics[scale=0.3245]{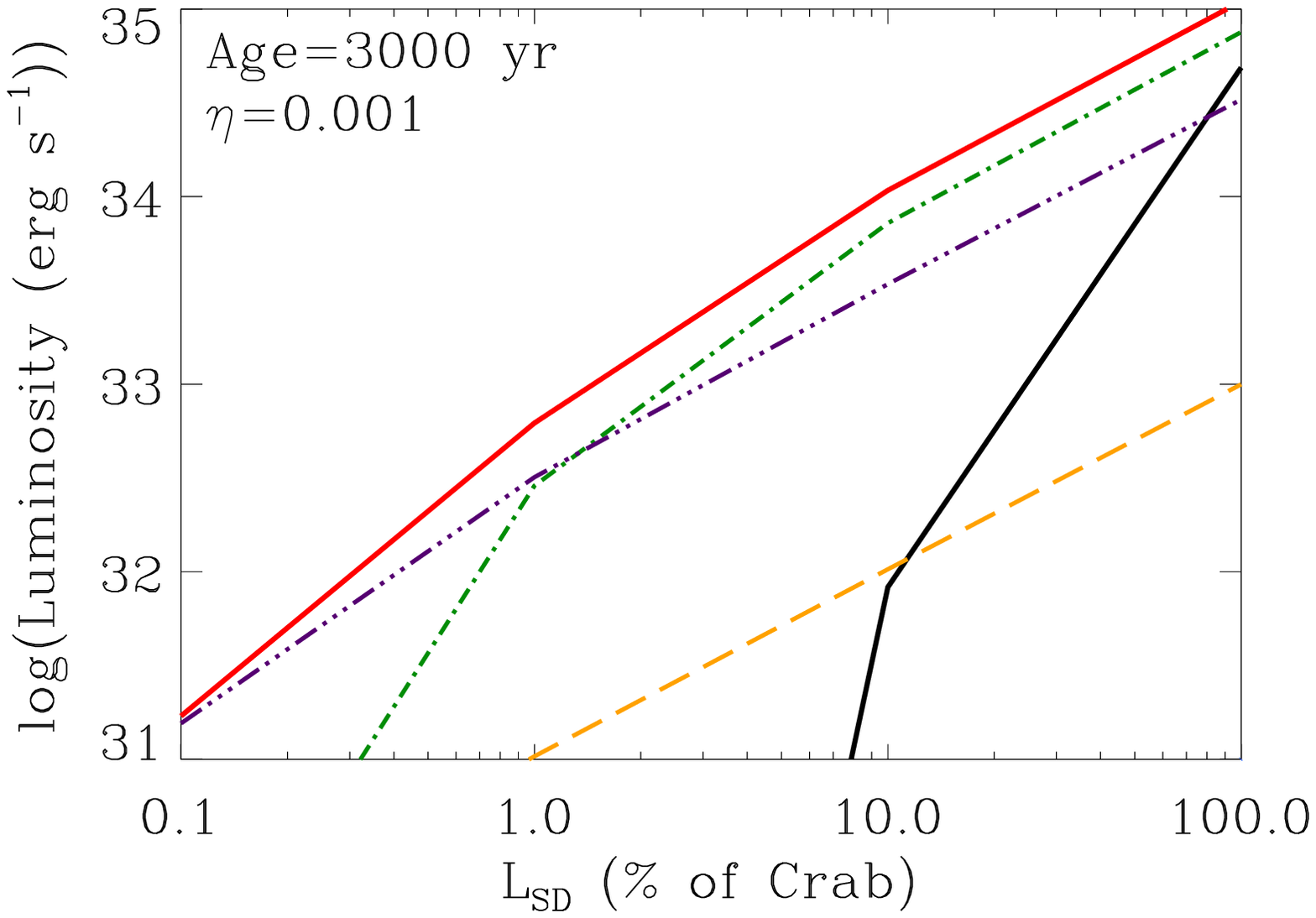}  \includegraphics[scale=0.3245]{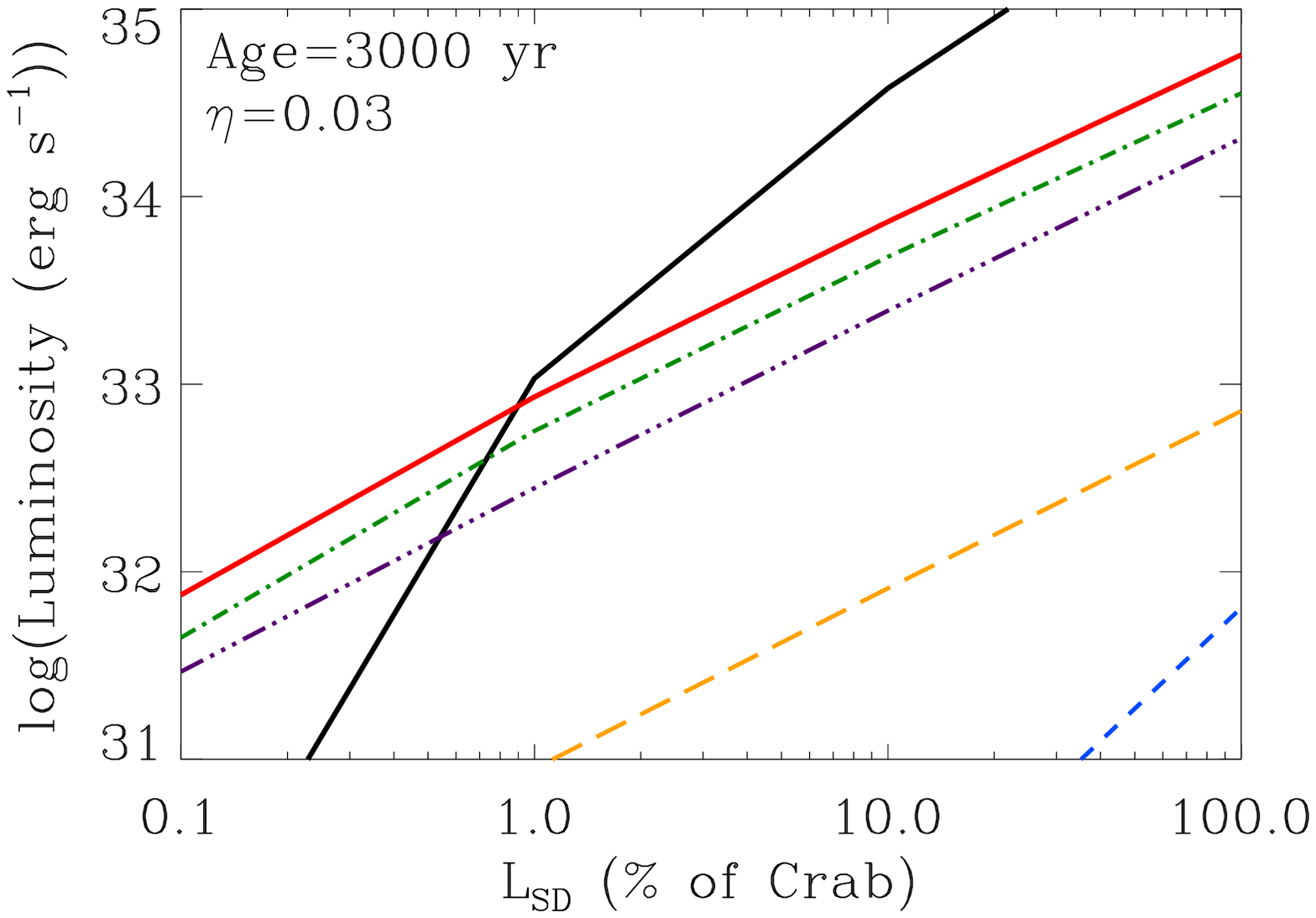} 
\includegraphics[scale=0.3245]{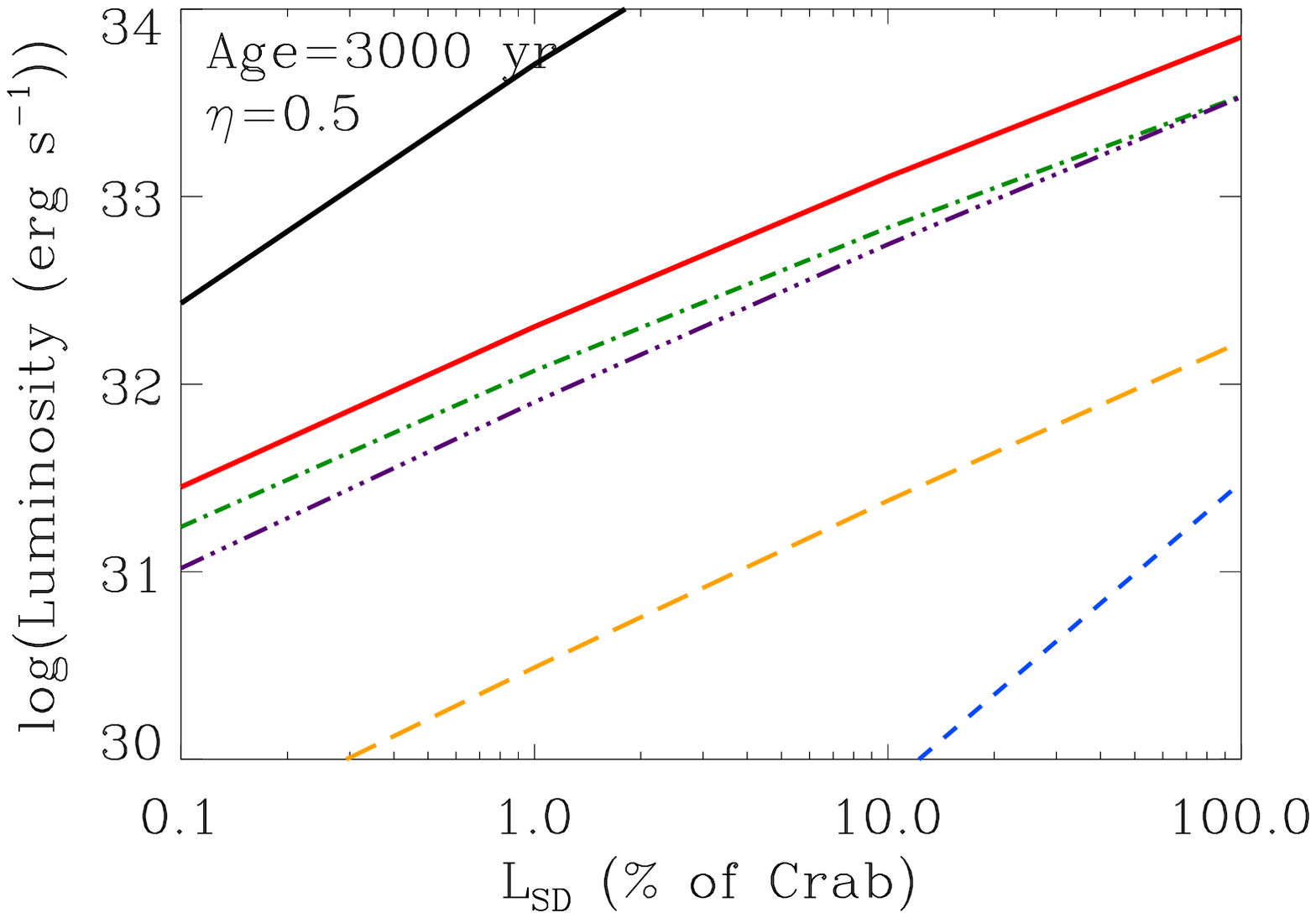}  \includegraphics[scale=0.3245]{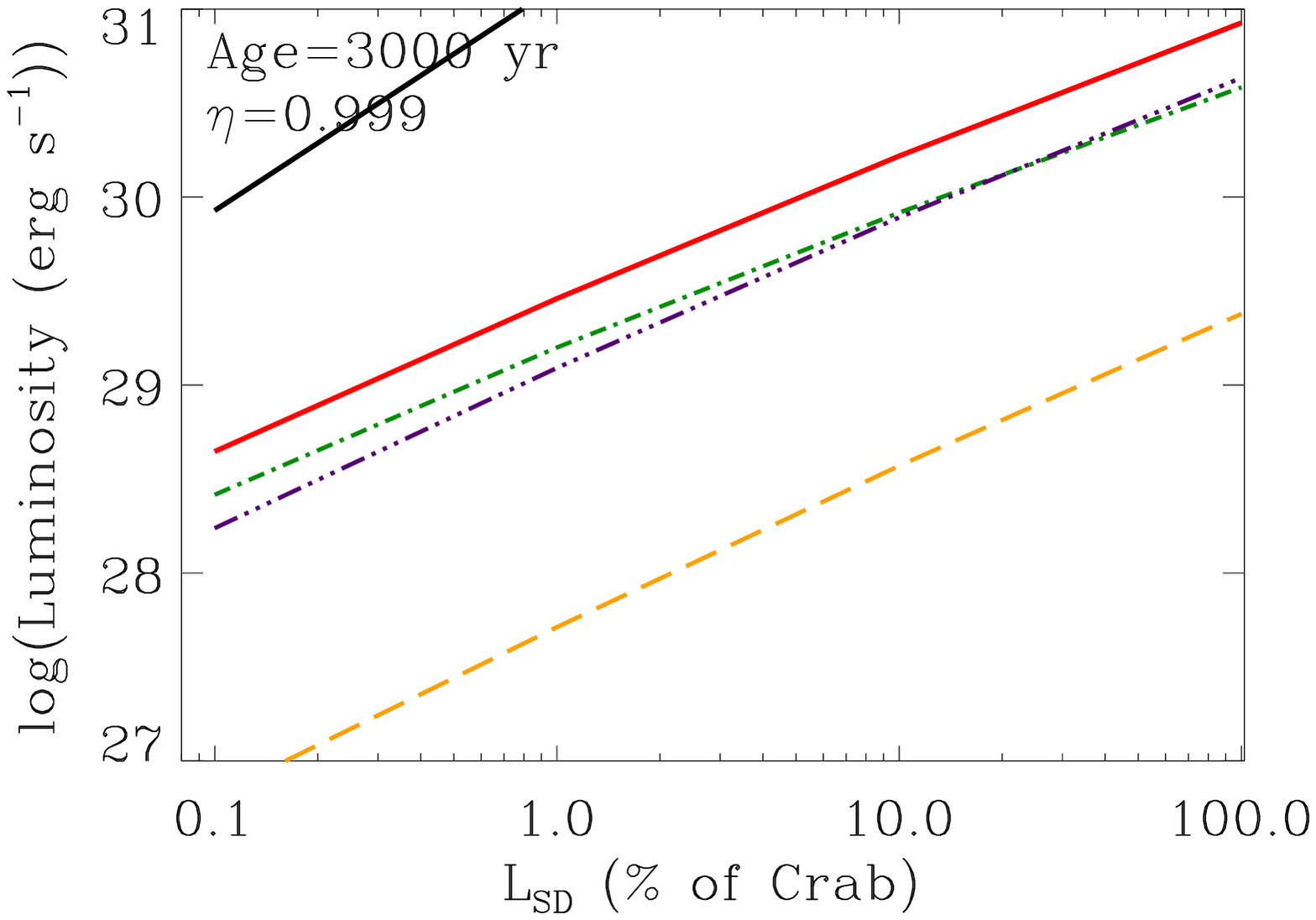}\\
\includegraphics[scale=0.3245]{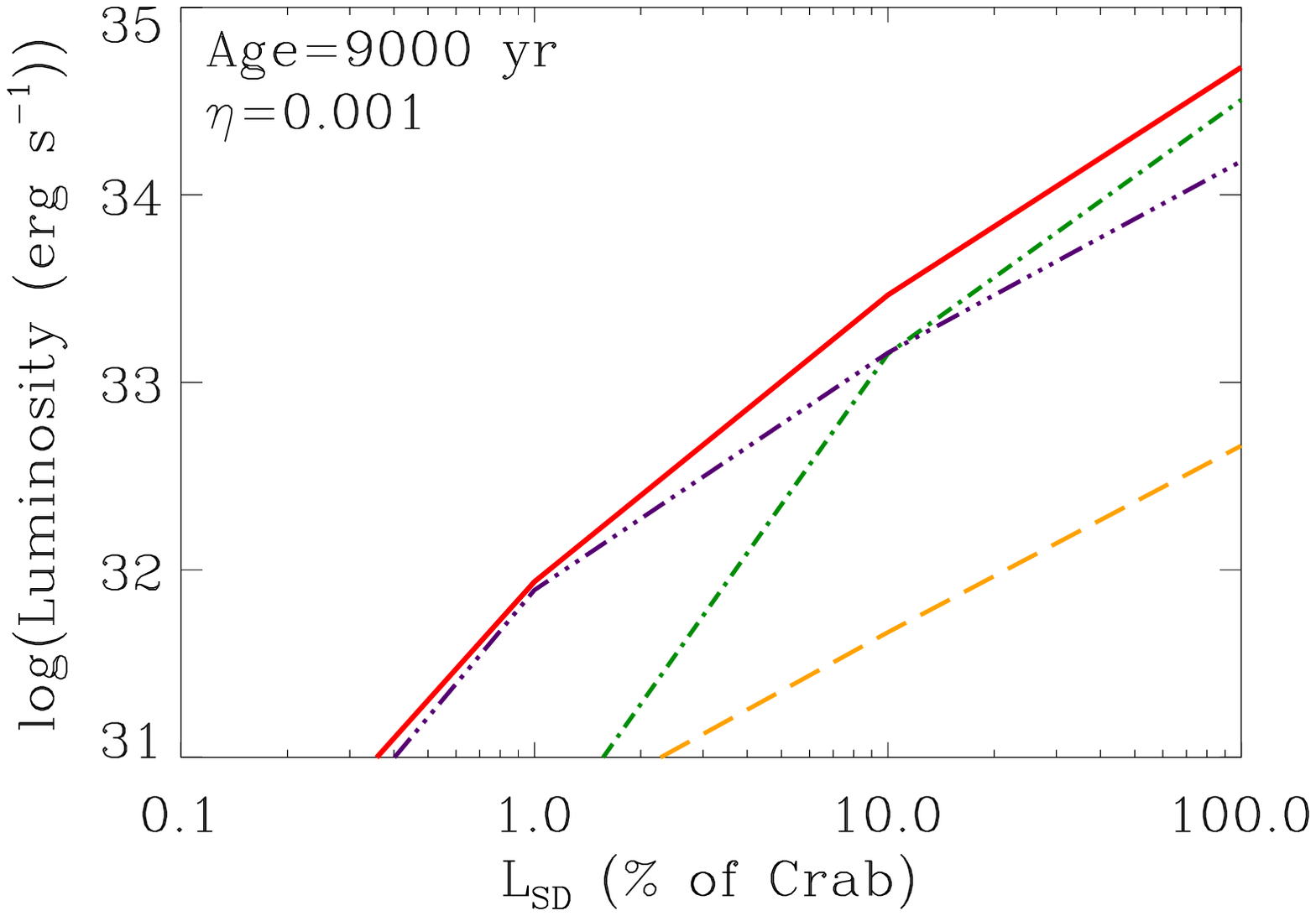}  \includegraphics[scale=0.3245]{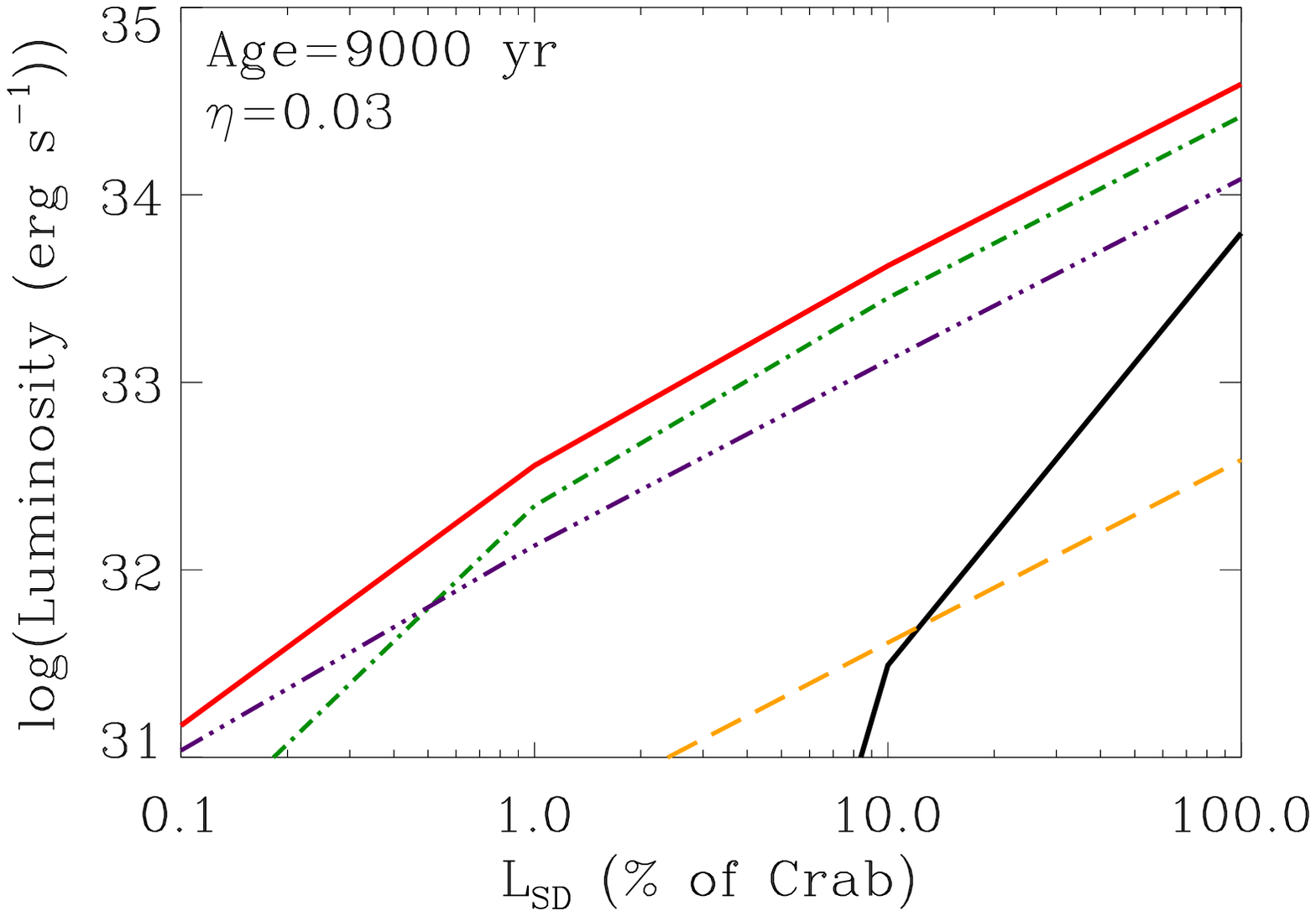} 
\includegraphics[scale=0.3245]{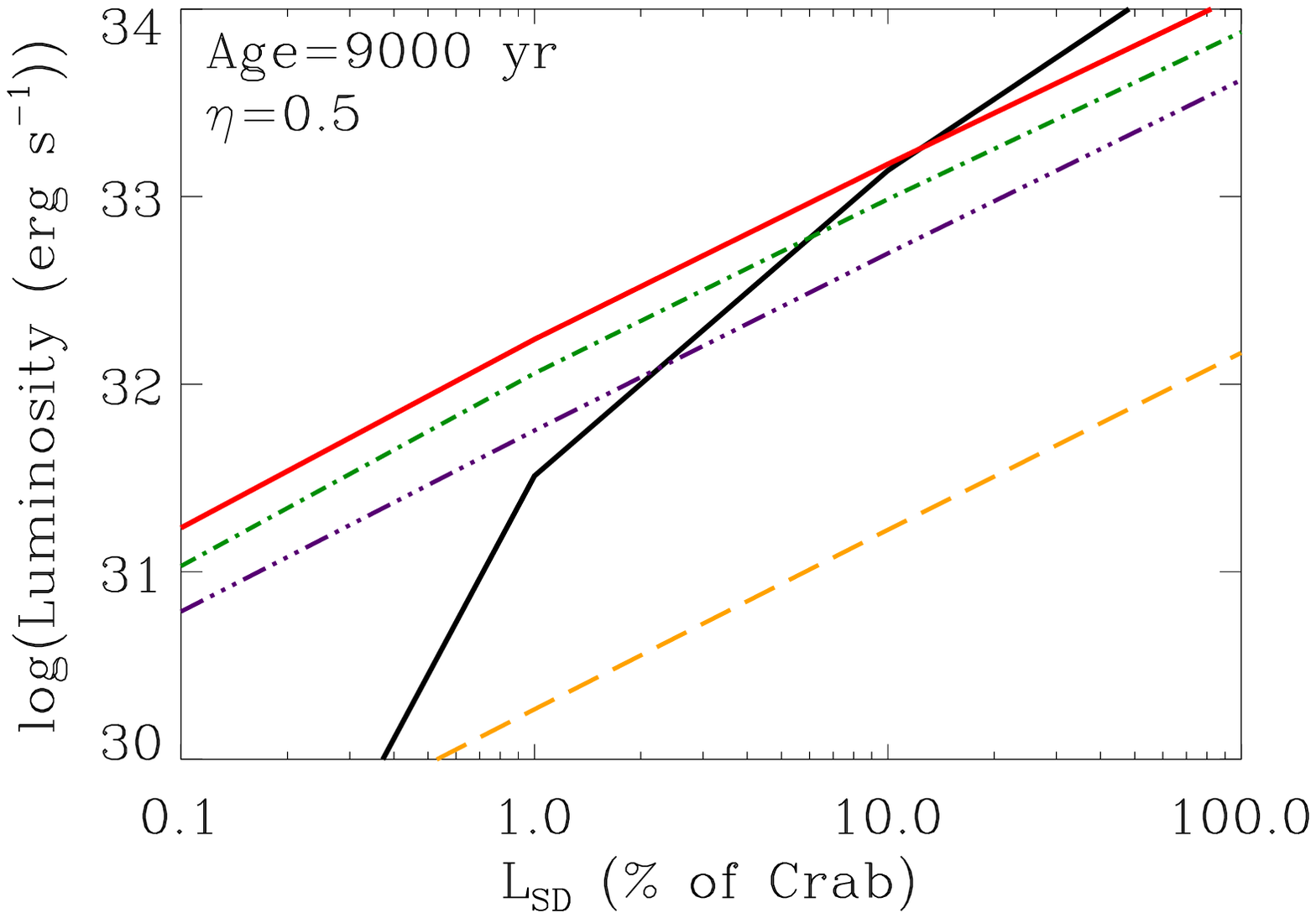}  \includegraphics[scale=0.3245]{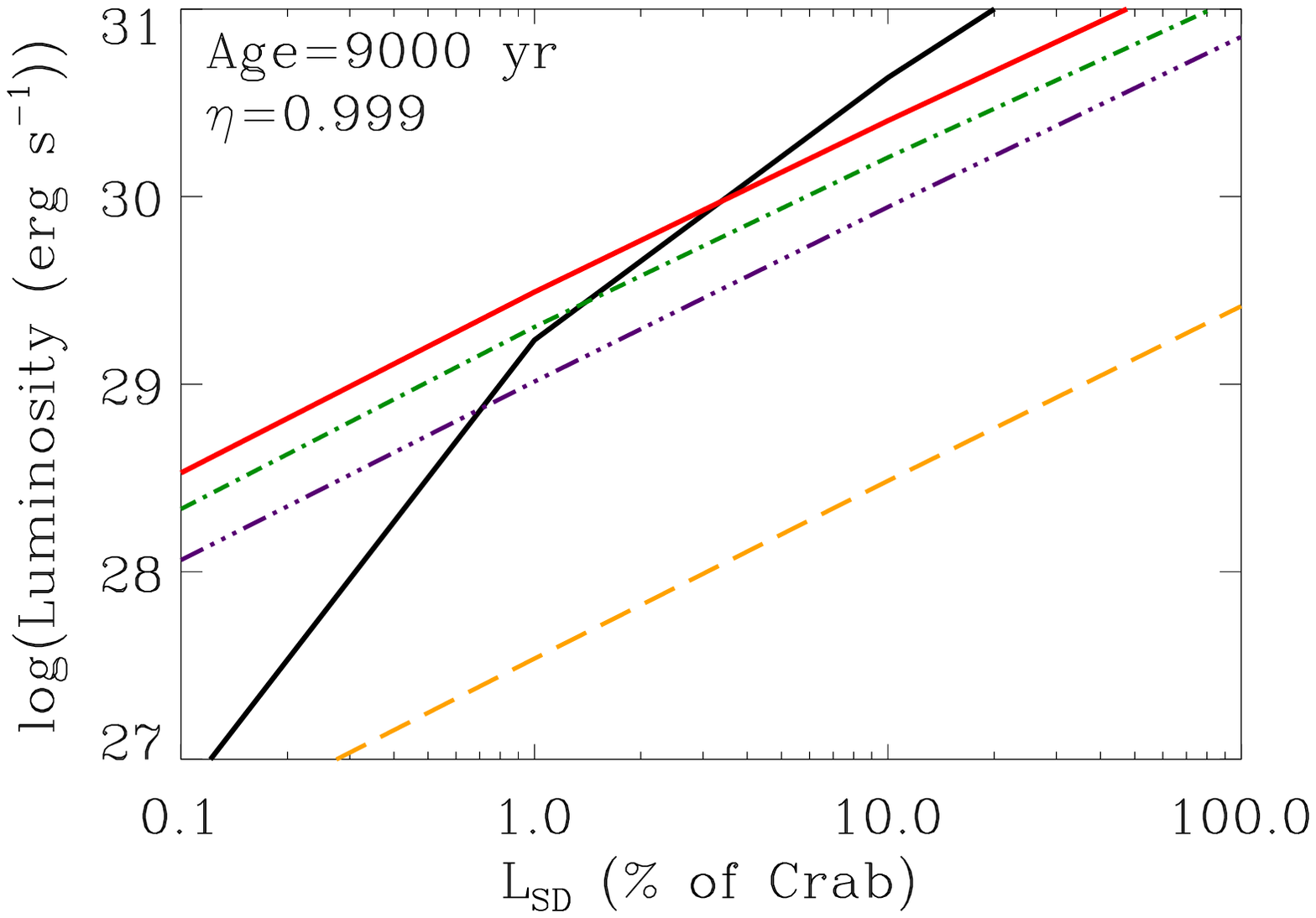}
\end{center}
\caption{Luminosities between 1 and 10 TeV of the contributions of the spectrum as a function of the spin-down luminosity. We fix an 
age of 940, 3000, and 9000 years (from top to bottom) and a magnetic fraction of 0.001, 0.03, 0.5 \& 0.999 (from left to right). The color coding is as in Fig. \ref{lum-age}.}
\label{lum-sd}
\end{figure}
\end{landscape}

To compare the TeV luminosities, we integrated the simulated gamma-ray emission between 1 and 10 TeV. For comparison, we have also computed   the synchrotron luminosity integrated between 1 and 10 keV. We compare the contributions of different photon backgrounds, namely SSC, FIR, NIR, and CMB, to the total IC yield of each of the nebulae. The results for the luminosity as a function of age are shown in Fig. \ref{lum-age} (for fixed $L_{SD}$=0.1, 1, 10, and 100\% of the Crab, from top to bottom, and a magnetic fraction of 0.001, 0.03, 0.5 \& 0.999, from left to right). The results for the luminosity as a function of spin-down power are shown in Fig. \ref{lum-sd} (for fixed increasing age, from top to bottom, and a magnetic fraction of 0.001, 0.03, 0.5 \& 0.999, from left to right).

The IC components have a very similar behavior one to another, with the exception of the SSC, which has a similar slope as the synchrotron contribution. This slope similarity between the synchrotron and the SSC luminosity  is seen for most of the plots in this section. There are some particular cases in which this is not the case, though. In the top-left panel of Fig. \ref{lum-age},  the CMB contribution decays with age  much more steeply than the FIR contribution to the total yield. This is the result of cutting the energy range in a small band, from 1 to 10 TeV, where, in this case, the IC contribution off the CMB is falling. The latter dominates the FIR contribution at 1 TeV in this case, where it starts to fall steeply; due to the value of $\gamma_{max}$ (see Table \ref{param}), there are not enough  electrons to generate higher energy photons interacting with the CMB background. 

If we consider the SSC contribution, depicted by the blue-dashed line, we note it is only visible in the y-axis scale of the different panels of Fig. \ref{lum-age} in only a few occasions. It is irrelevant for $L_{SD}$=0.1, 1, and 10\% of the Crab power, disregarding the age and the magnetic fraction of the nebulae. On the contrary, it only becomes relevant for highly energetic (Crab-like) particle dominated nebulae at low ages (of less than a few thousand years).
The  Crab Nebula today corresponds to the bottom row, second column plot of Fig. \ref{lum-age} when the age (in the x-axis) is taken as 940 years. 
%That is, the set of parameters of this model (given also in Table \ref{param}) stand for a perfect fit to the Crab Nebula data.
It is seen there how uncommon the SSC domination is:
A lower or higher magnetic fraction (left or right panels), or a higher age (movement along the x-axis), and the SSC contribution 
would quickly be sub-dominant to the IC-FIR or even to the IC-CMB components.

Fig. \ref{lum-sd} shows the IC contributions of the spectrum as a function of spin-down. 
As we increase the spin-down power, all the IC contributions increase their luminosity due to the presence of additional high-energy electrons, but the SSC depends also on the power of the synchrotron emission, which is increasing too due to the higher magnetic field. This effect makes the SSC a steeper function of the spin-down power compared to the other contributions. Consistently with the results of Fig. \ref{lum-age}, the SSC contribution requires a young age and $\sim 70\%$
of the Crab's power to become relevant. 
Fully magnetized nebulae ($\eta = 0.999$), if they exist, are never SSC-dominated no matter the age or pulsar spin-down power. This is partly also a result of the increased synchrotron losses produced by the very high magnetic field, which diminishes the relative importance of all IC components.
Finally, we note that --mimicking the SSC behaviour-- for lower spin-down luminosities and older ages than that of the Crab Nebula, the synchrotron luminosity falls down very quickly. This is partly because the energy range where we are integrating the luminosity is in the synchrotron cutoff regime produced by the electron population cut at high energies. The former results clarify why the Crab Nebula, and only it, is SSC dominated: 
There are no other pulsars we know of, young and powerful enough so that 
SSC could play any role against the comptonization of FIR, or CMB photons.

\begin{figure*}
\begin{center}
\includegraphics[scale=0.45]{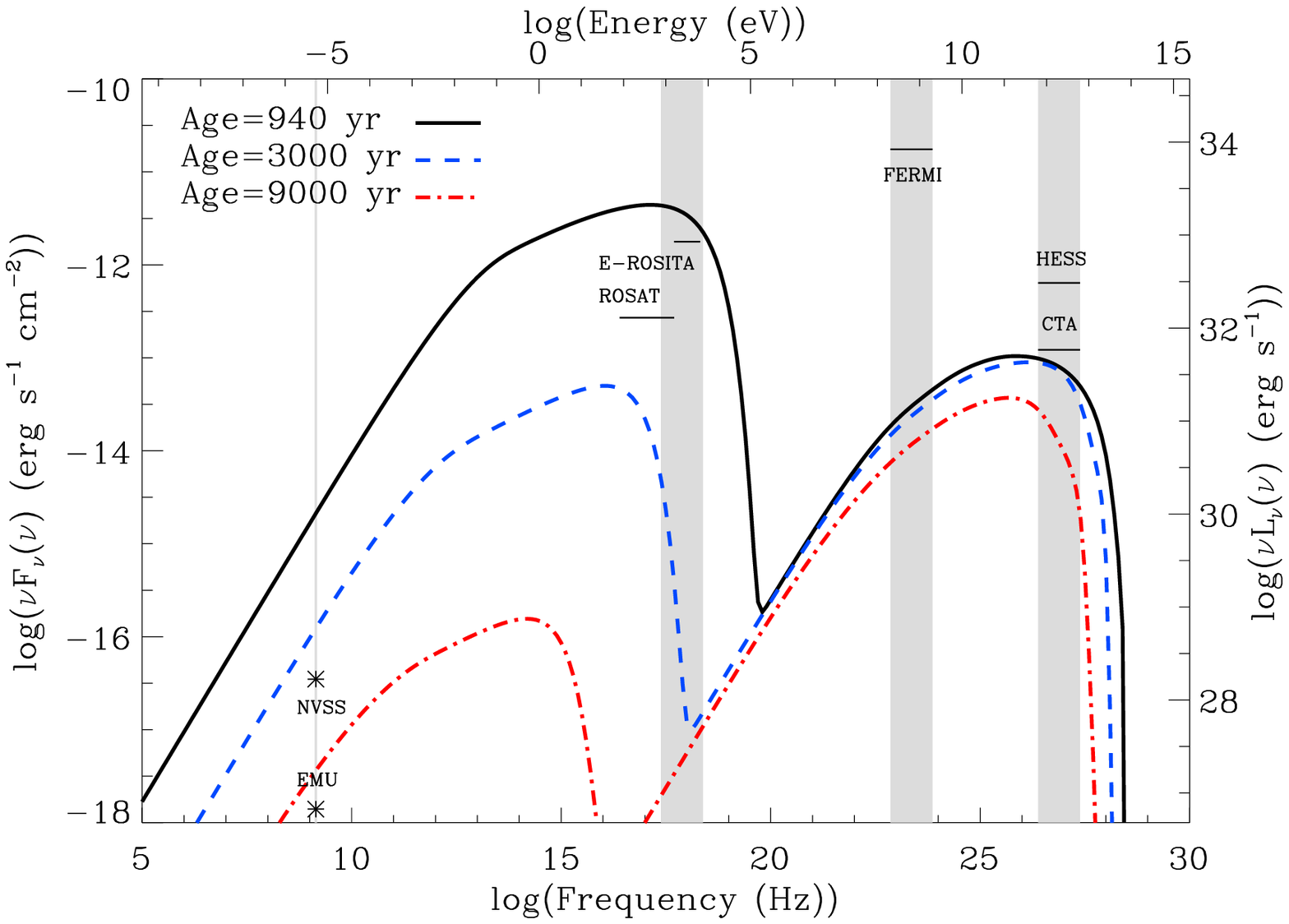}  \hspace{0.2cm} \includegraphics[scale=0.45]{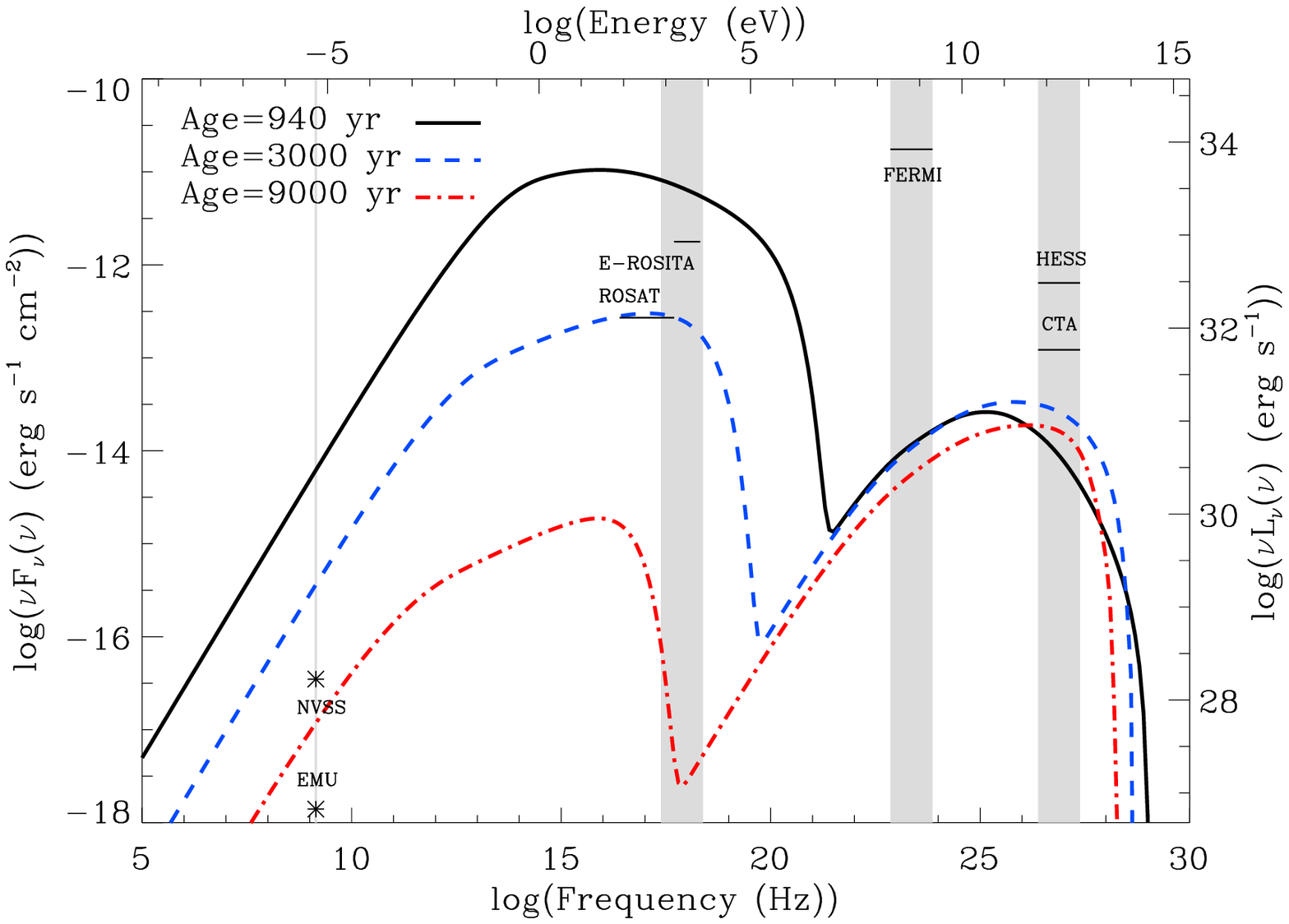} \\
\includegraphics[scale=0.45]{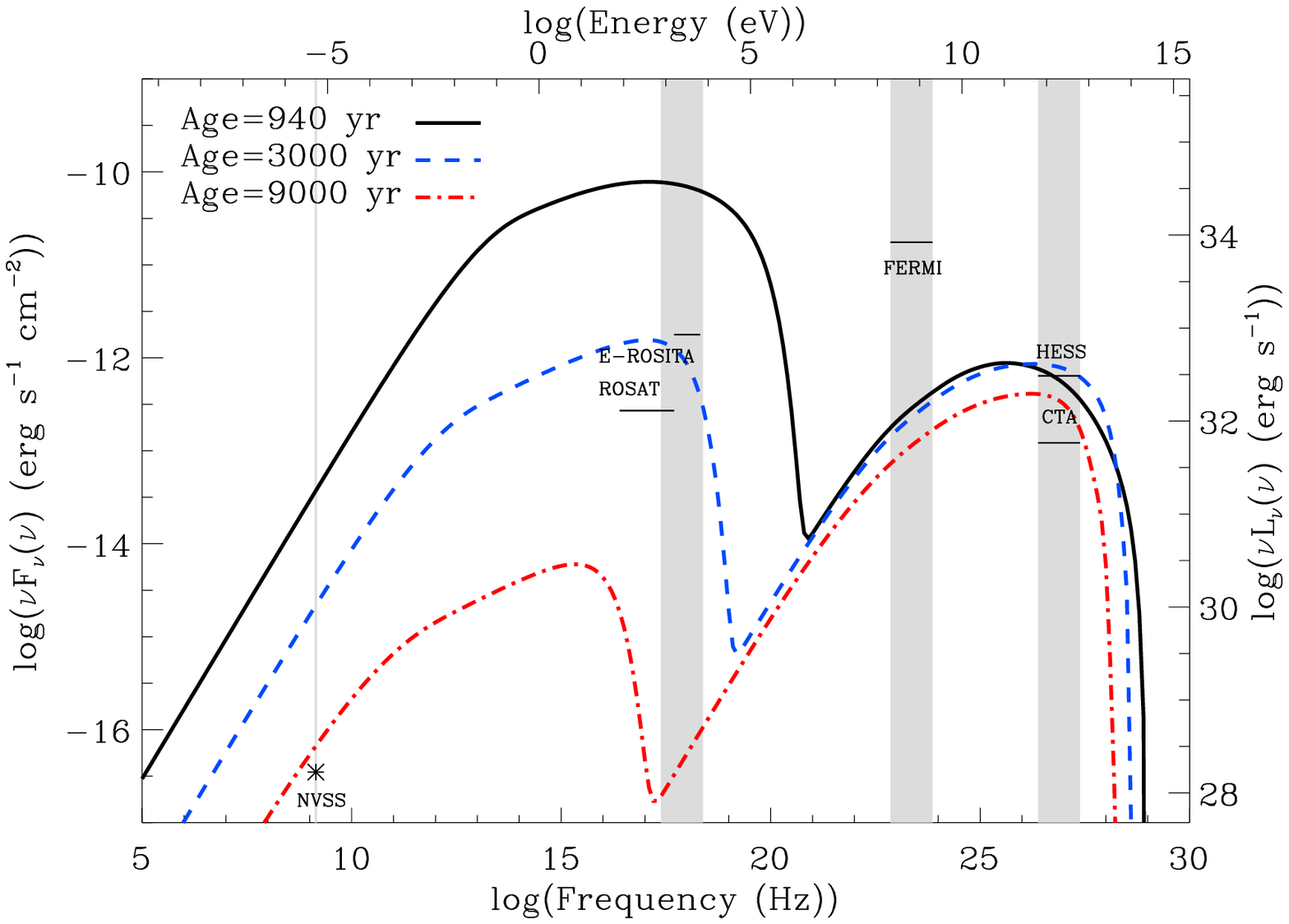}  \hspace{0.2cm} \includegraphics[scale=0.45]{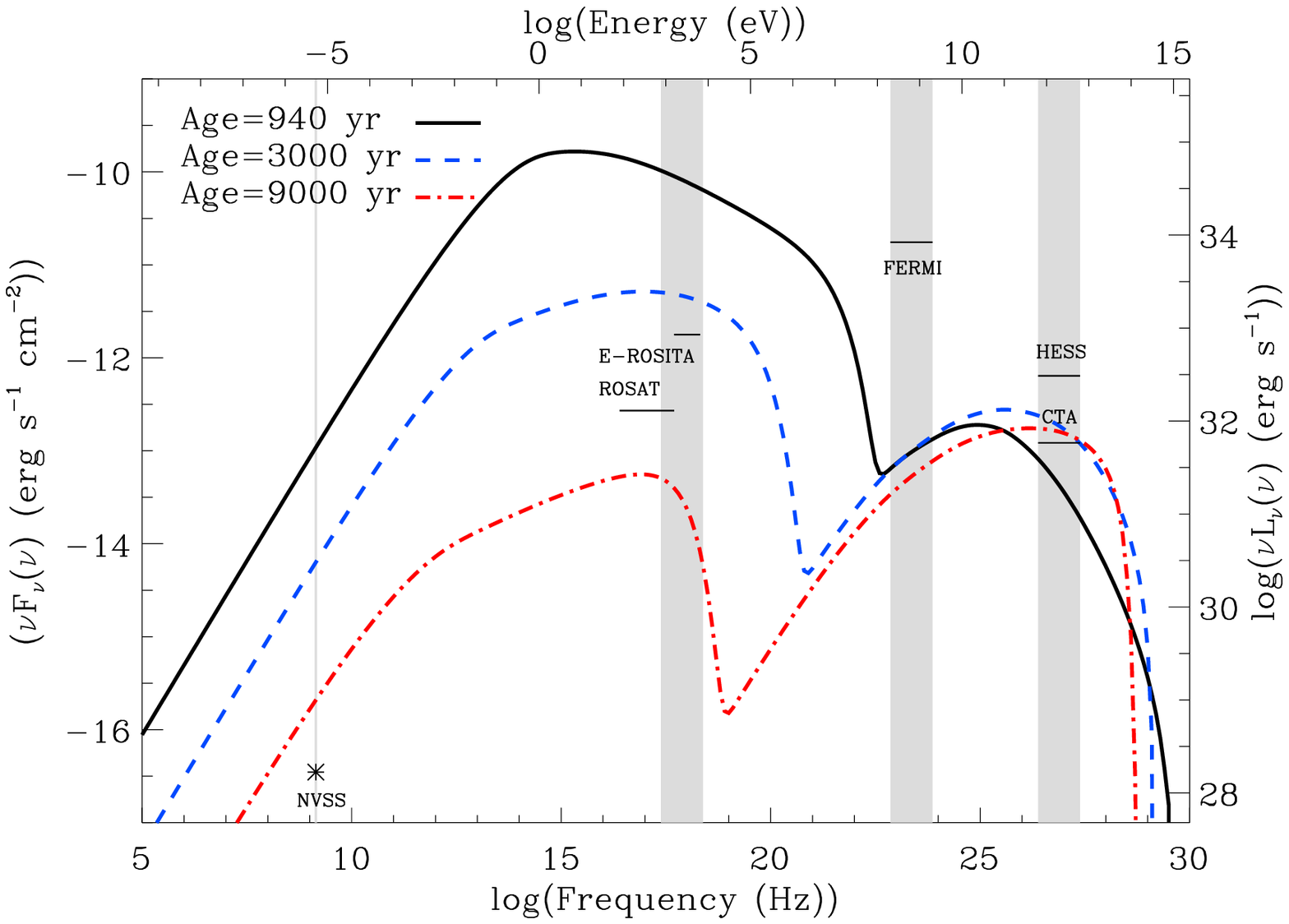}\\
\includegraphics[scale=0.45]{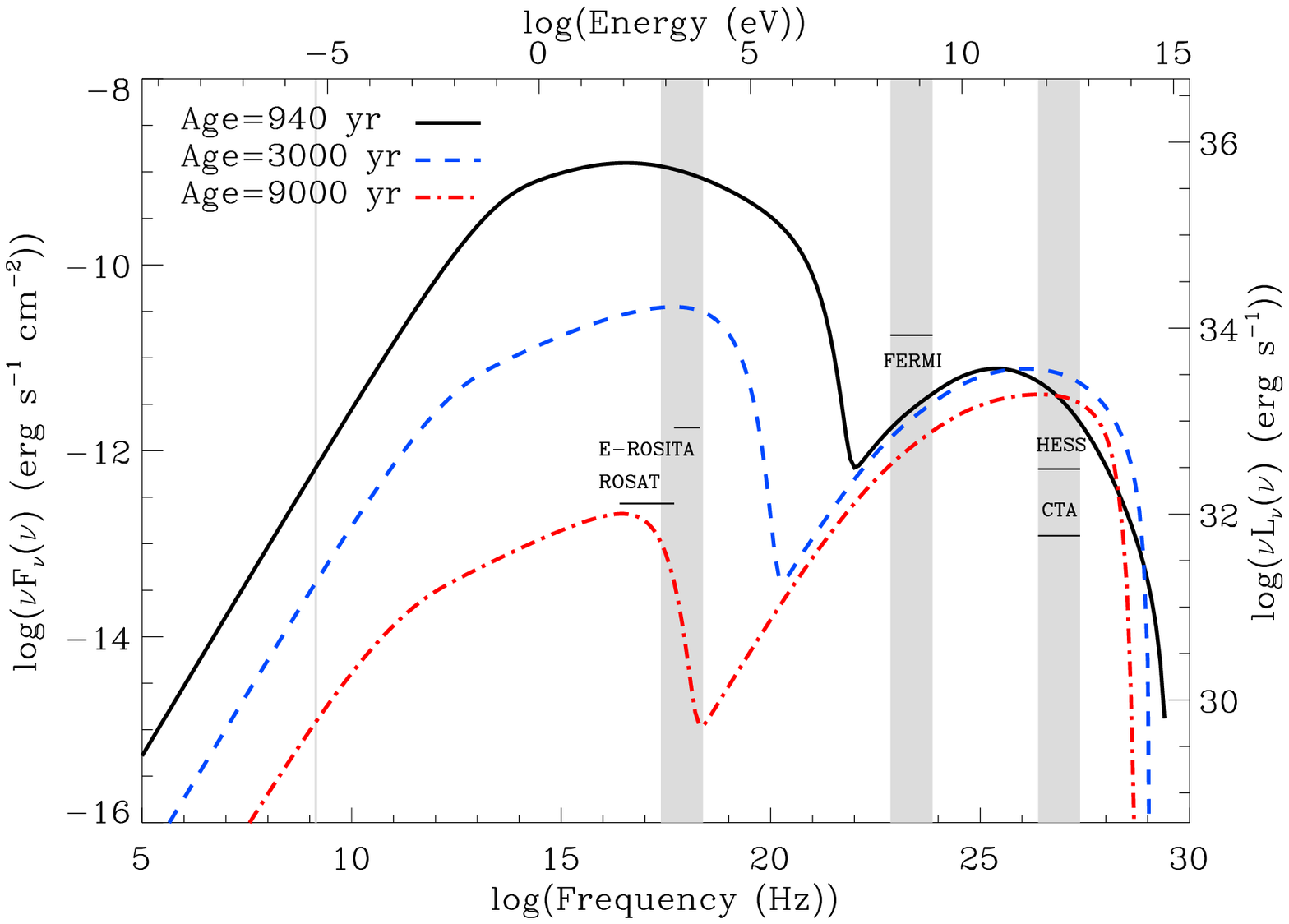}  \hspace{0.2cm} \includegraphics[scale=0.45]{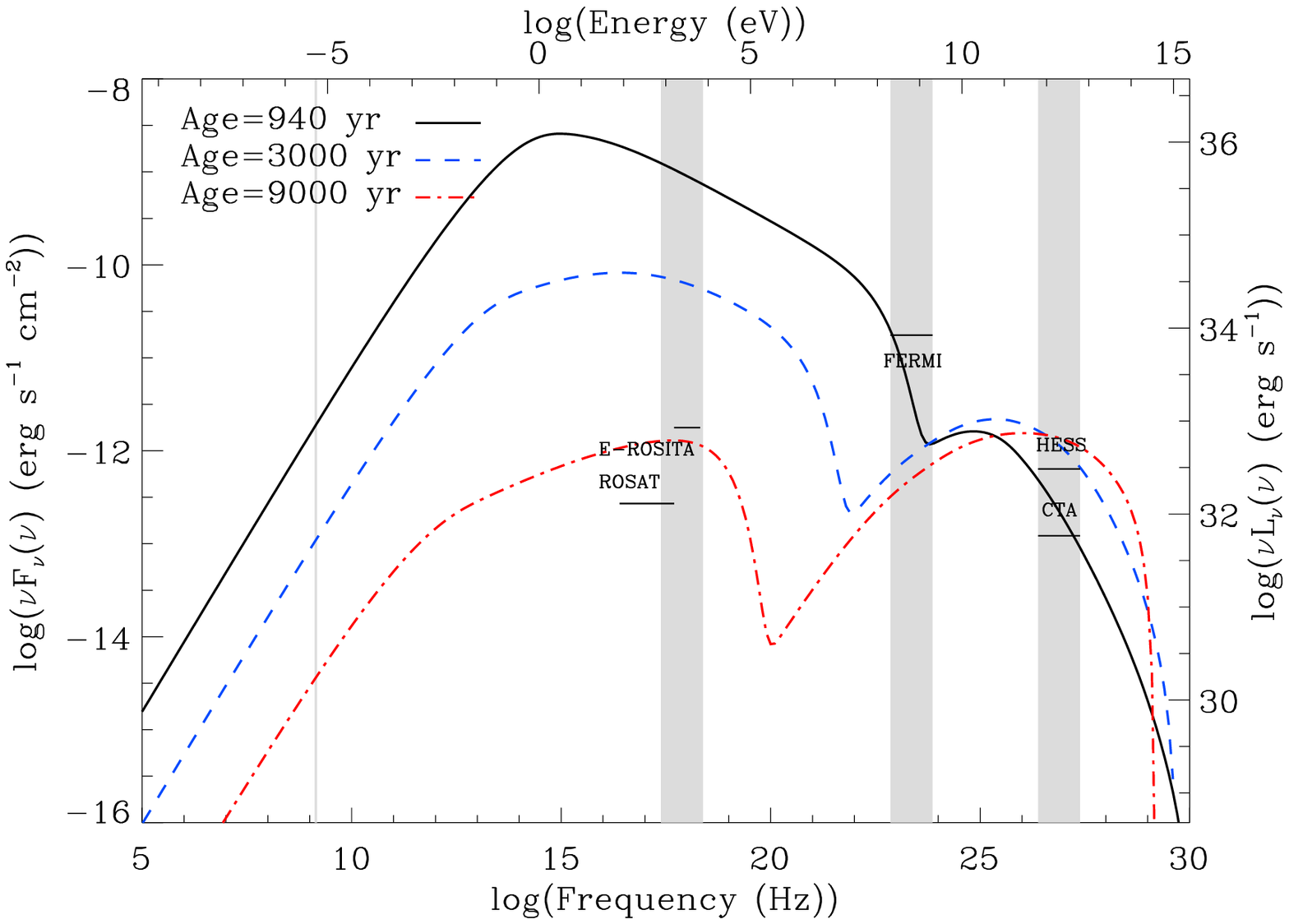}\\
\end{center}
\caption{Comparison of the SEDs for $L_{SD}$=0.1, 1, and 10\% of the Crab (from top to bottom)  
as a function of the age. The magnetic fraction is fixed at 0.03 (left) and 
0.5 (right). The shadowed columns correspond to the frequency intervals in radio, X-rays, the GeV, and the TeV bands where we have integrated the luminosity. The sensitivity of some surveys and telescopes in these energy ranges are shown by 
thin black lines.}
\label{SED-1}
\end{figure*}

\begin{figure*}
\begin{center}
\includegraphics[scale=0.45]{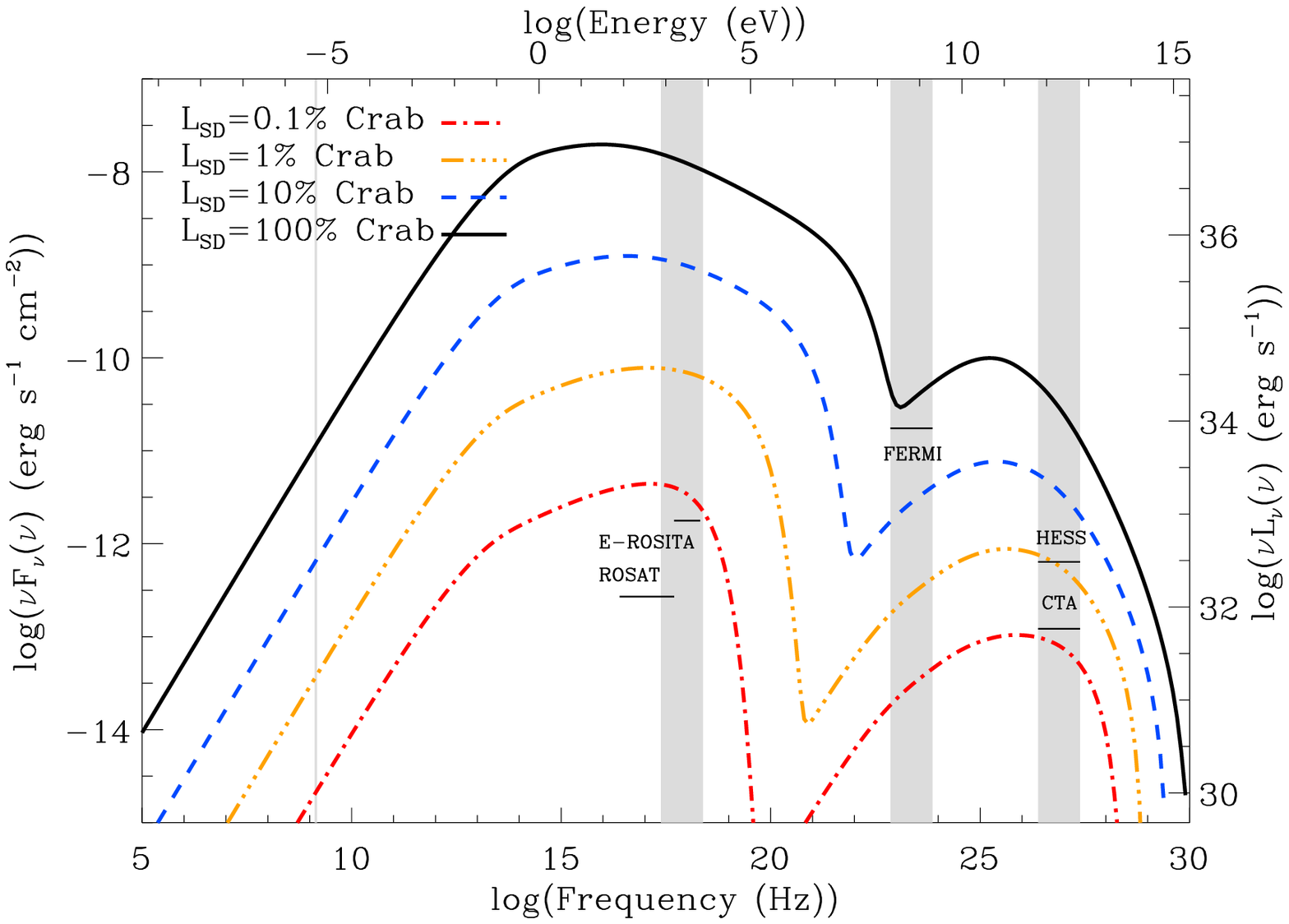}  \hspace{0.2cm} \includegraphics[scale=0.45]{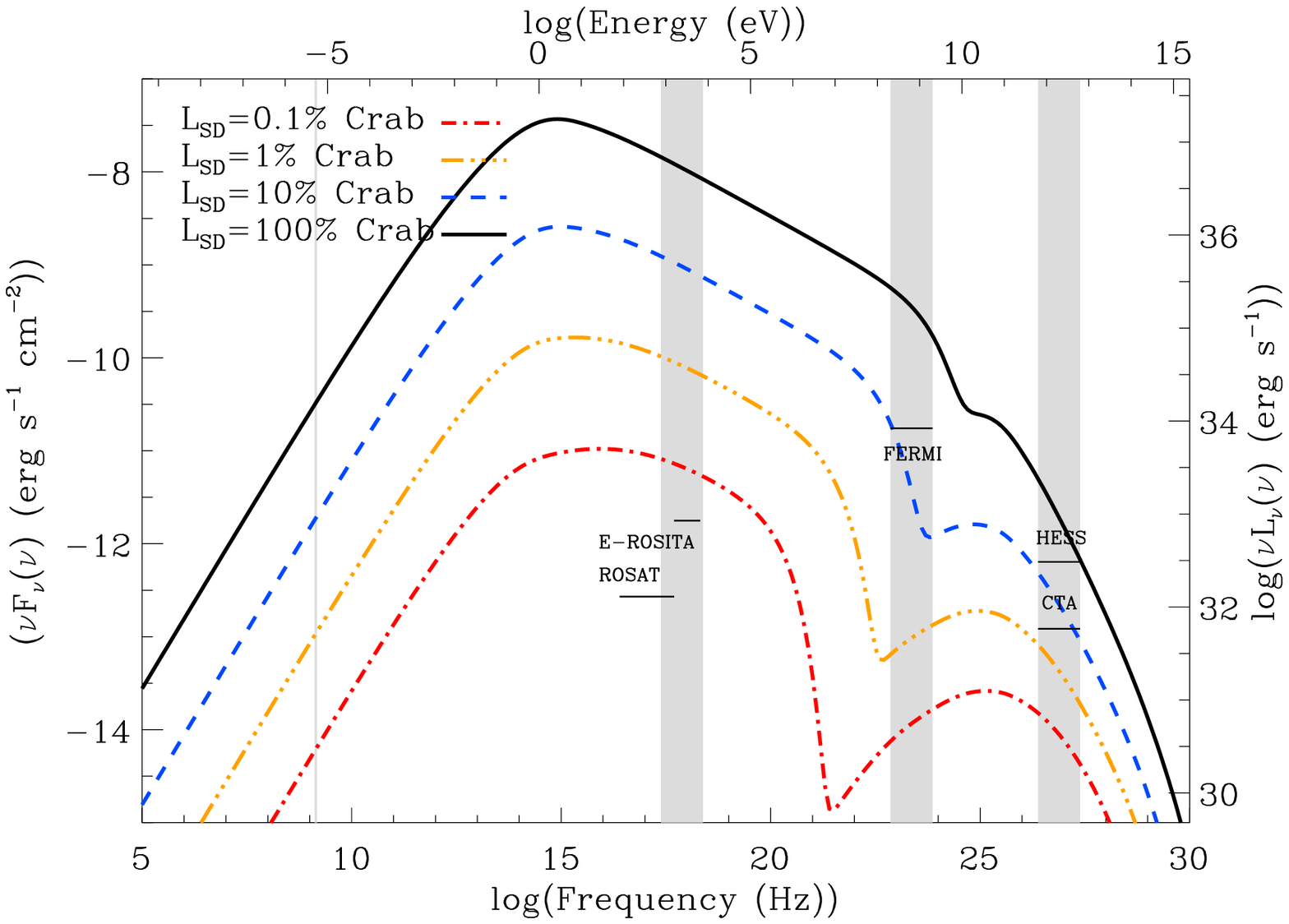} \\
\includegraphics[scale=0.45]{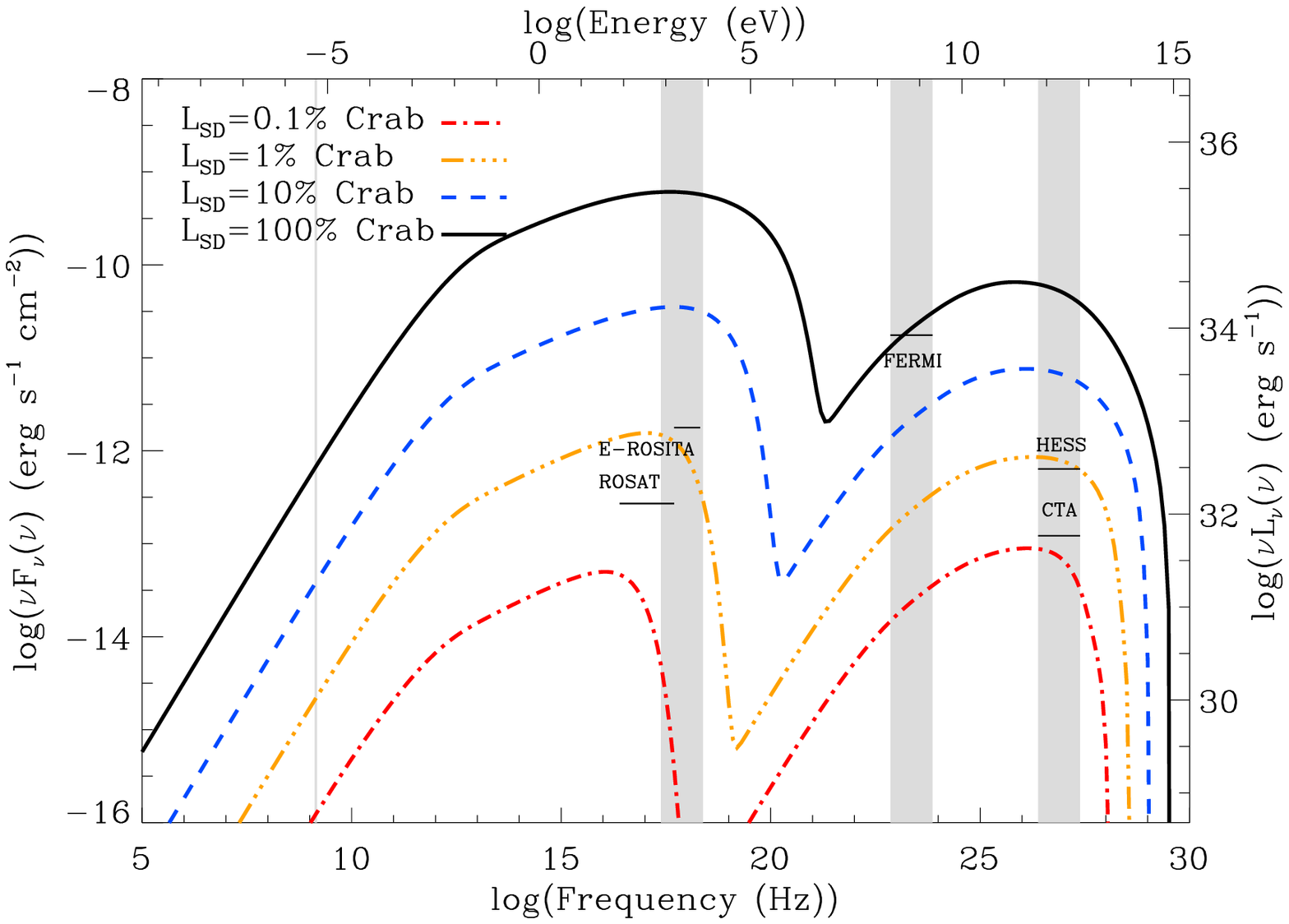} \hspace{0.2cm}  \includegraphics[scale=0.45]{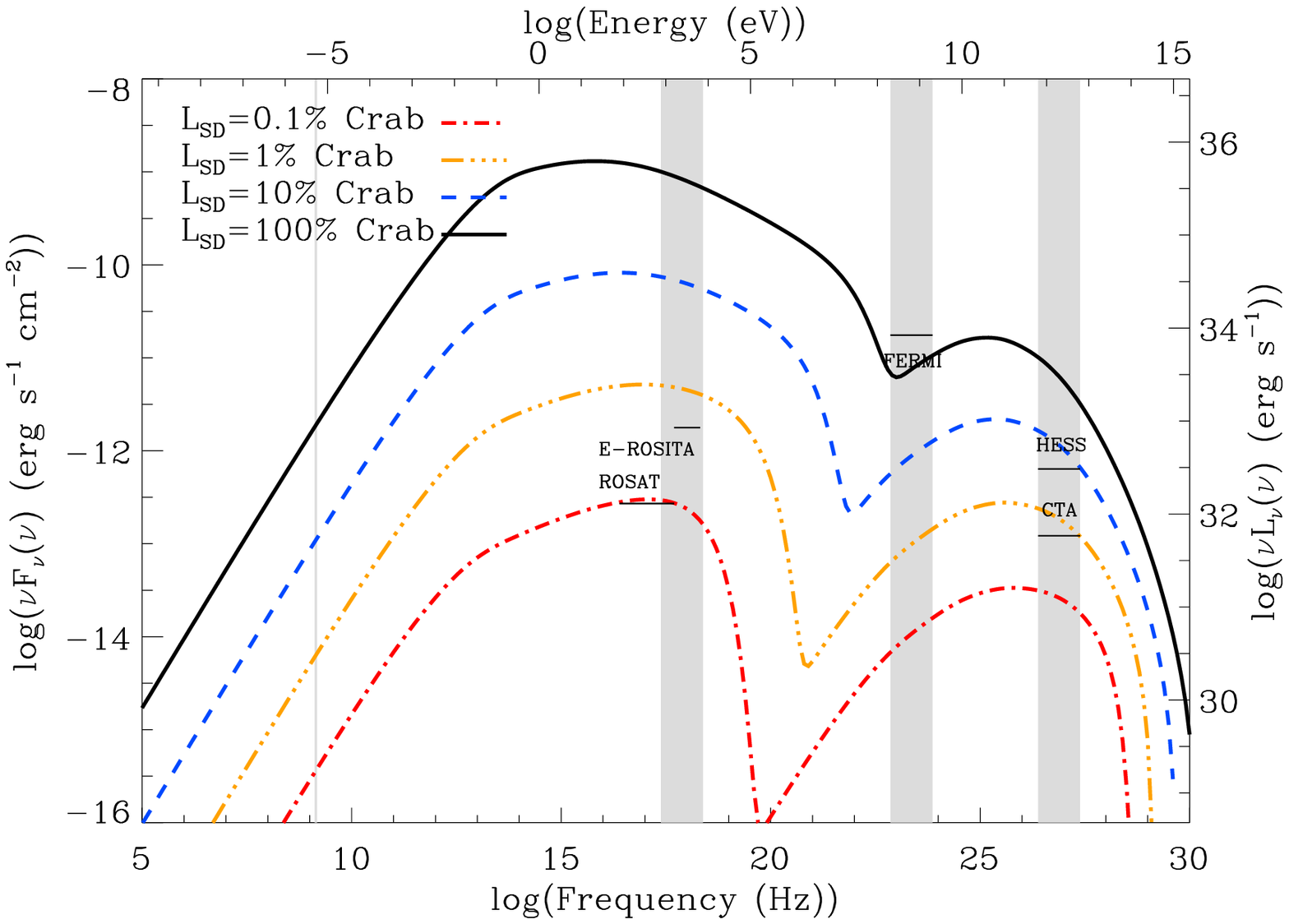}\\
\includegraphics[scale=0.45]{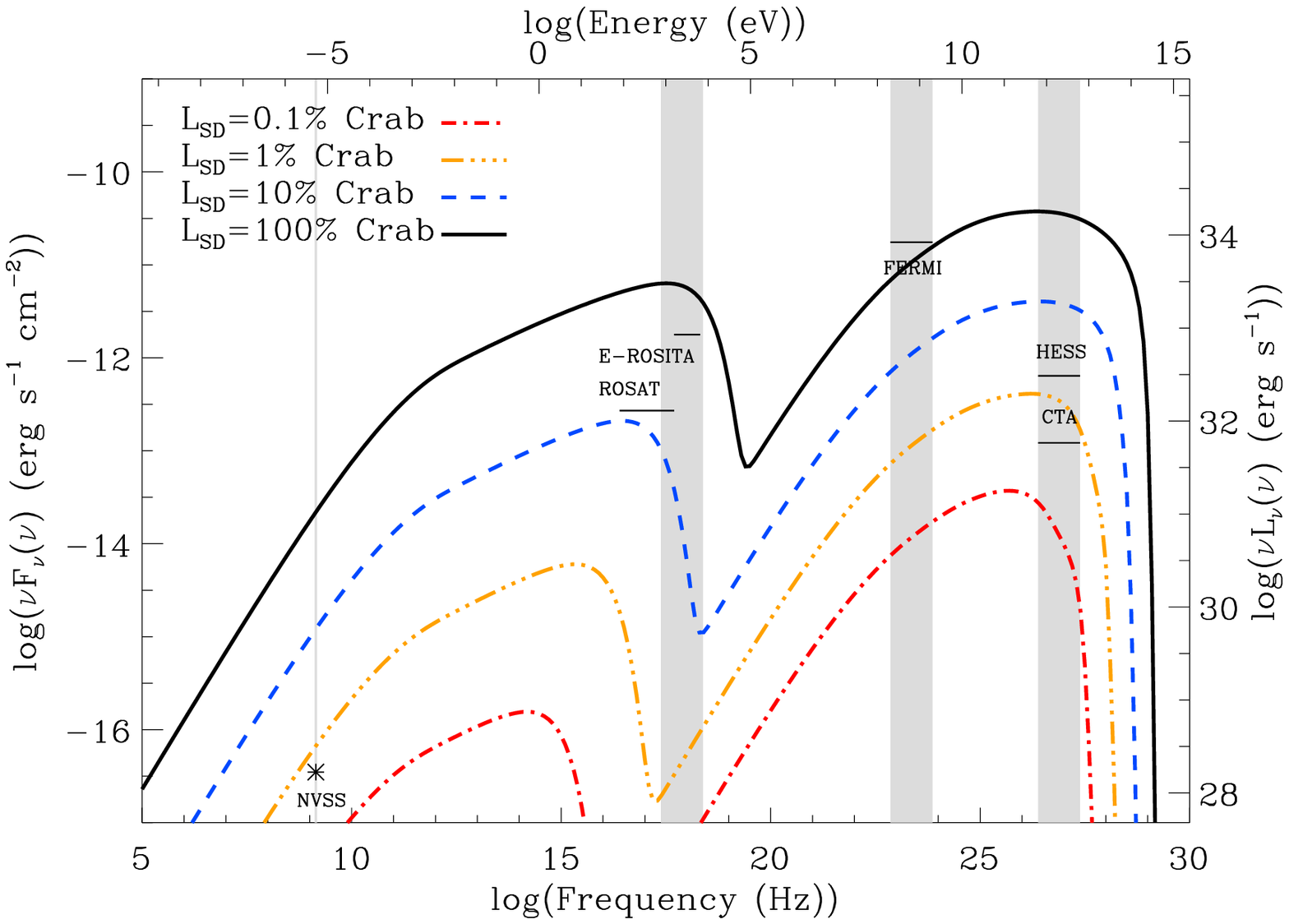}   \hspace{0.2cm} \includegraphics[scale=0.45]{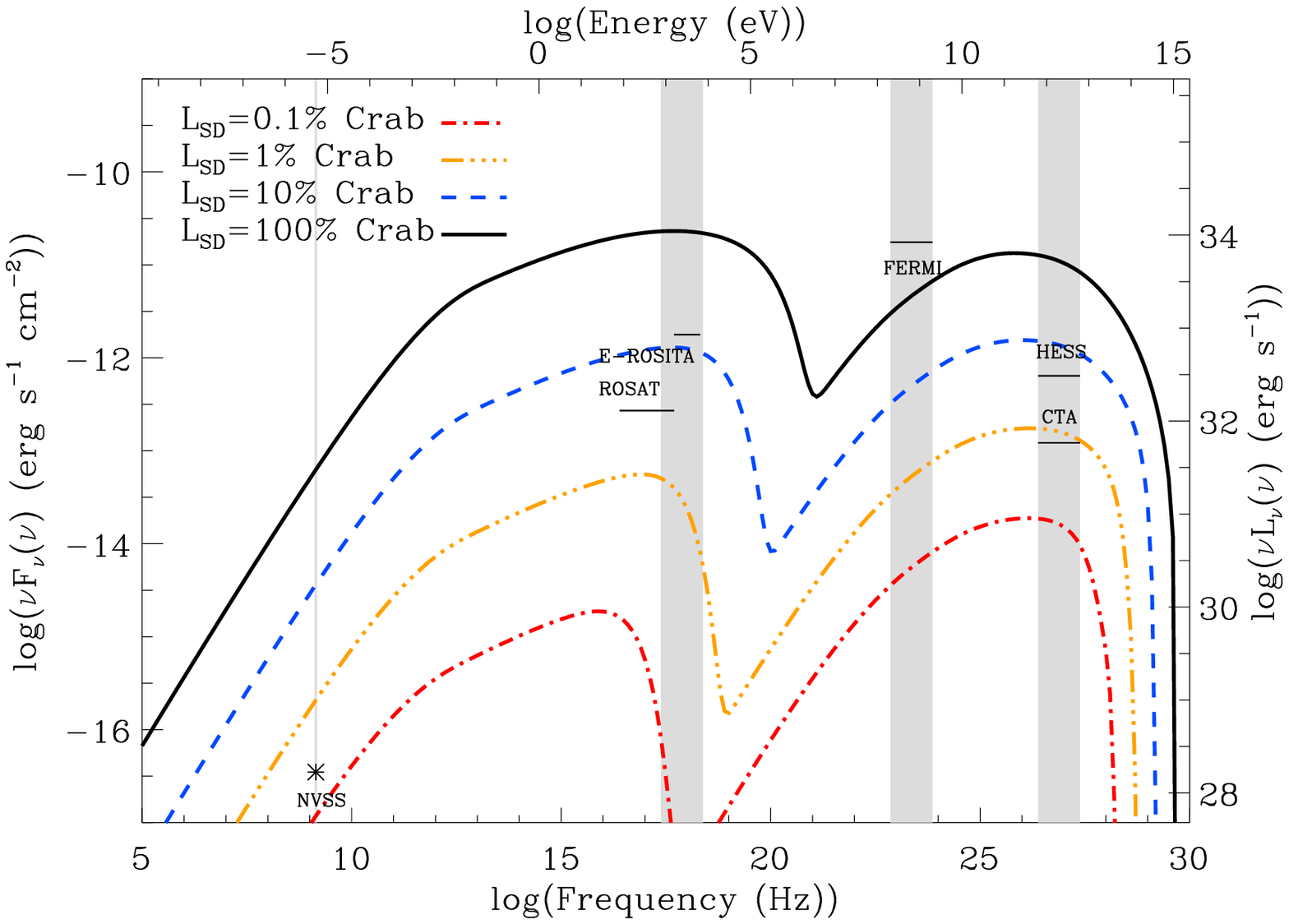}
\end{center}
\caption{Comparison of the SEDs for an age of 940, 3000, and 9000 years (from top to bottom) as a function of the spin-down luminosity. The magnetic fraction is fixed at 0.03 (left)
and 0.5 (right). }
\label{SED-2}
\end{figure*}

Similar considerations can be done by inspecting the SED as a function of age and spin-down power (Fig. \ref{SED-1} and \ref{SED-2}). In each Figure, the SED showed in the left panel is calculated for a particle-dominated nebula ($\eta$=0.03) whereas the right one is computed for a nebula in equipartition ($\eta$=0.5). The shadowed areas correspond to the frequency intervals in radio, X-rays, GeV, and TeV bands where we integrate the luminosity to compare their ratios (see below). Several instrument sensitivities (in survey mode) are also shown, corresponding to the NVSS and EMU in radio\footnote{See for instance \url {http://askap.pbworks.com/w/page/14049306/RadioSurveys} or \url {http://www.atnf.csiro.au/people/rnorris/emu/science_goals/index.htm}}, e-Rosita and ROSAT in X-rays\footnote{See \url {http://www.mpe.mpg.de/455799/instrument}}, Fermi (3-yr Galactic) in the GeV band\footnote{From the Fermi-LAT performance \url {http://www.slac.stanford.edu/exp/glast/groups/canda/lat_Performance.htm}}, and the current (H.E.S.S.) and future (CTA) experiments in the TeV band (for 50 hours and 5$\sigma$ detection) (e.g., Gast et al., 2011, Actis et al. 2012).

It is interesting to note that, for the considered sensitivities, no young PWN at TeV energies (for any age or magnetic fraction, top row in Fig. \ref{SED-1}) would be detectable if the pulsar's spin-down power is 0.1\% Crab or lower (and under the caveats of the assumptions discussed in the previous section, e.g., assuming the same spectral slope in the injection than those in the Crab Nebula). This conclusion is particularly stable for H.E.S.S.-like telescopes;  the youngest of the pulsar's considered is more than one order of magnitude below the sensitivity considered. The effect of using different injection or FIR energy density to this conclusion is discussed below. 

For more energetic pulsars (1\% of Crab, middle row) distinctions in age and magnetic fraction appears to reflect strongly on the TeV flux and therefore on the detectability of the nebulae. For instance, only low magnetic fraction, i.e., particle dominated nebulae, can be detected by H.E.S.S.-like telescopes if young enough (a few thousand years). If the same nebulae were in equipartition, the TeV luminosity would be very much suppressed and the detection even with CTA would require a deep observation. Which exact ages of the nebulae will CTA detect in these conditions will ultimately depend on the injection and environmental parameters. For the ones we have assumed, larger ages are preferred, when enough electrons are available for interaction. 
On the contrary, nebulae powered by pulsars with spin-down of 10\% Crab or more are all detectable by H.E.S.S.-like telescopes if they are particle dominated, no matter the age (bottom-left panel of Fig. \ref{SED-1}). The possibility of detection is less clear in case of larger magnetic fractions (bottom-right panel, same Figure). 
We recall that 2 kpc is assumed for the distance in the scale of the left-axis of Figs. \ref{SED-1} and \ref{SED-2}, but the general trend of these conclusions should scale with it, worsening the chances of detection the farther the nebula is located. This already points to an interesting observational bias, which we discuss further below: if there are magnetically dominated PWNe, similar to the ones simulated here, it would be hard  to detect them with the current generation of TeV telescopes.

To illustrate the effect of the initial spin-down power injected, the SED is shown in Fig. \ref{SED-2} for each of the 4 $L_0$ proposed. Similarly to Fig. \ref{SED-1} the figures on the left panels are calculated for a particle dominated nebula ($\eta$=0.03) and the right ones for equipartition ($\eta$=0.5).
The trends noted above are more clearly shown here. In particular, for relatively old PWNe, at ages of 9000 years (bottom row), the increase of the X-ray nebula to detectable levels with the current instruments has a strong dependence on the magnetic fraction considered. For instance, a relatively bright pulsar with a spin-down energy of 10\% of the Crab, at 9000 years would be detectable by ROSAT if in equipartition, but not for smaller magnetic fields.

%%%%%%%%%%%%%%%%%%%%%%%%%%%%%%%%%%%%%%%%%%%%%%%%%%
\subsection{The effect of magnetic fraction on the X-ray and TeV luminosity}
%%%%%%%%%%%%%%%%%%%%%%%%%%%%%%%%%%%%%%%%%%%%%%%%%%

\begin{landscape}
\begin{figure}
\begin{center}
\includegraphics[scale=0.3245]{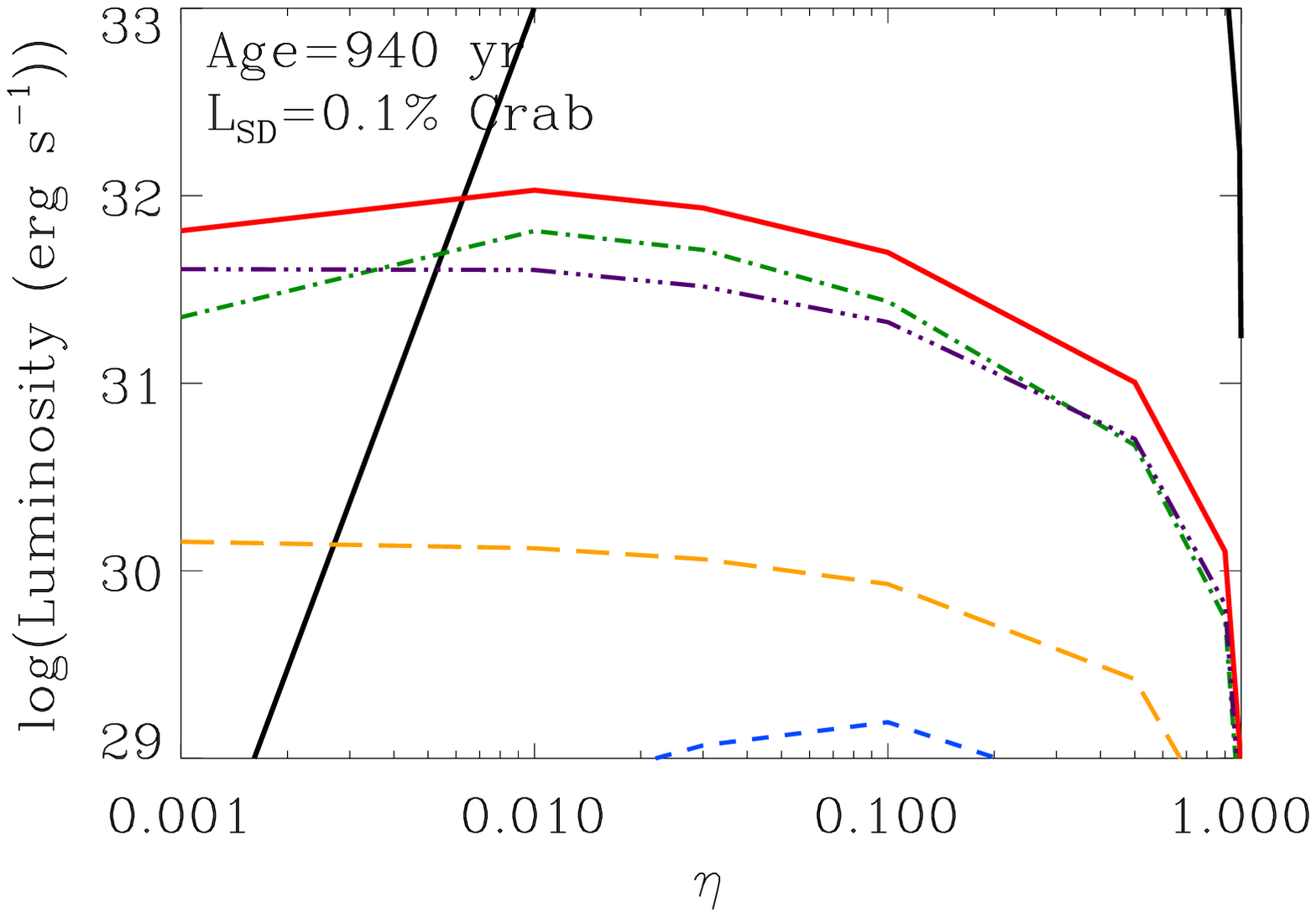}  \includegraphics[scale=0.3245]{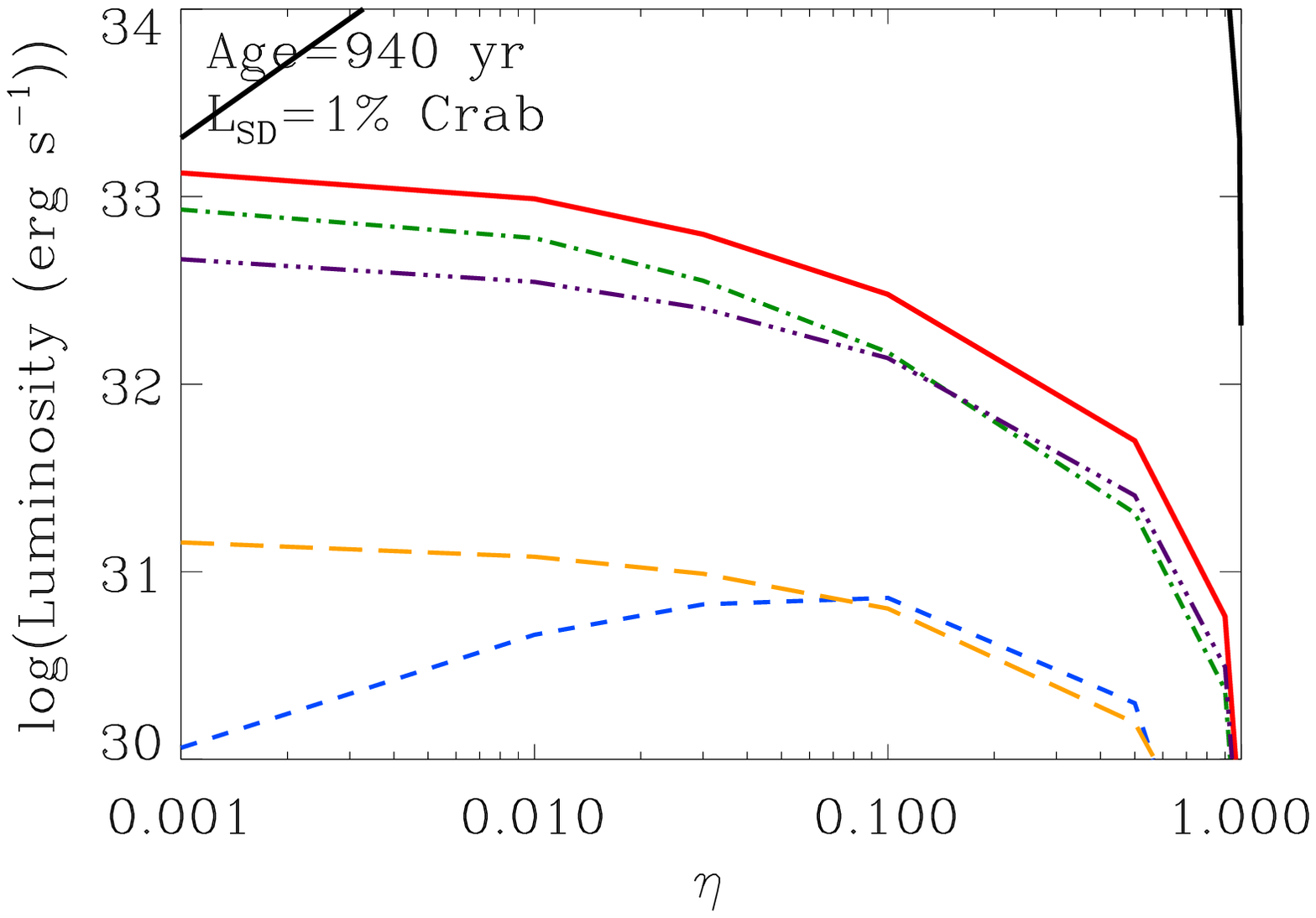} 
\includegraphics[scale=0.3245]{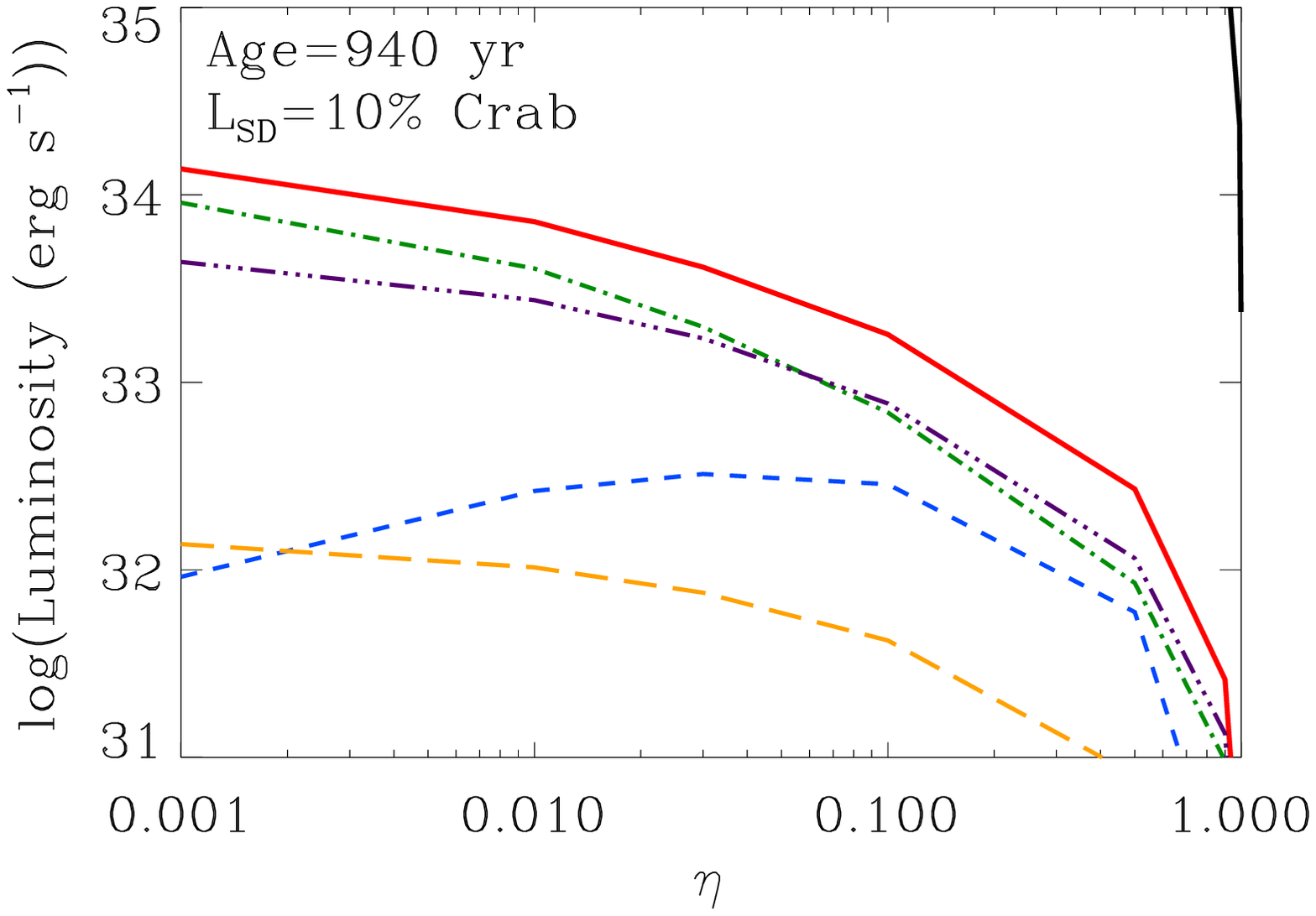}  \includegraphics[scale=0.3245]{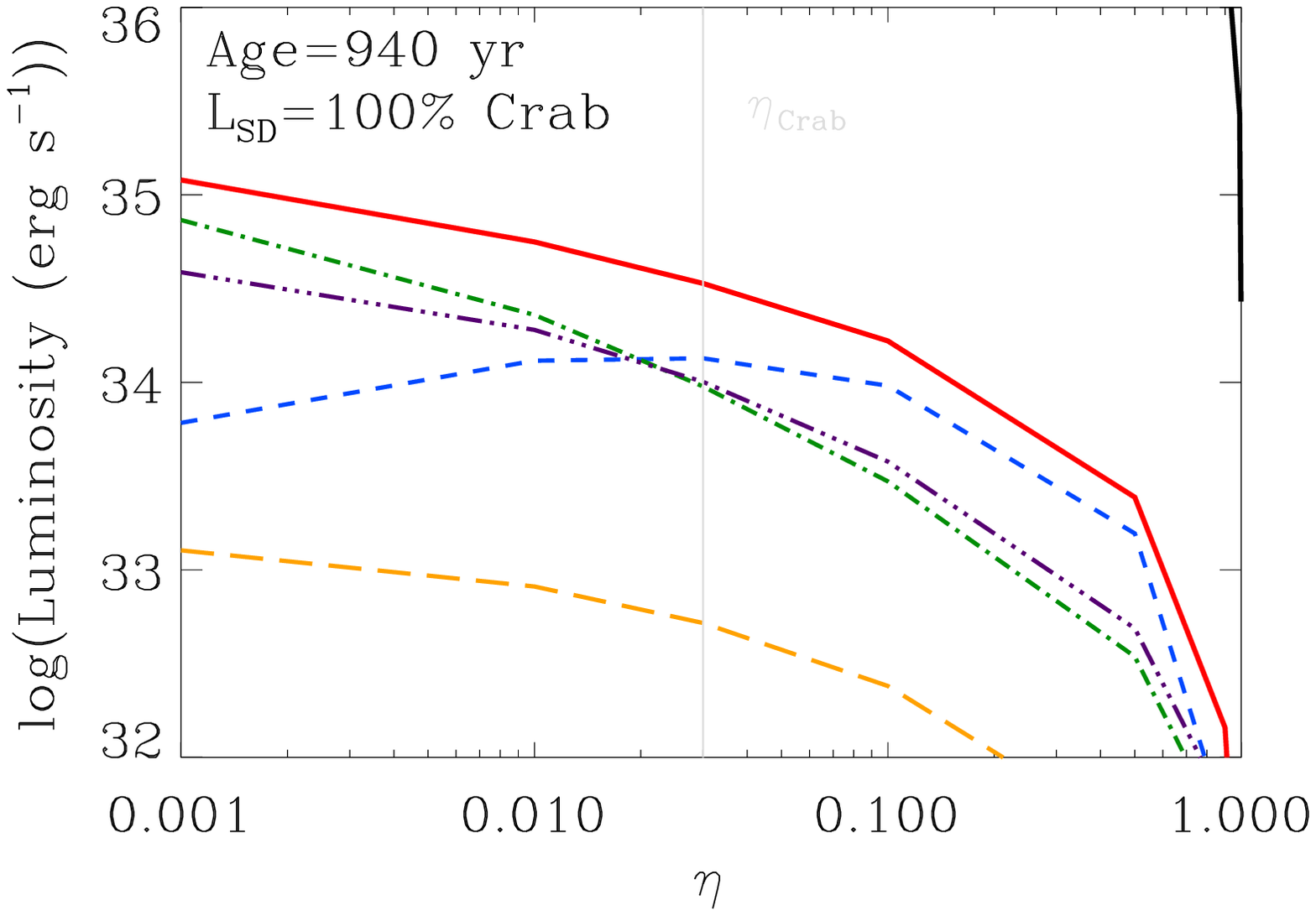} \\
\includegraphics[scale=0.3245]{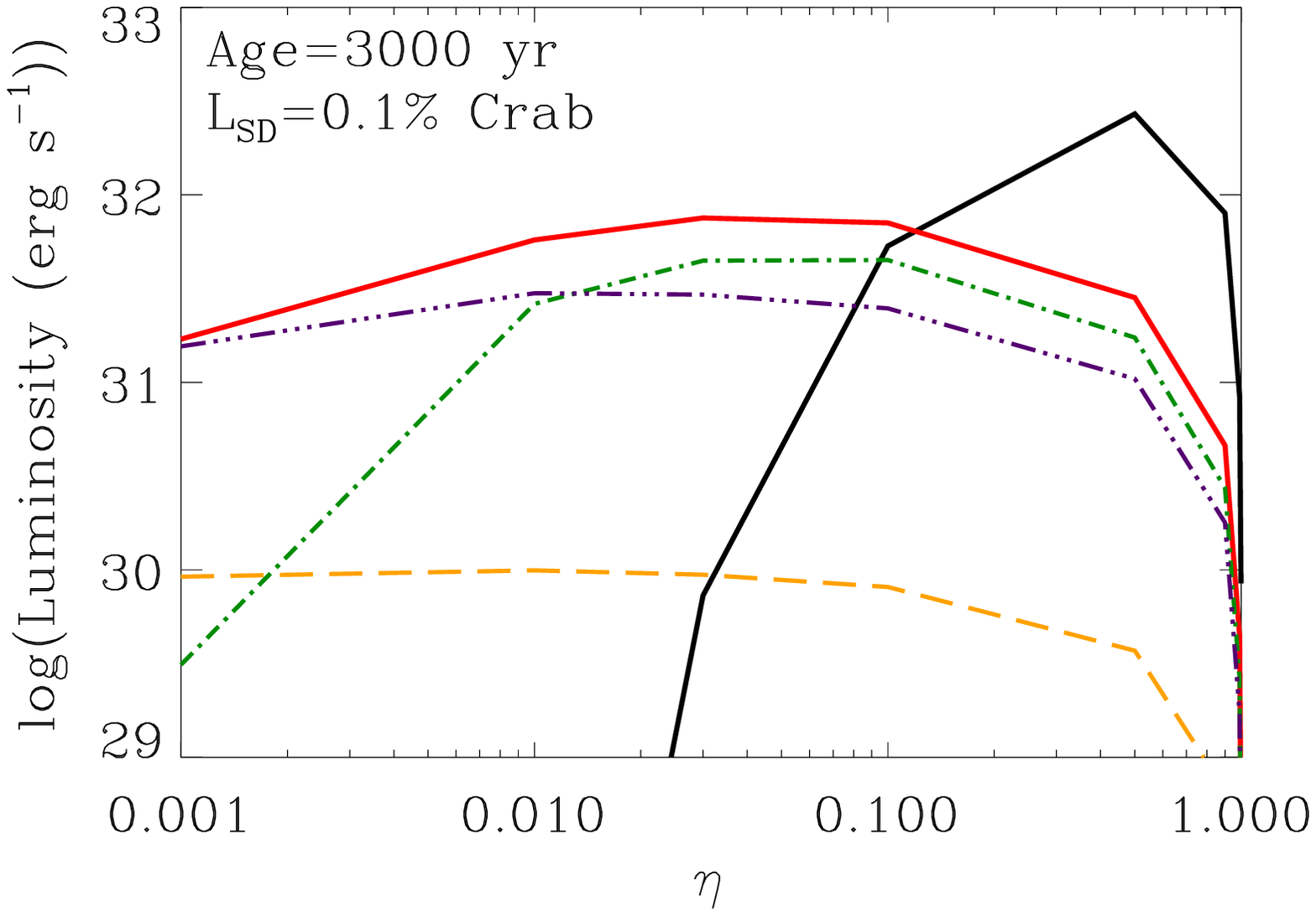} \includegraphics[scale=0.3245]{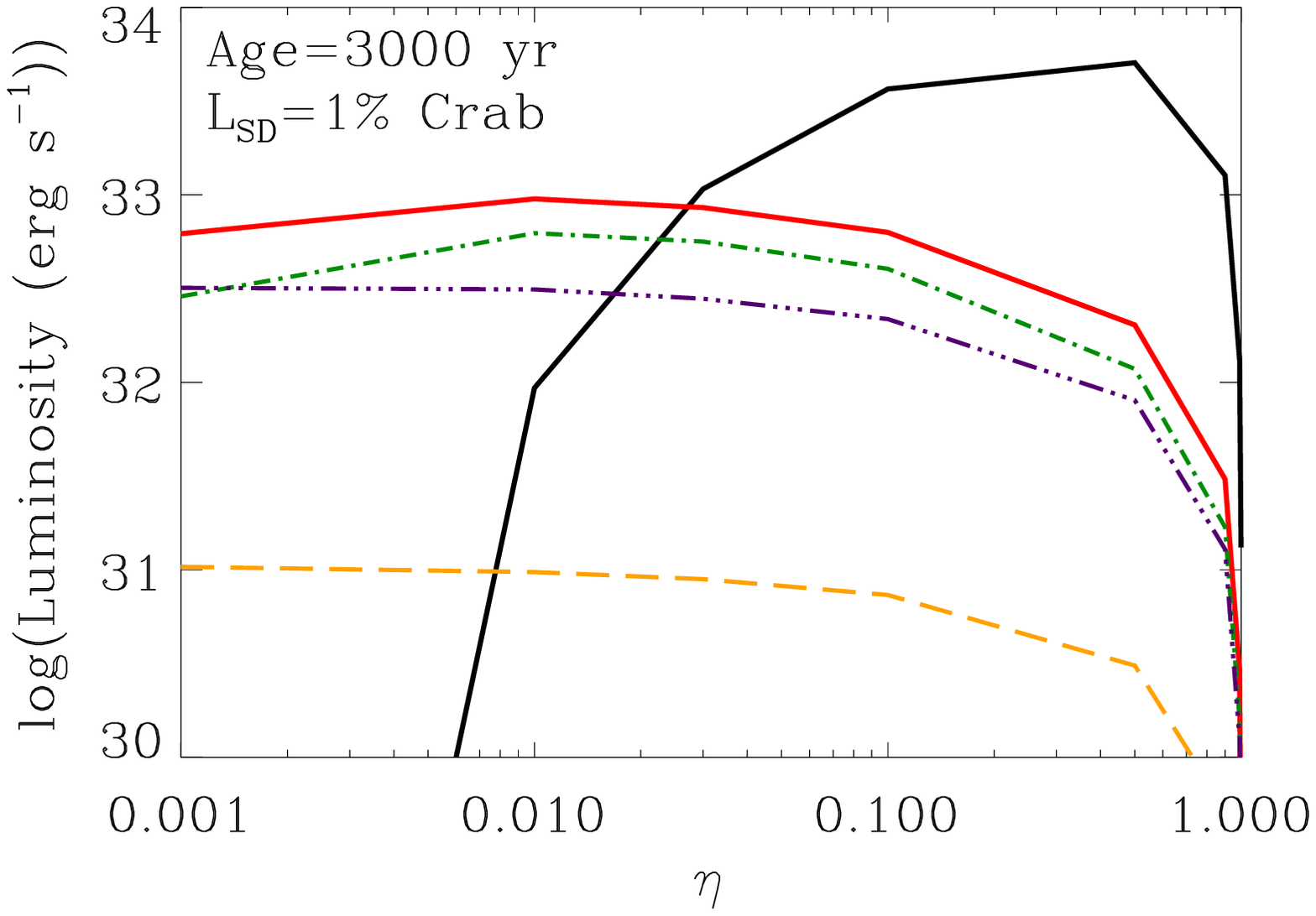} 
\includegraphics[scale=0.3245]{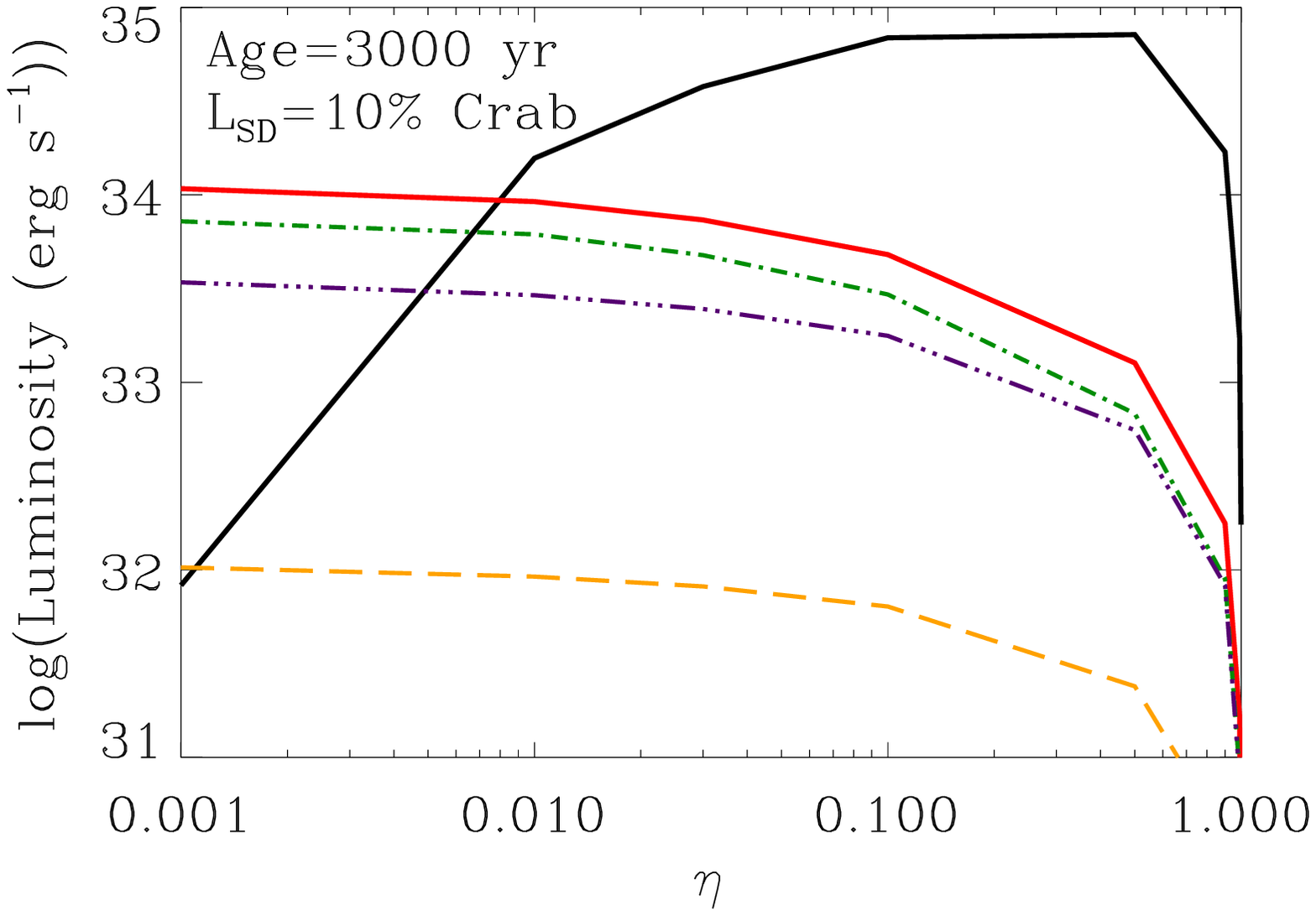}  \includegraphics[scale=0.3245]{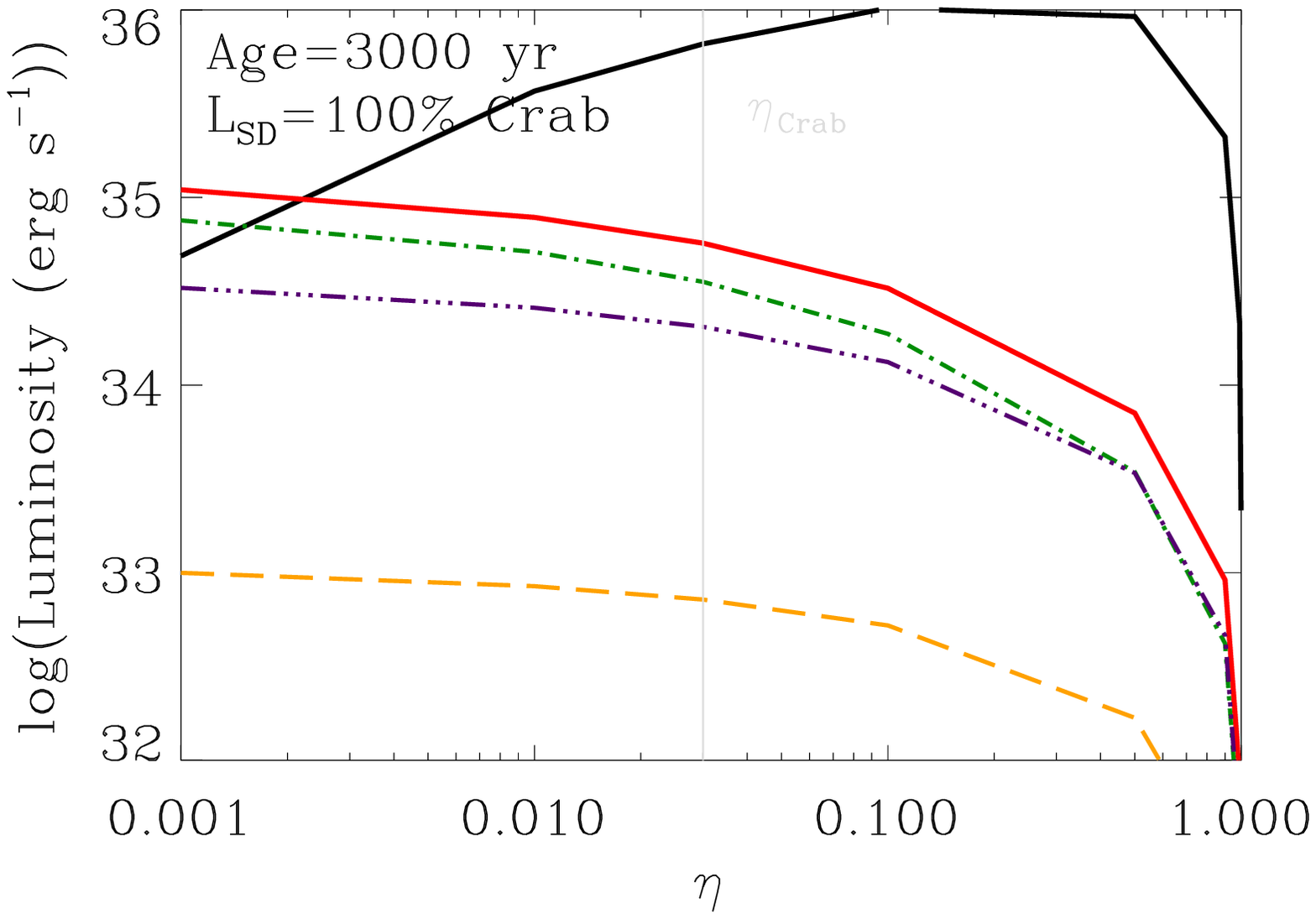} \\
\includegraphics[scale=0.3245]{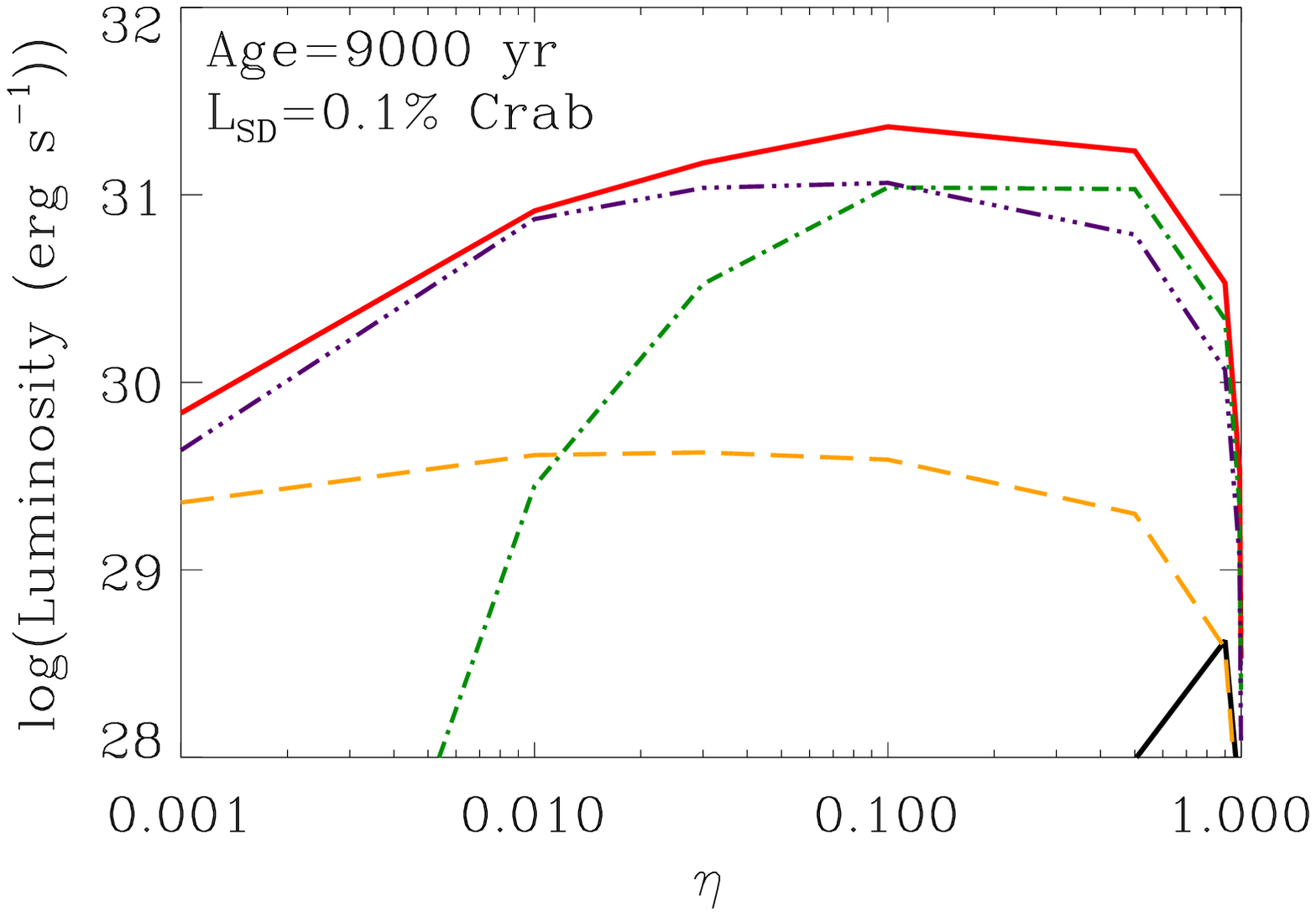}  \includegraphics[scale=0.3245]{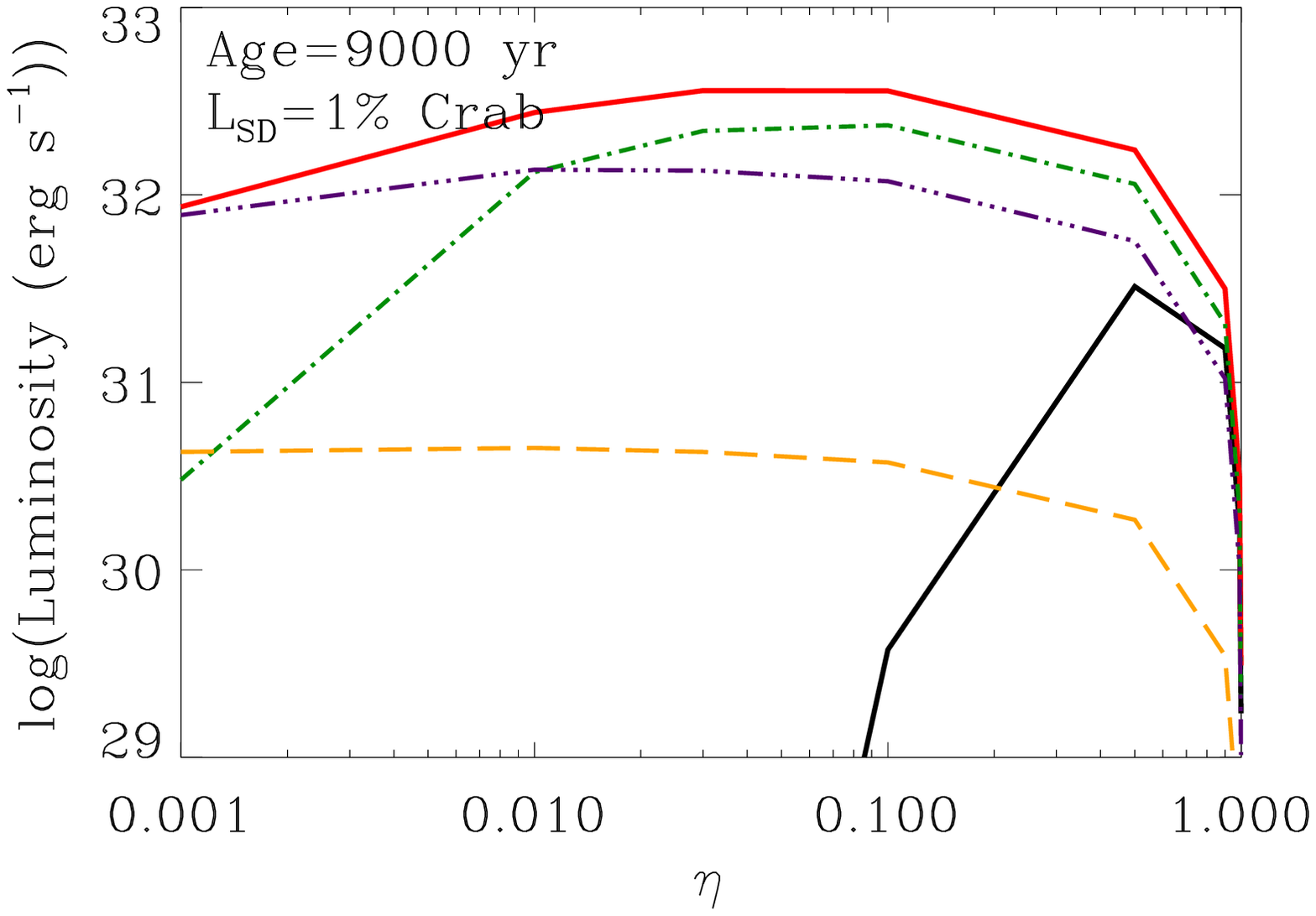} 
\includegraphics[scale=0.3245]{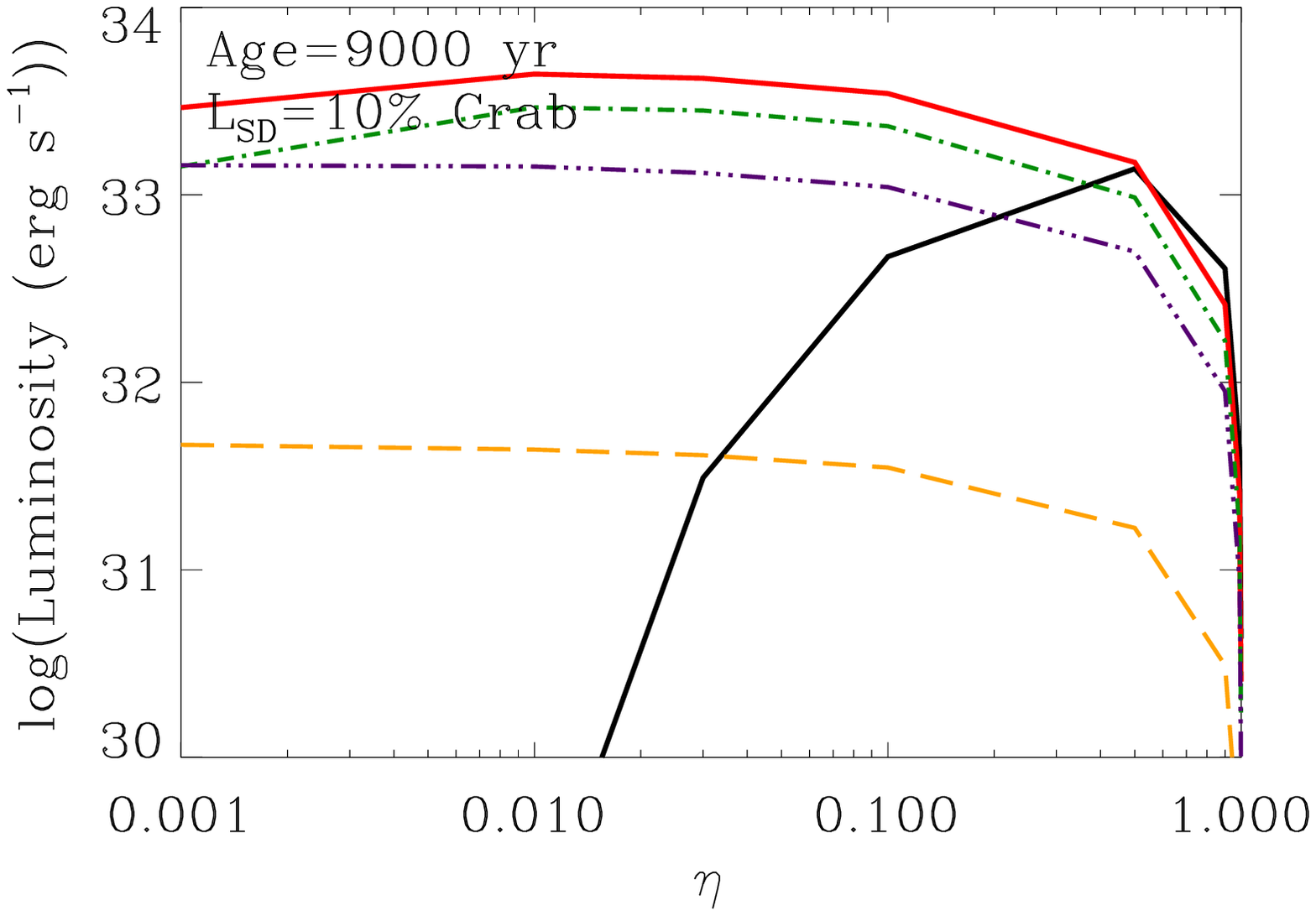}  \includegraphics[scale=0.3245]{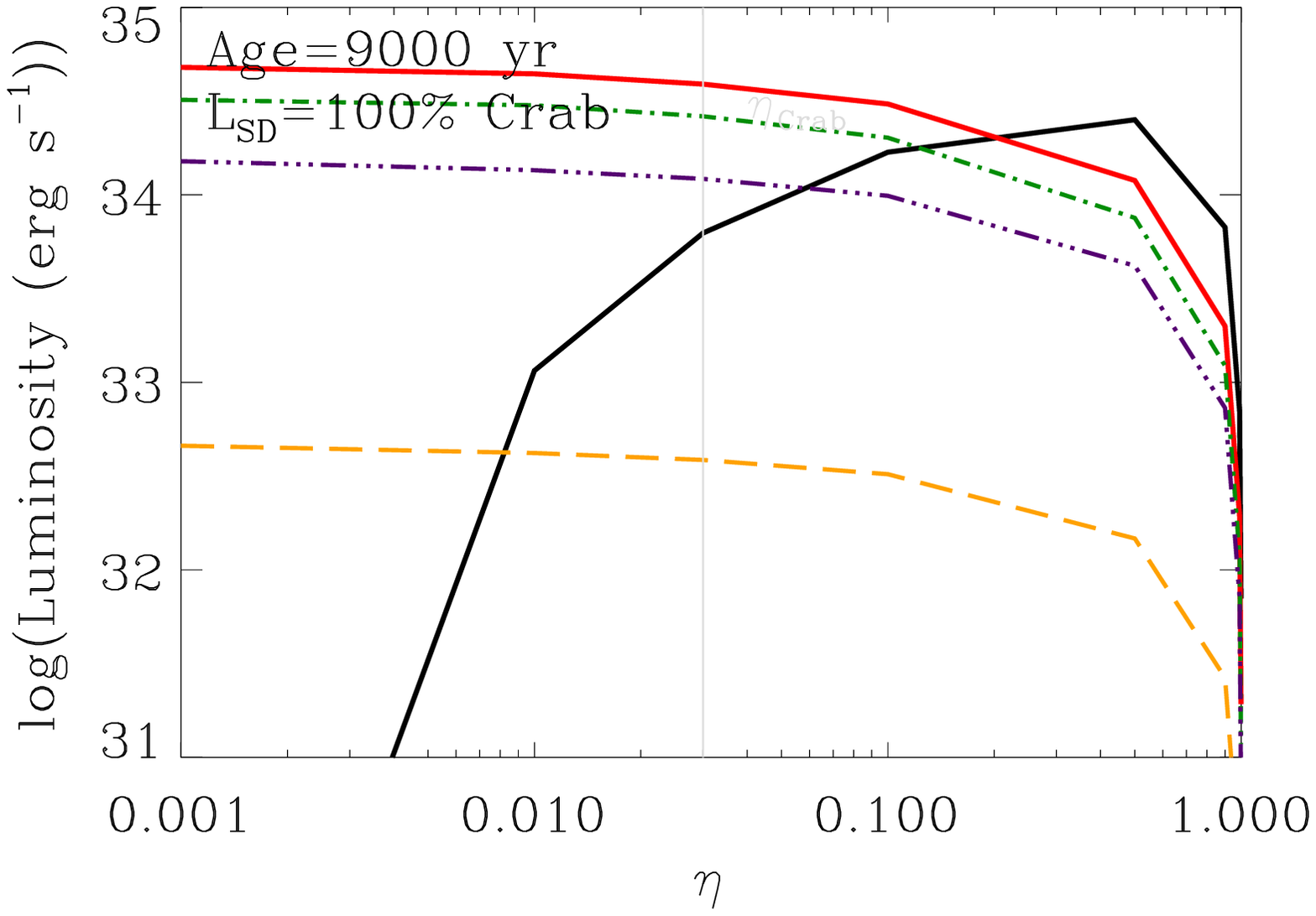}
\end{center}
\caption{Luminosities between 1 and 10 TeV of the IC, and between 1 and 10 keV of the synchrotron contributions to the spectrum as a function of the magnetic fraction. We fix an 
age of 940, 3000, and 9000 years (from top to bottom) and vary the spin-down luminosity (0.1\%, 1\%, 10\% \& 100\% of Crab, from left to right). The color coding is as in Fig. \ref{lum-age}. }
\label{lum-eta}
\end{figure}
\end{landscape}

Generally, magnetic equipartition is assumed when discussing X-ray nebulae, but recent TeV observations have shown that many (if not all) of these PWNe are particle dominated. 
To analyze in more detail the impact of the magnetic fraction parameter on the detectability of the nebulae (or on their flux level), we represent the IC contribution to the spectral flux between 1 and 10 TeV as a function of the magnetic fraction (Fig. \ref{lum-eta}). As discussed before, large spin-down and very young ages (top right panels) are required to observe a relevant contribution of SSC. The rightmost top panel corresponds to a pulsar such as Crab, having its age but different magnetic fraction. The contribution of SSC dominates for $\eta > 0.02$ whereas for lower $\eta$-values the total luminosity would be dominated by FIR even for a 940 years pulsar.

\begin{landscape}
\begin{figure}
\begin{center}
\includegraphics[scale=0.4]{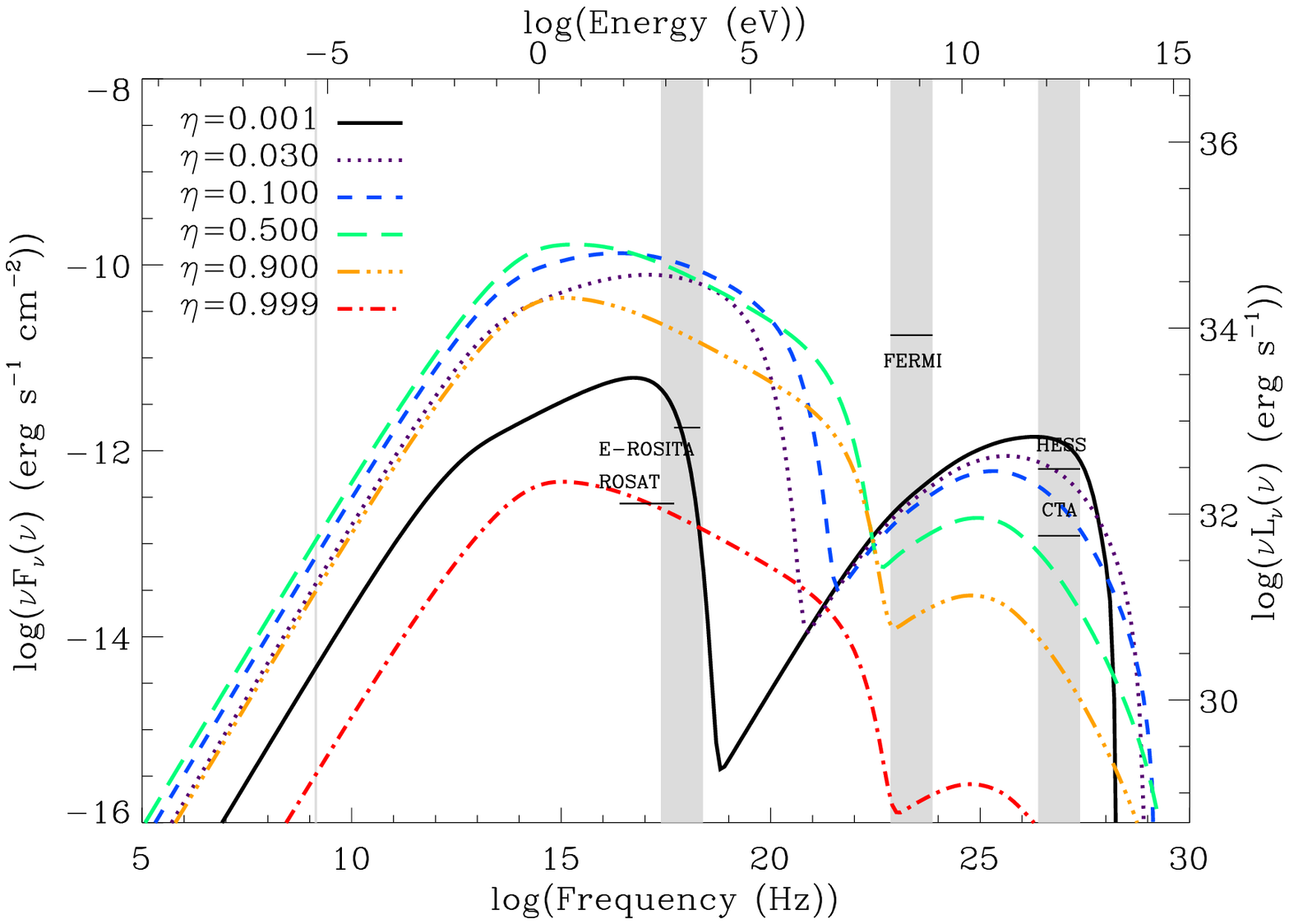}  \hspace{0.2cm} \includegraphics[scale=0.4]{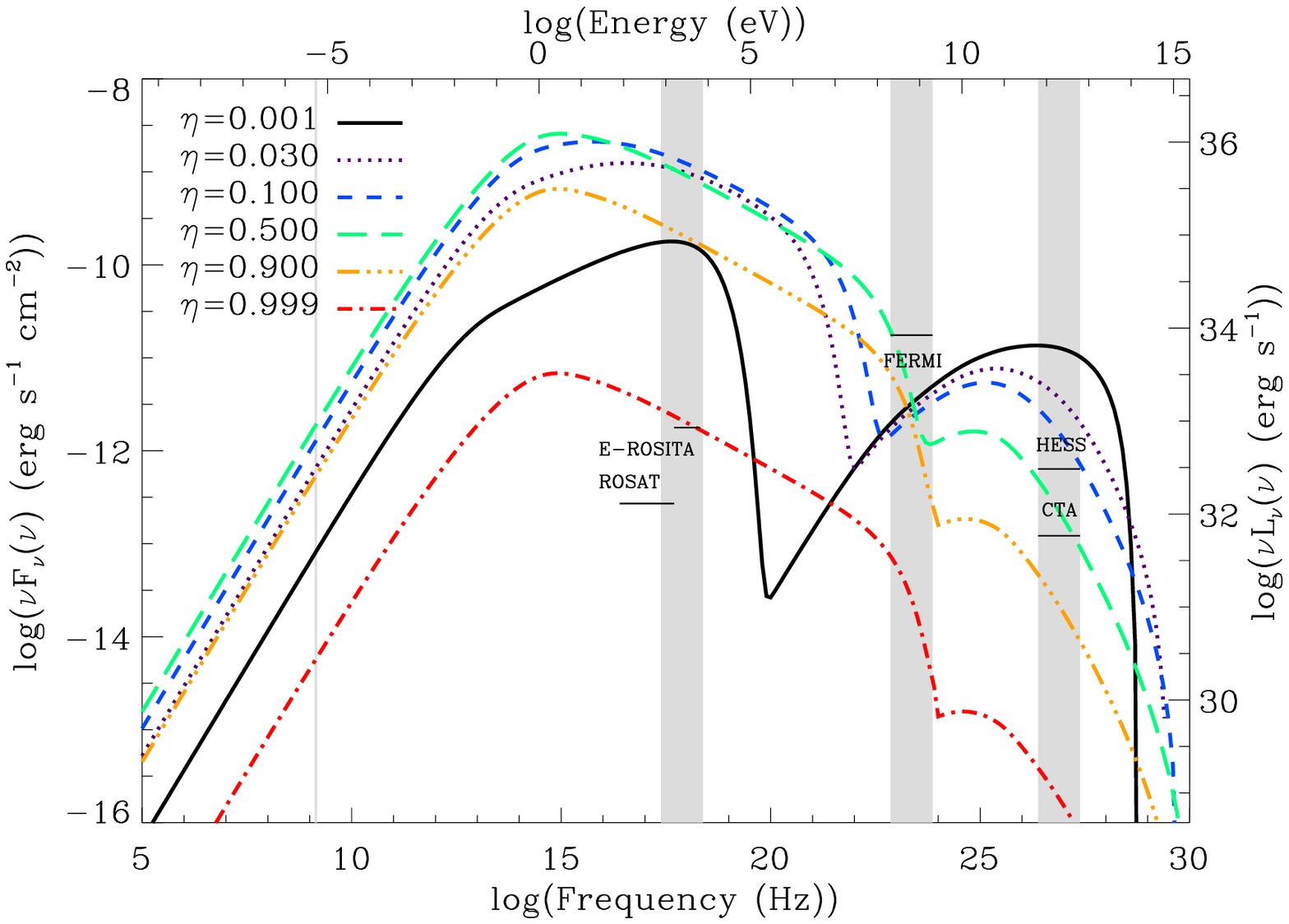} \hspace{0.2cm} \includegraphics[scale=0.4]{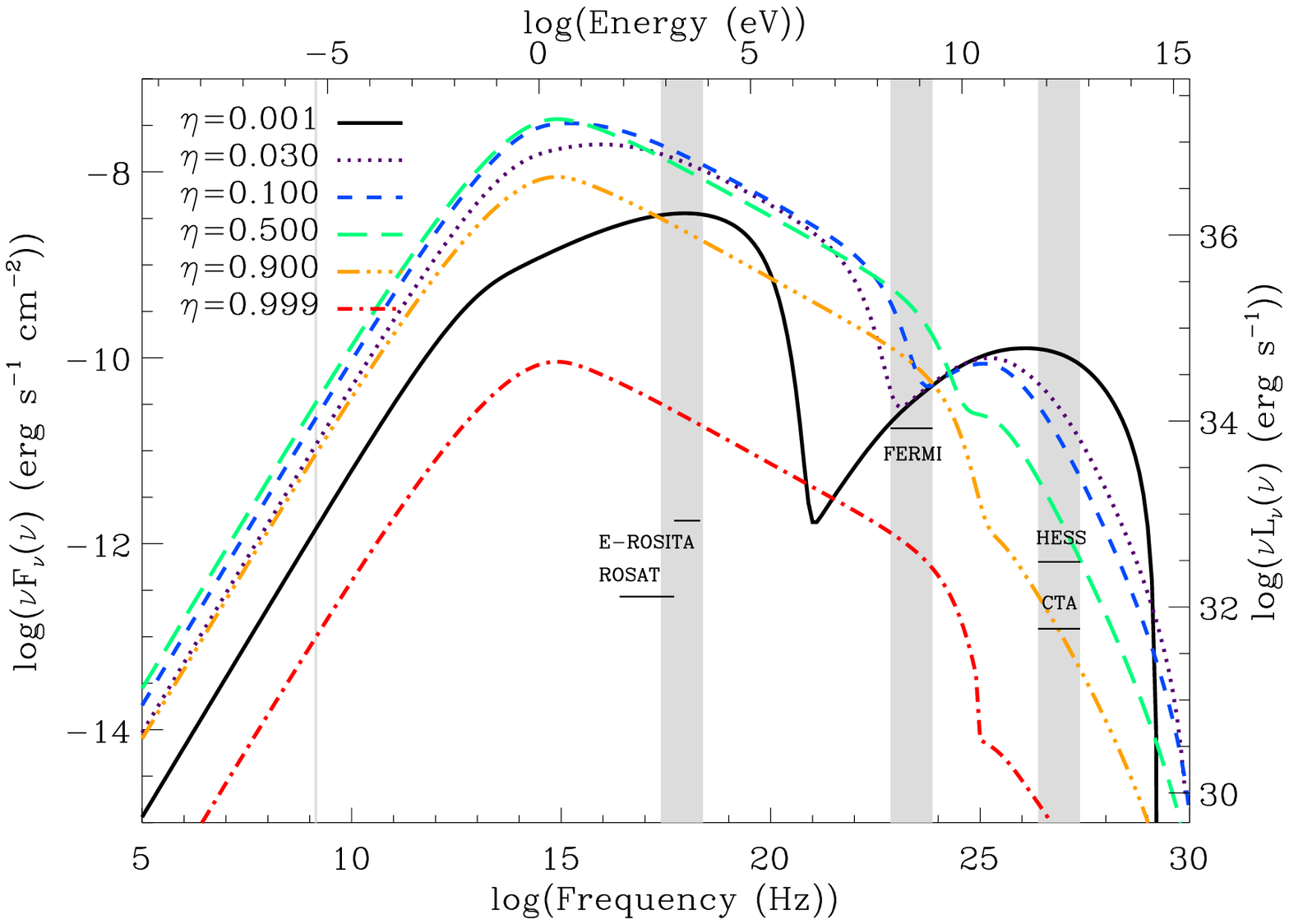} \\
\includegraphics[scale=0.4]{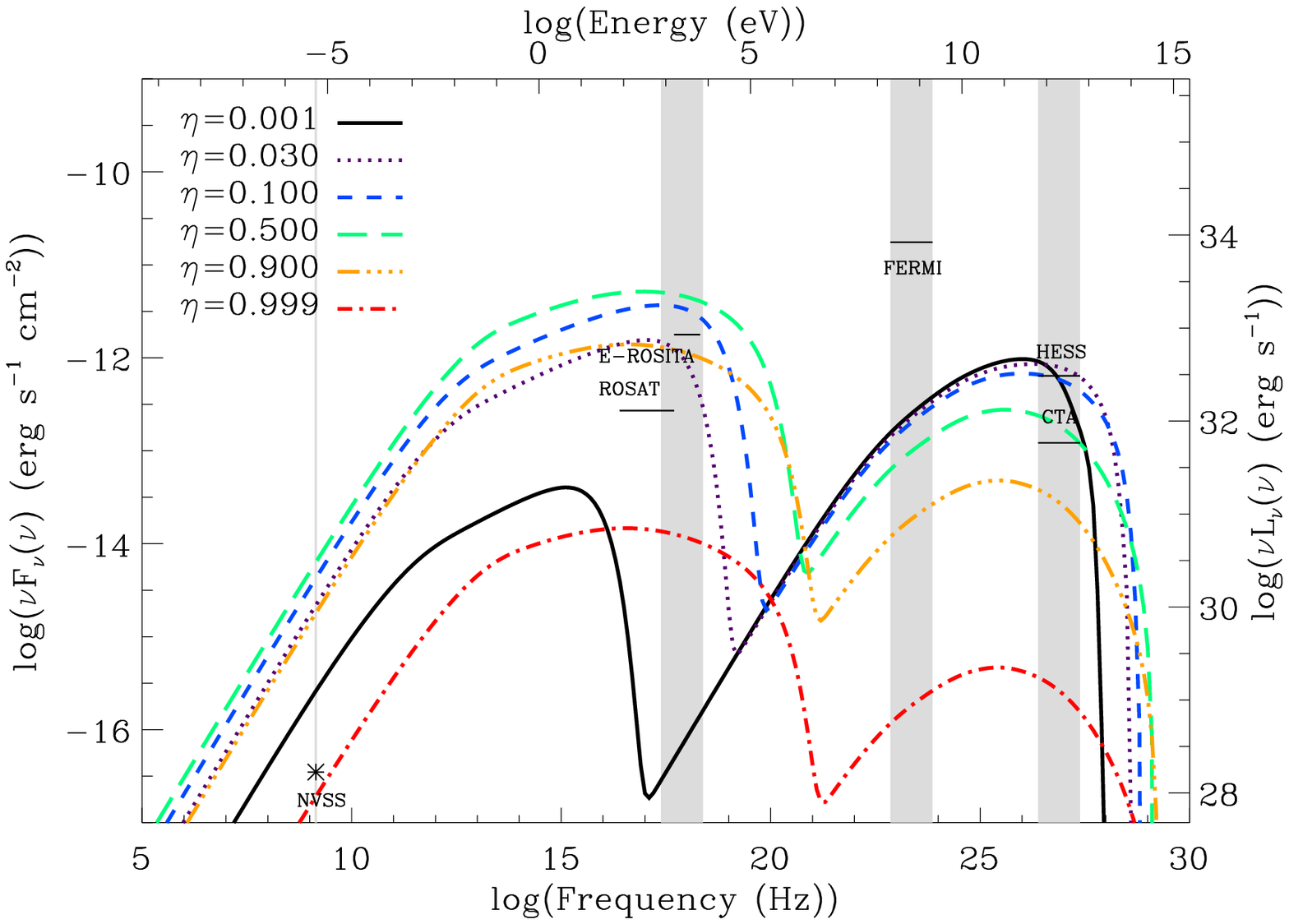} \hspace{0.2cm} \includegraphics[scale=0.4]{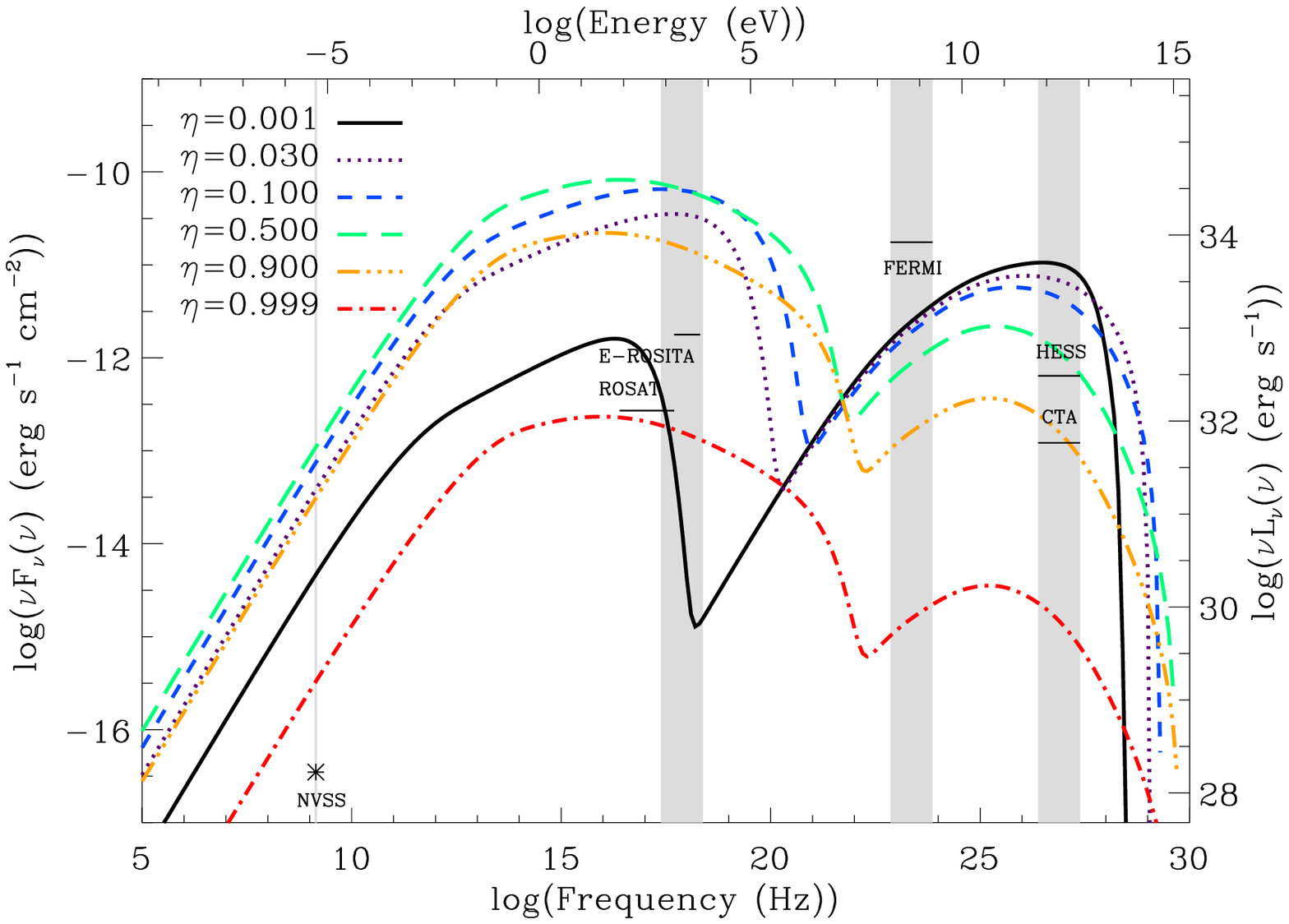} \hspace{0.2cm} \includegraphics[scale=0.4]{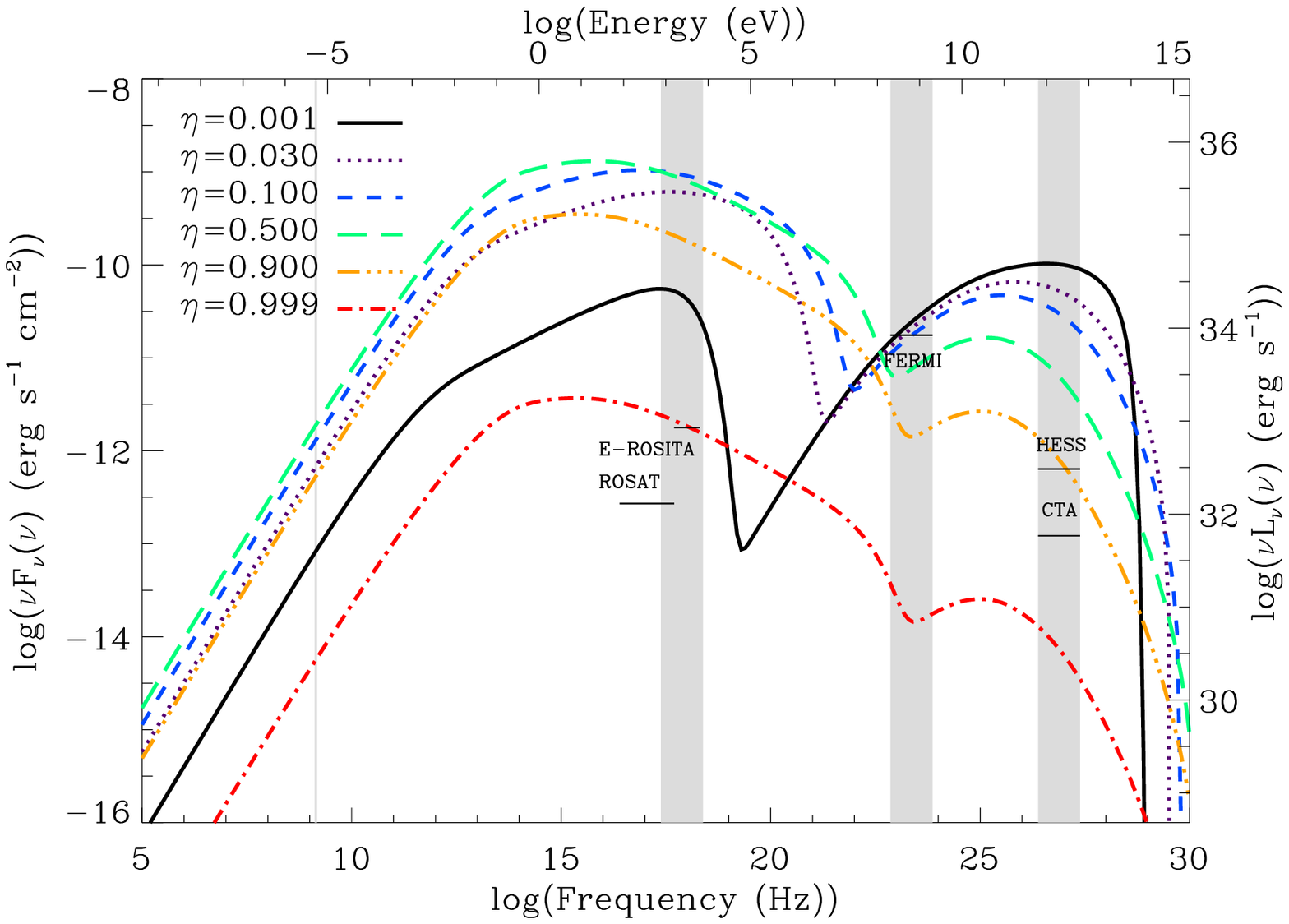}\\
\includegraphics[scale=0.4]{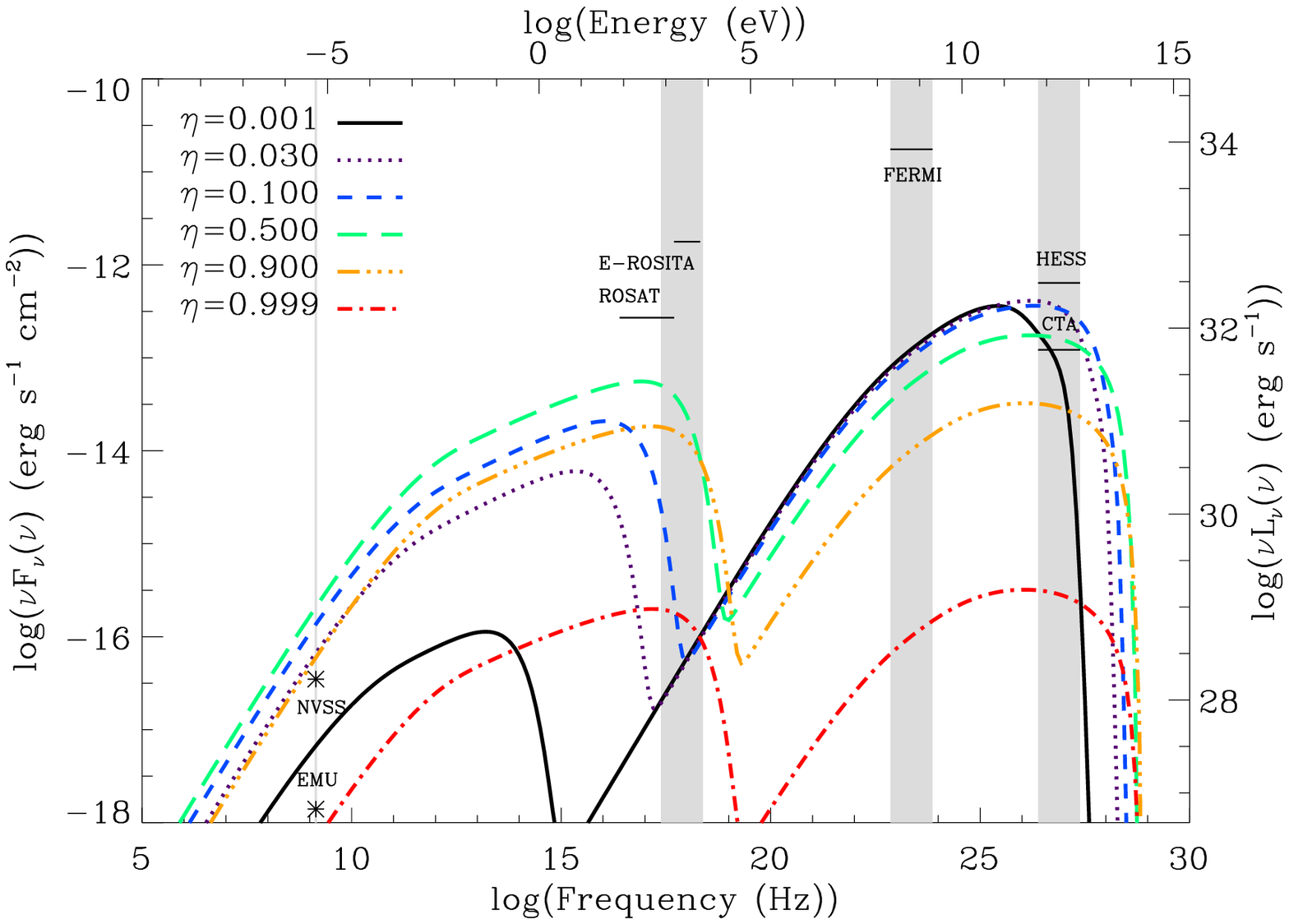} \hspace{0.2cm} \includegraphics[scale=0.4]{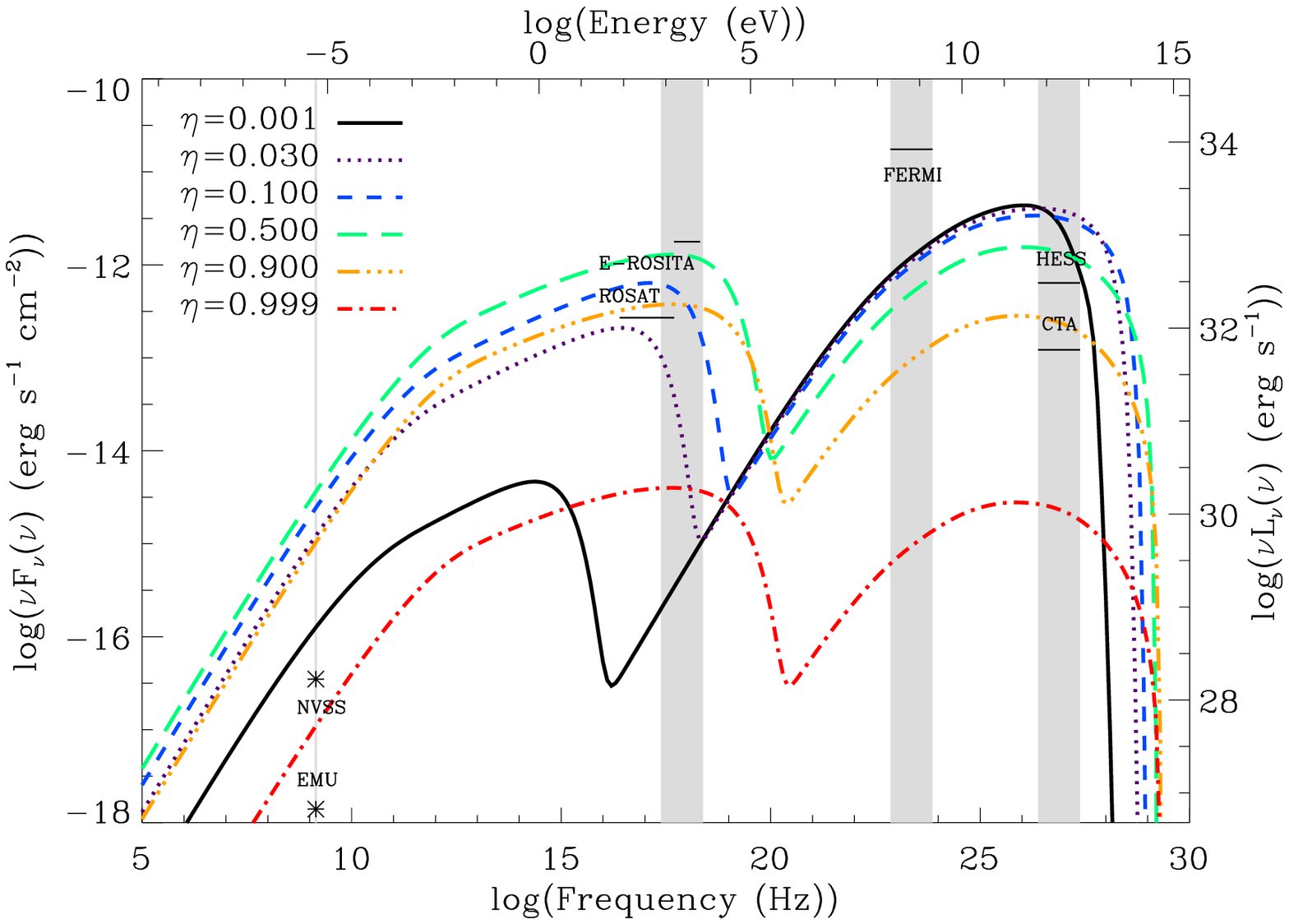} \hspace{0.2cm} \includegraphics[scale=0.4]{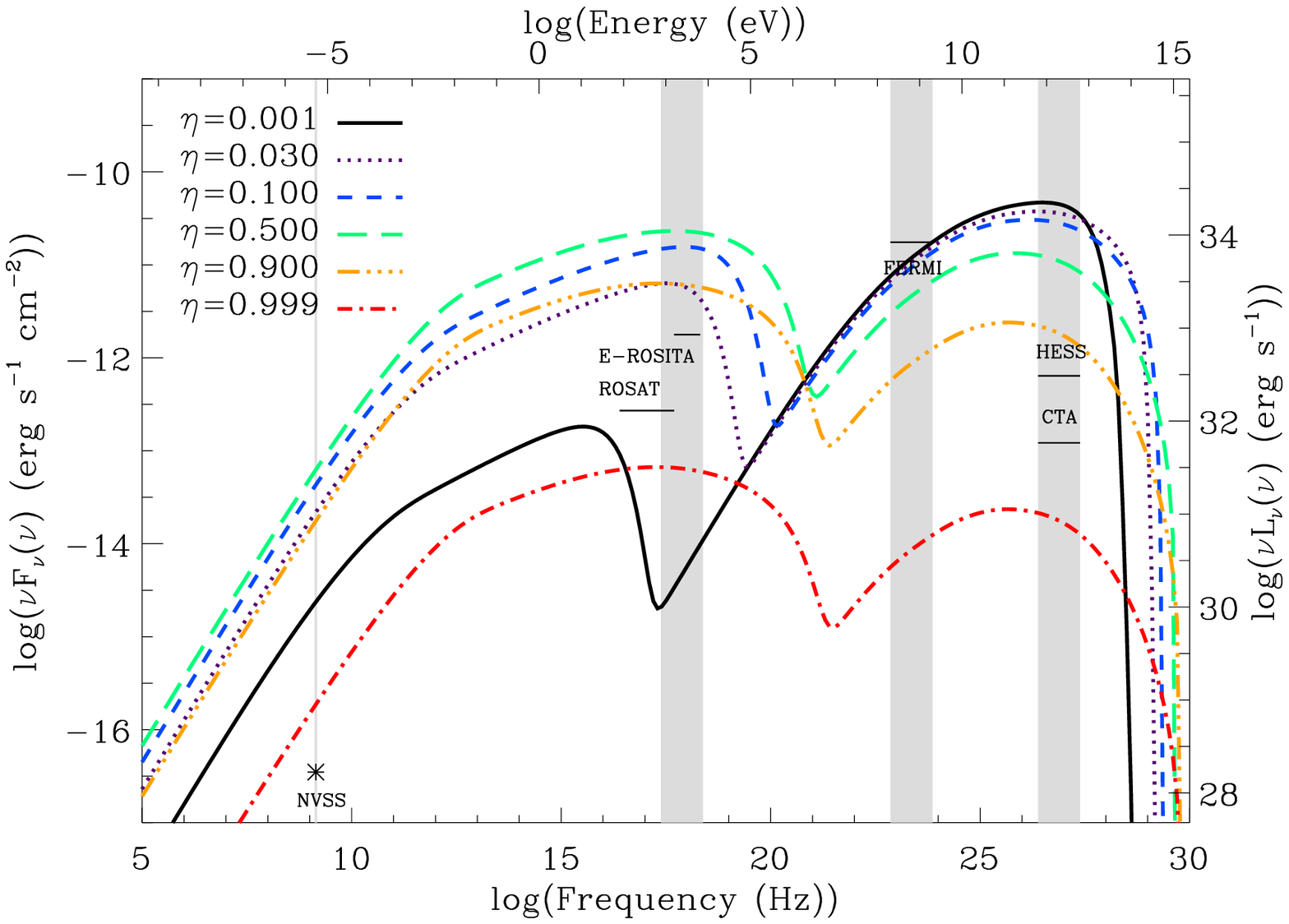}
\end{center}
\caption{Comparison of the SEDs for an age of 940, 3000, and 9000 years (from top to bottom) as a function of the magnetic fraction. The spin-down luminosity is fixed at 1\% 
10\%, and 100\% (panels from left to right) of Crab. }
\label{lum-eta-sed}
\end{figure}
\end{landscape}

Fig. \ref{lum-eta-sed} shows the corresponding SEDs for the three ages under consideration, and three spin-down powers (1, 10, and 100\% of Crab's). The impact of the magnetic fraction on the final SED is large, generating orders of magnitude variations in the luminosity even when keeping all other system's parameters fixed (same spin-down, $P$, $\dot P$, injection, and environment). At TeV energies, the simulations show (for Crab-like photon field background and injection parameters) that H.E.S.S.-like telescopes would not be sensitive enough to fully explore the $\eta>$0.5 regime, independently from the pulsar age (when lower than $10^4$ years) or spin-down power. Even for CTA, a complete coverage of the phase space of young nebulae (assuming that strongly magnetic field-dominated nebulae exist, of course) can only be partially achieved for up to $\eta < 0.9$ and $L_0>10\%$ Crab, for near PWNe.
Below, we give details on the impact that a different injection or a different FIR background energy density have
on this conclusion.

Finally, we calculate the total bolometric power integrating the total luminosity $L(\nu)$, corresponding to the spectra in Fig. \ref{lum-eta-sed}. The results for 1\% and 10\% of the Crab luminosity are shown in Table \ref{ratios-bol-sd}. The total radiated power is in all cases less than the injected spin-down (see Table \ref{param}) at the age considered, amounting a few percent for young PWNe (with $\eta \sim 0.01-0.1$). 
Equipartition naturally produces the maximum
of the radiated power in all cases. The integrated-in-time spin-down power ranges from $\sim 4 \times 10^{47}$ erg, for 1\% of Crab, to $\sim 5 \times 10^{48}$ erg,  for 10\% of Crab.

\begin{table}
\centering
\scriptsize
\caption{Ratio of the bolometric 
radiated power (erg/s) of the spectra in Fig. \ref{lum-eta-sed} divided by the spin-down power (erg/s) at the given age.
Two examples are shown for 1\% and 10\% of Crab's spin-down.}
\vspace{0.2cm}
\begin{tabular}{l l l l l l l l }
\hline

	$\eta$	&	940 yr	&	3000 yr	&	9000 yr \\
\hline
1\% Crab \\
\hline	

	0.001	&	0.00580	&	0.00471	&	0.01480 \\
	0.01	&	0.04356	&	0.00794	&	0.01880 \\
	0.03	&	0.09022	&	0.01350	&	0.02013 \\
	0.1	&	0.16356	&	0.02757	&	0.02000 \\
	0.5	&	0.18889	&	0.04229	&	0.01480 \\
	0.9	&	0.04733	&	0.01169	&	0.00292 \\
	0.99	&	0.00491	&	0.00123	&	0.00030 \\
	0.999	&	0.00049	&	0.00012	&	0.00003 \\

\hline
10\% Crab \\
\hline
	0.001	&	0.01689	&	0.00640	&	0.01960 \\
	0.01	&	0.08822	&	0.01457	&	0.02280 \\
	0.03	&	0.15733	&	0.02800	&	0.02320 \\
	0.1	&	0.25778	&	0.05443	&	0.02307 \\
	0.5	&	0.26667	&	0.07157	&	0.01747 \\
	0.9	&	0.06467	&	0.01900	&	0.00427 \\
	0.99	&	0.00667	&	0.00199	&	0.00044 \\
	0.999	&	0.00067	&	0.00020	&	0.00004 \\

\hline
\hline
\label{ratios-bol-sd}
\end{tabular}
\end{table}

%%%%%%%%%%%%%%%%%%%%%%%%%%%%%%%%%%%%%%%%%%%%%%%%%%
\subsection{Luminosity ratios for different wavelengths}
%%%%%%%%%%%%%%%%%%%%%%%%%%%%%%%%%%%%%%%%%%%%%%%%%%

%Fig. \ref{lum-eta-sed} allows us to compare the expected luminosity at different wavelengths. For instance, while a 1\% of Crab's spin-down power nebula at 2 kpc in equipartition ($\eta$=0.5) (leftmost panels) would be barely detected by CTA, the flux at X-rays would be enough to be detected with a typical X-ray survey sensitivity if the pulsar is young enough (a few thousand years at most). For older sources ($\sim$9000 years), the X-ray flux would quickly decrease while staying more or less constant in the TeV range. The radio instruments are sensitive enough to detect all bright PWNe (1 or 10\% $L_0$) provided that their angular extension is suitable to the antennas. On the contrary, the detection of GeV emission seems difficult to achieve, being the predictions orders of magnitude lower than the Fermi-LAT sensitivity. The only exception is the Crab-like realization, which at young ages shows a bright-enough emission above 100 MeV to allow detection. 

\begin{landscape}
\begin{figure}
\begin{center}
\includegraphics[scale=0.3245]{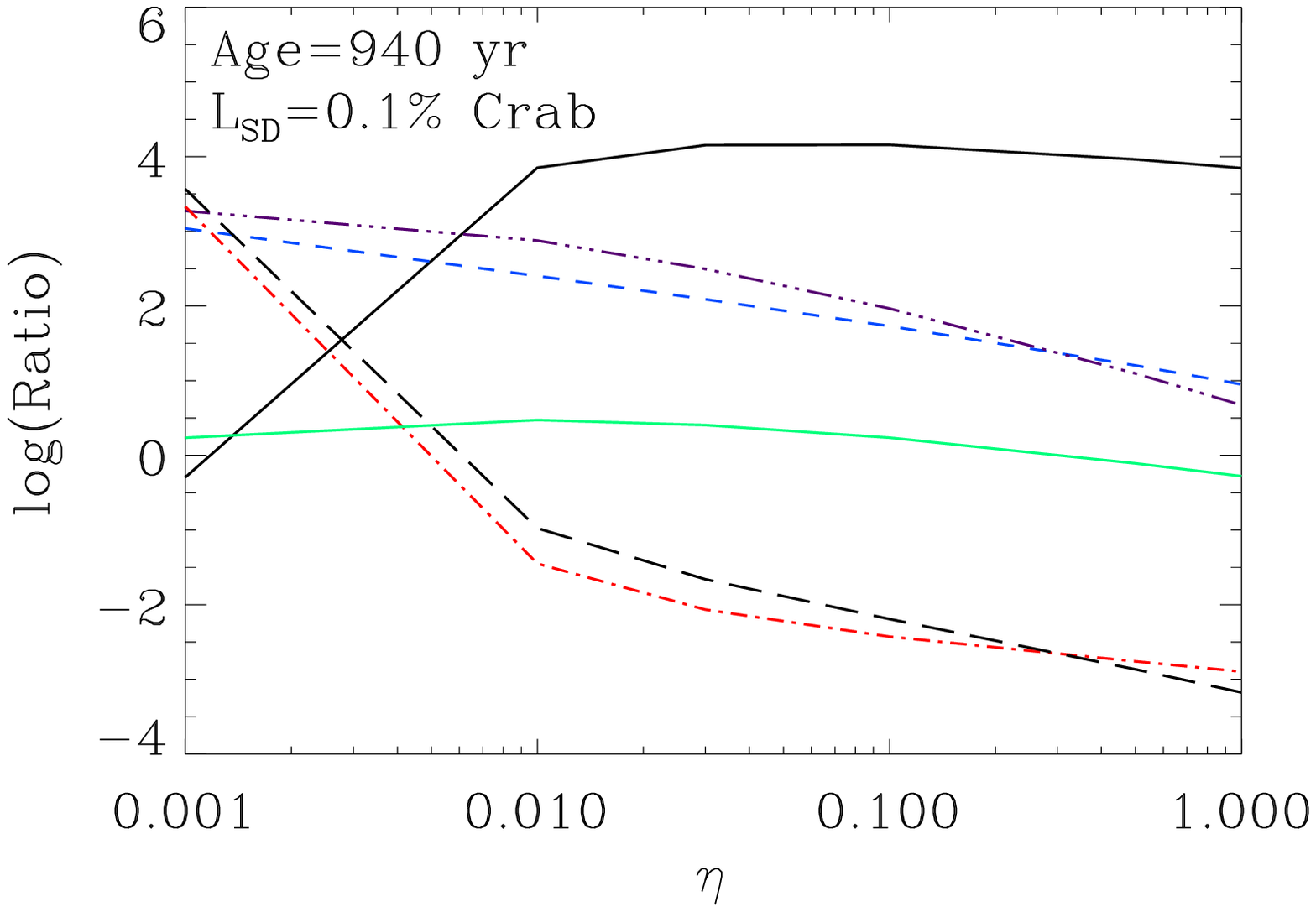}  \includegraphics[scale=0.3245]{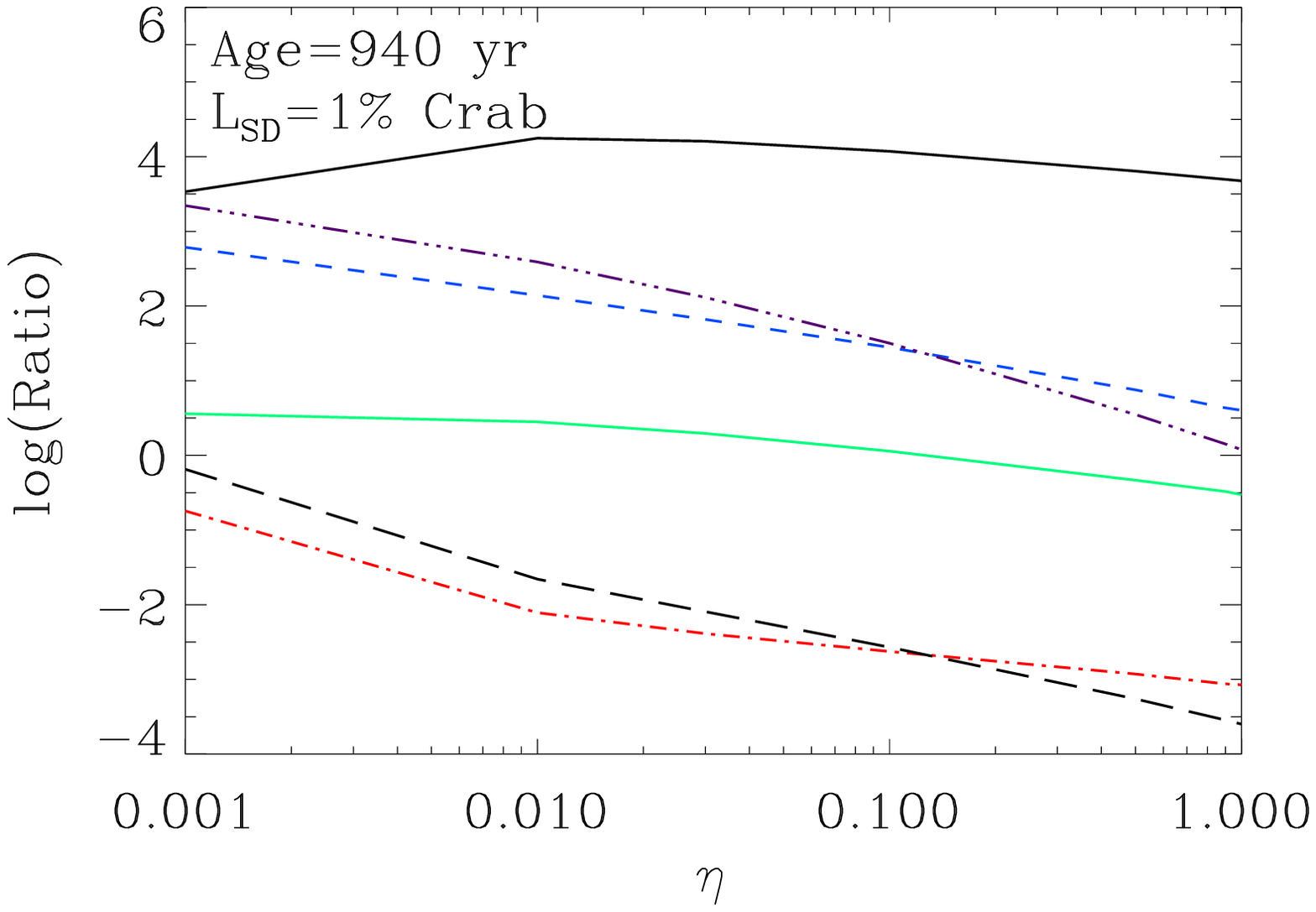} 
\includegraphics[scale=0.3245]{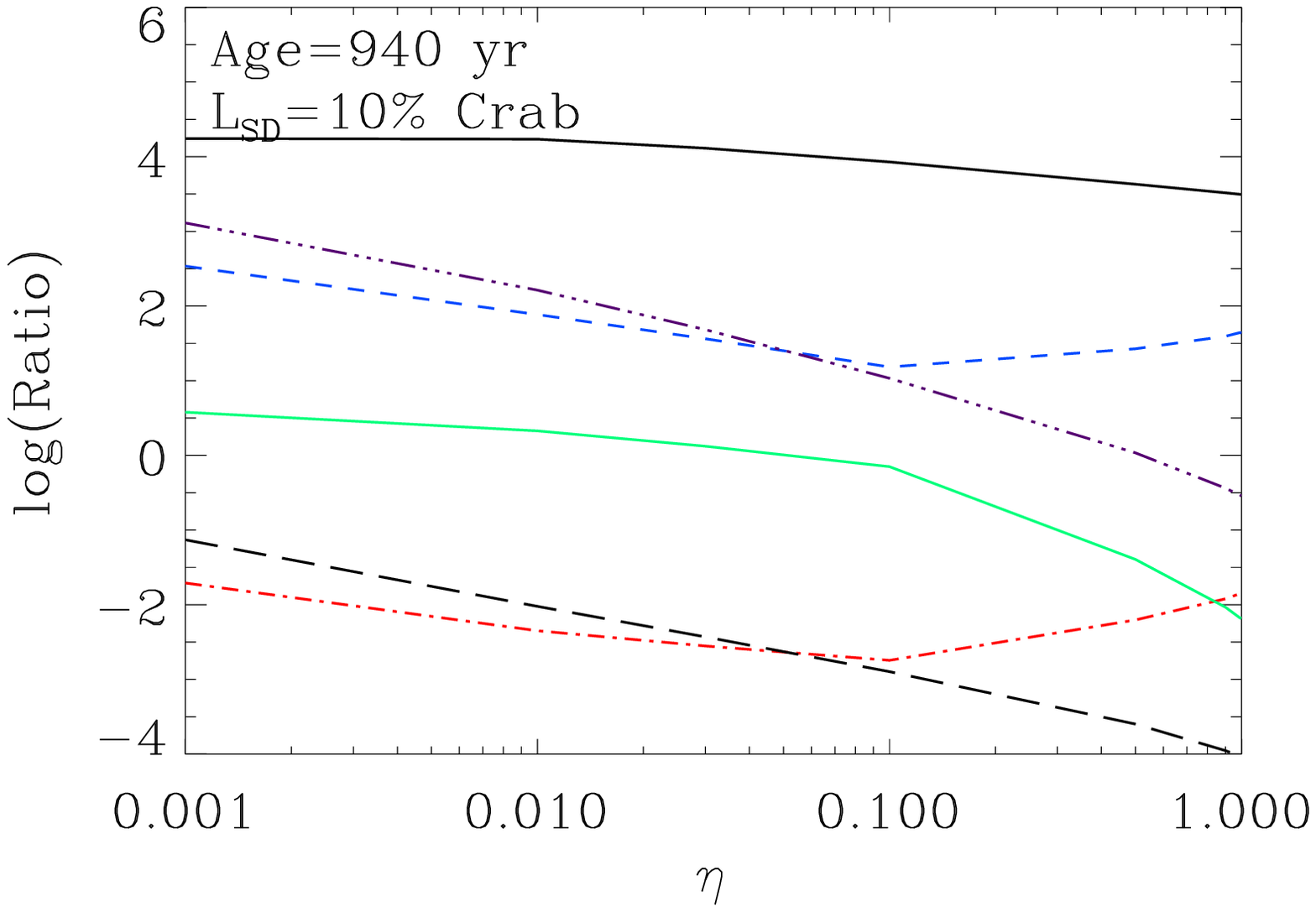}  \includegraphics[scale=0.3245]{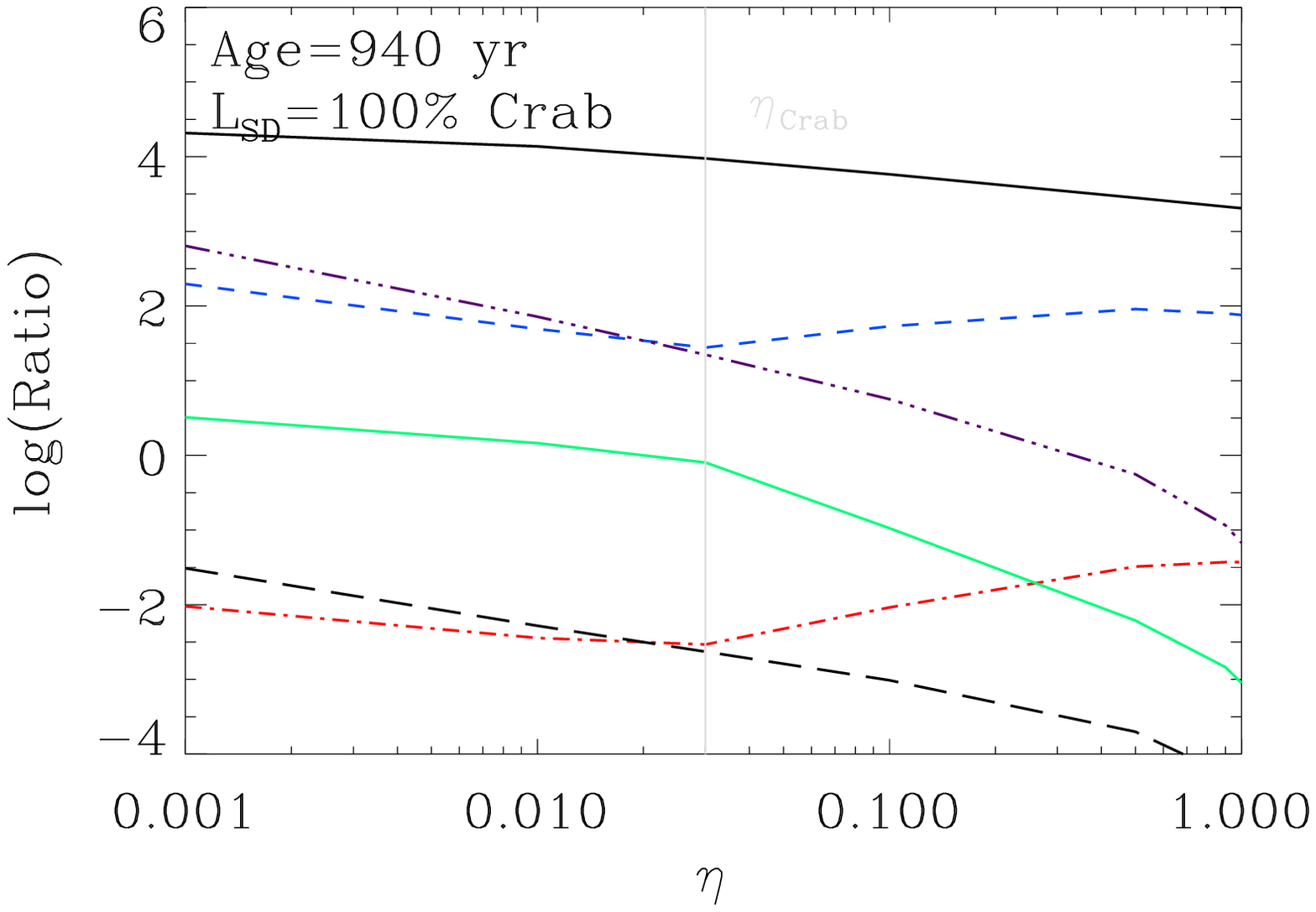} \\
\includegraphics[scale=0.3245]{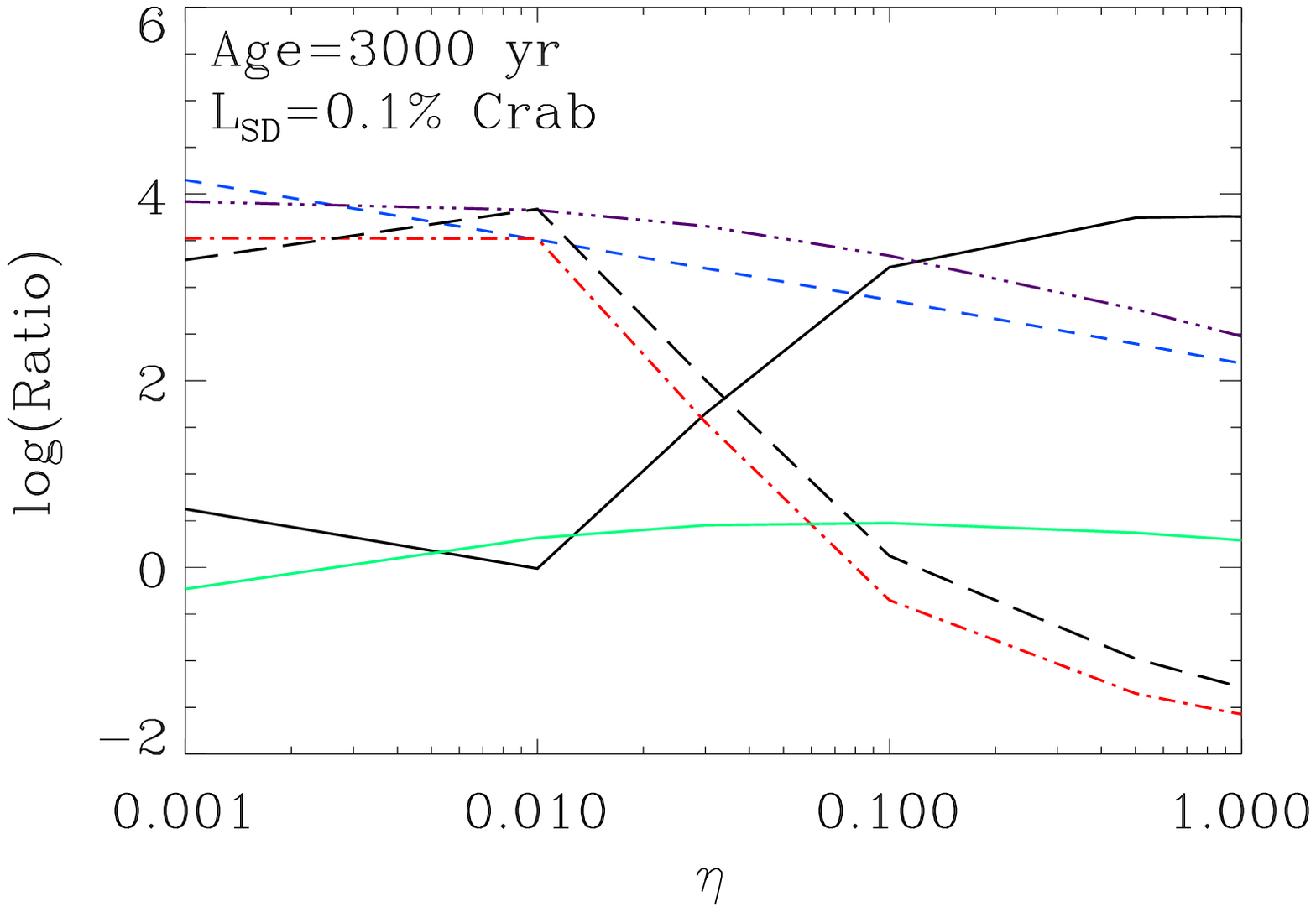}  \includegraphics[scale=0.3245]{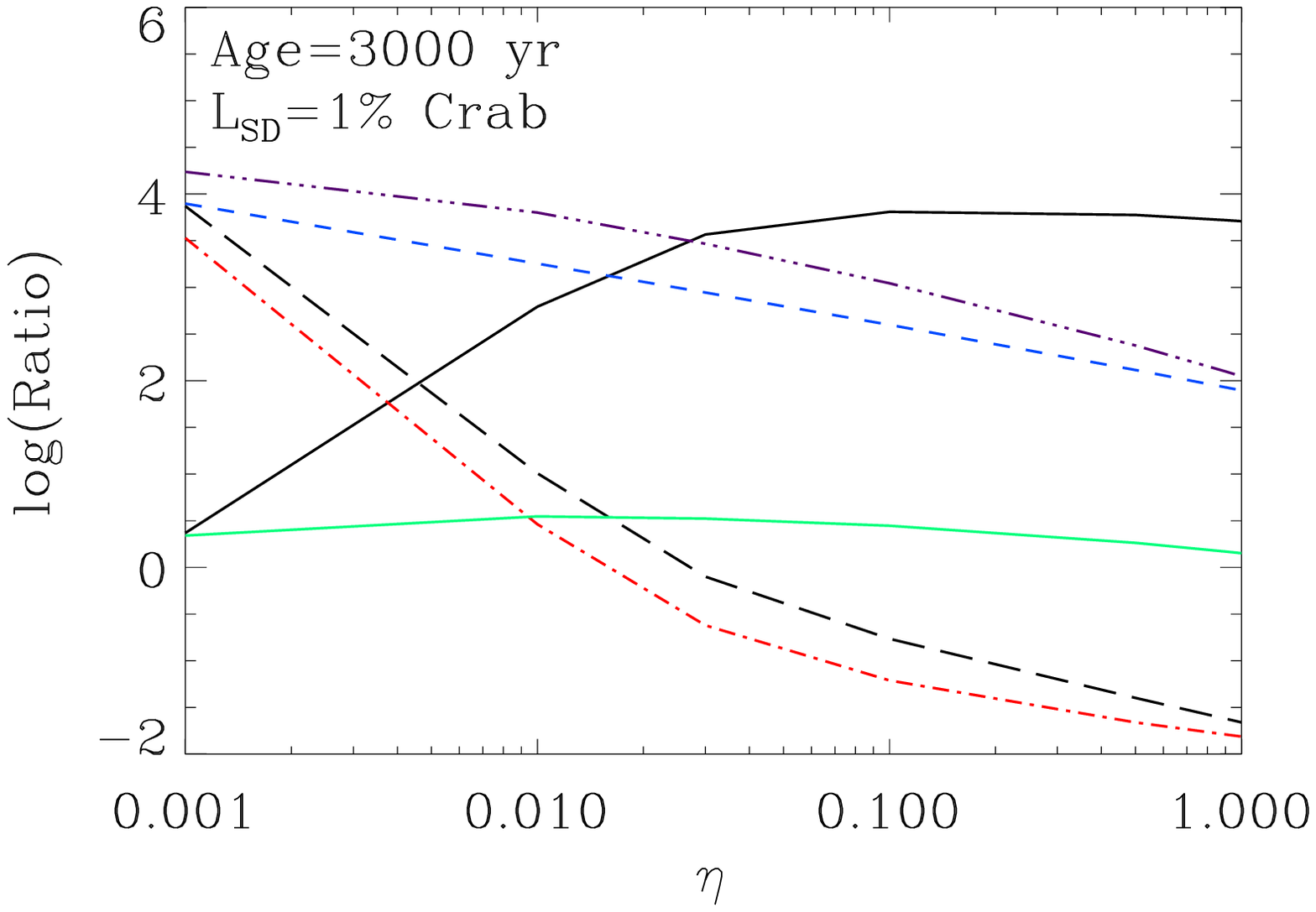} 
\includegraphics[scale=0.3245]{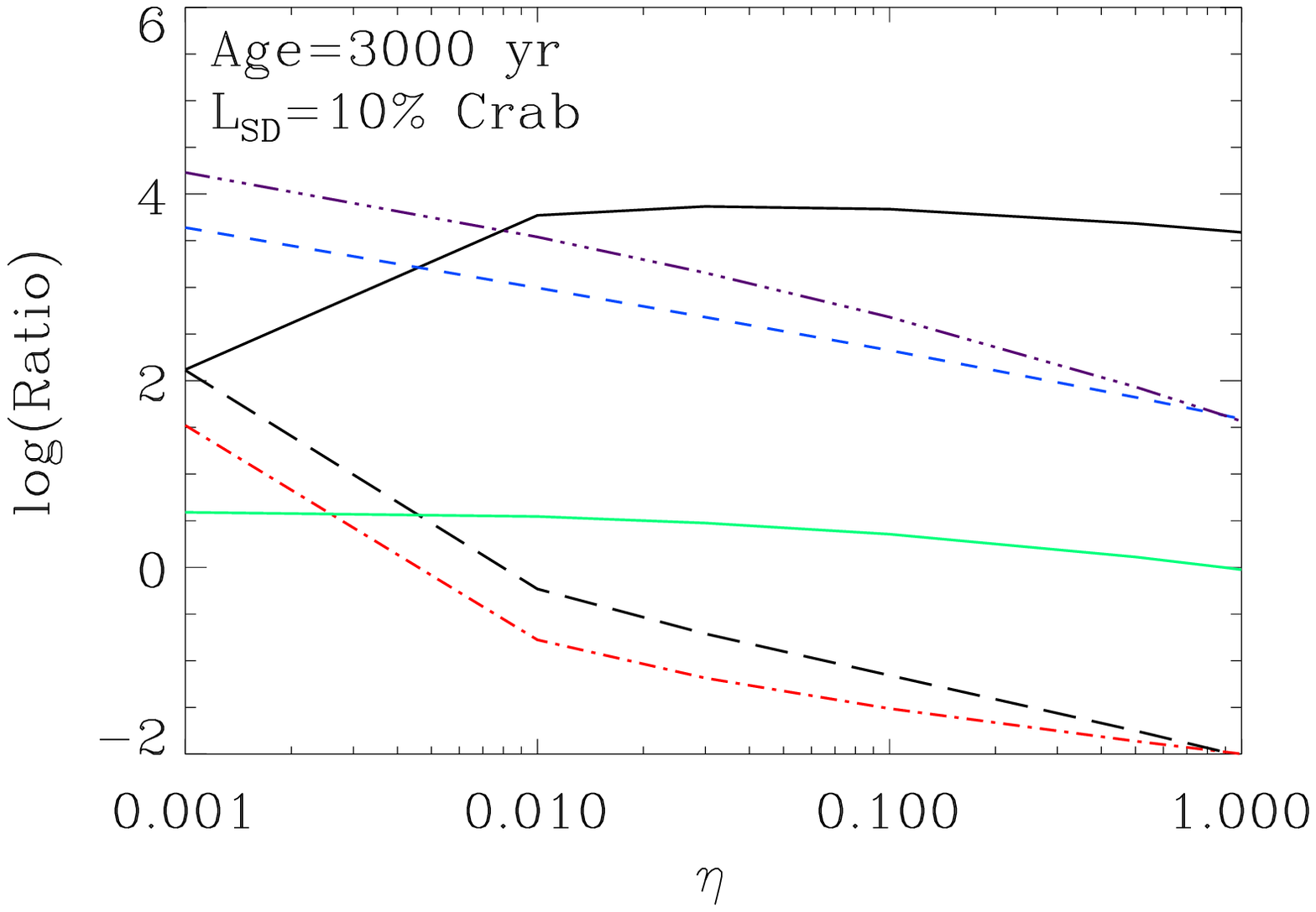}  \includegraphics[scale=0.3245]{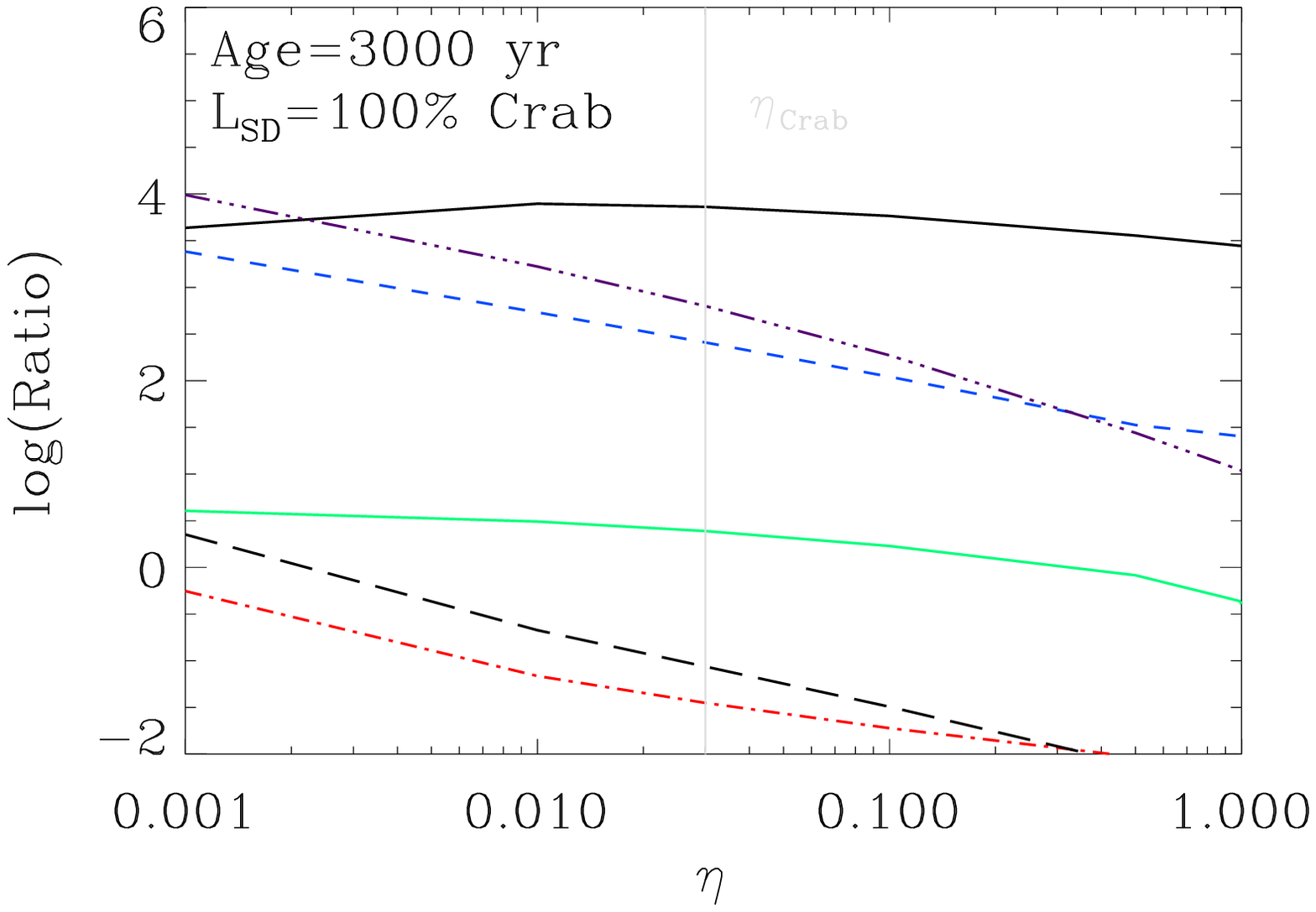} \\
\includegraphics[scale=0.3245]{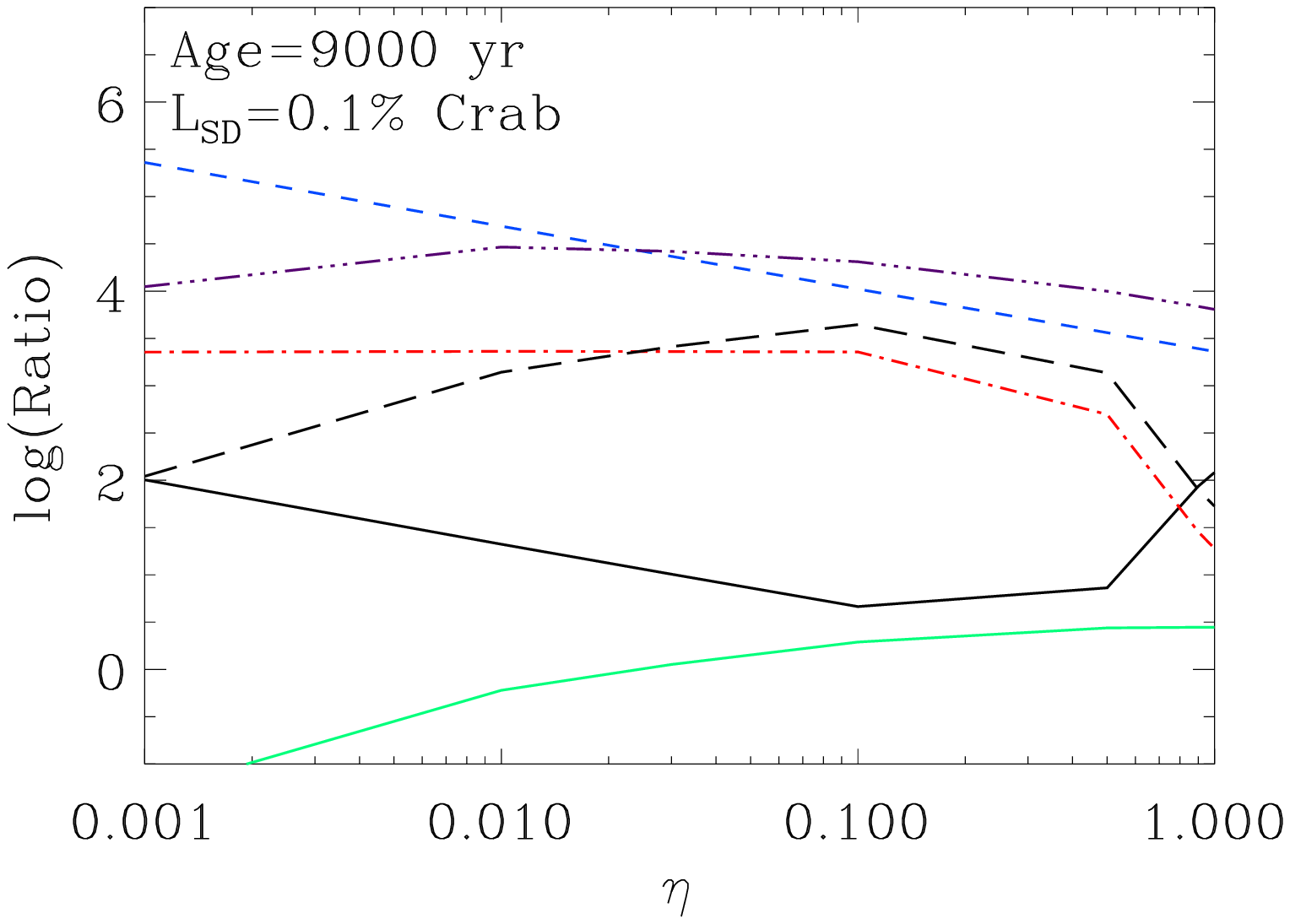}  \includegraphics[scale=0.3245]{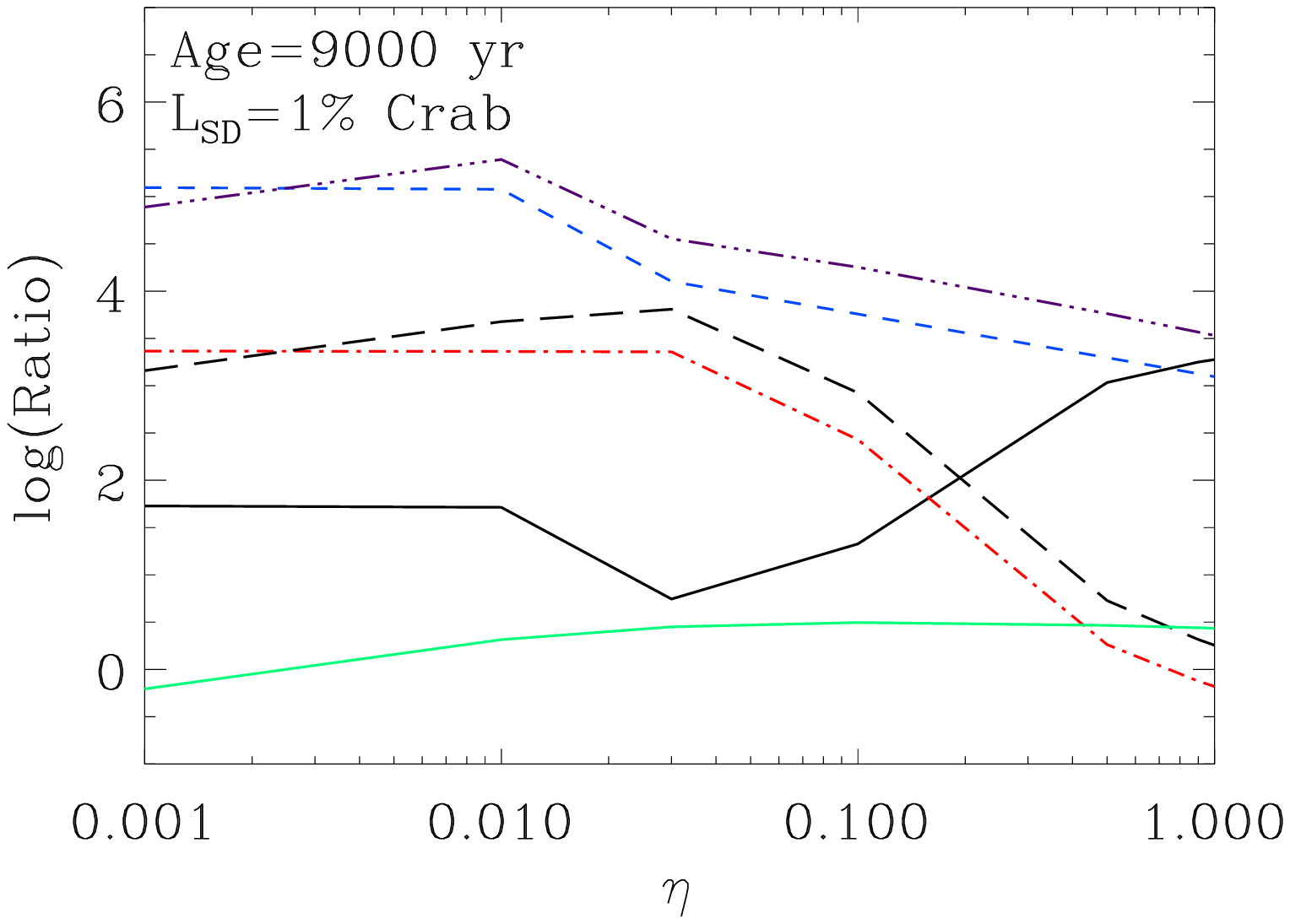} 
\includegraphics[scale=0.3245]{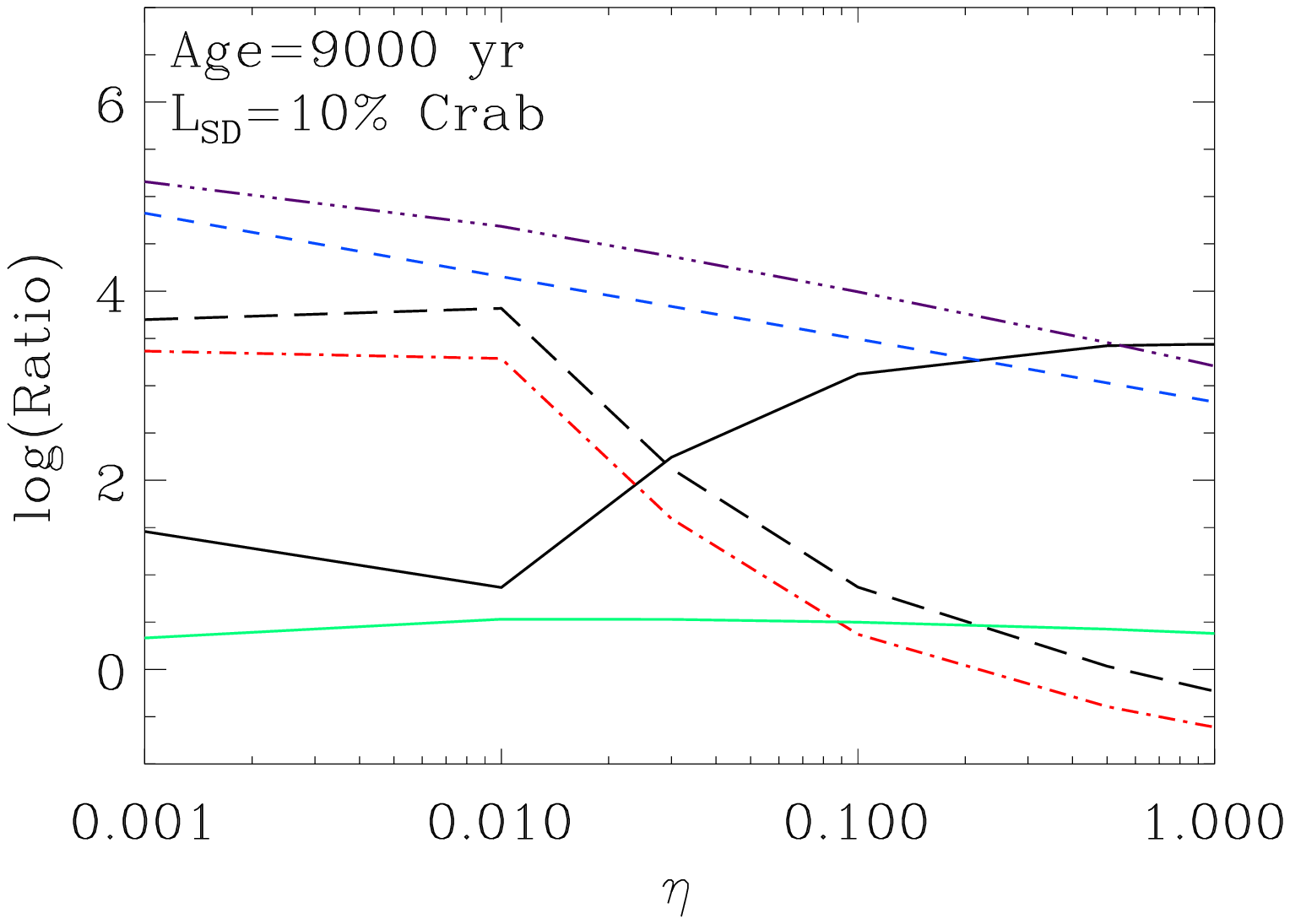}  \includegraphics[scale=0.3245]{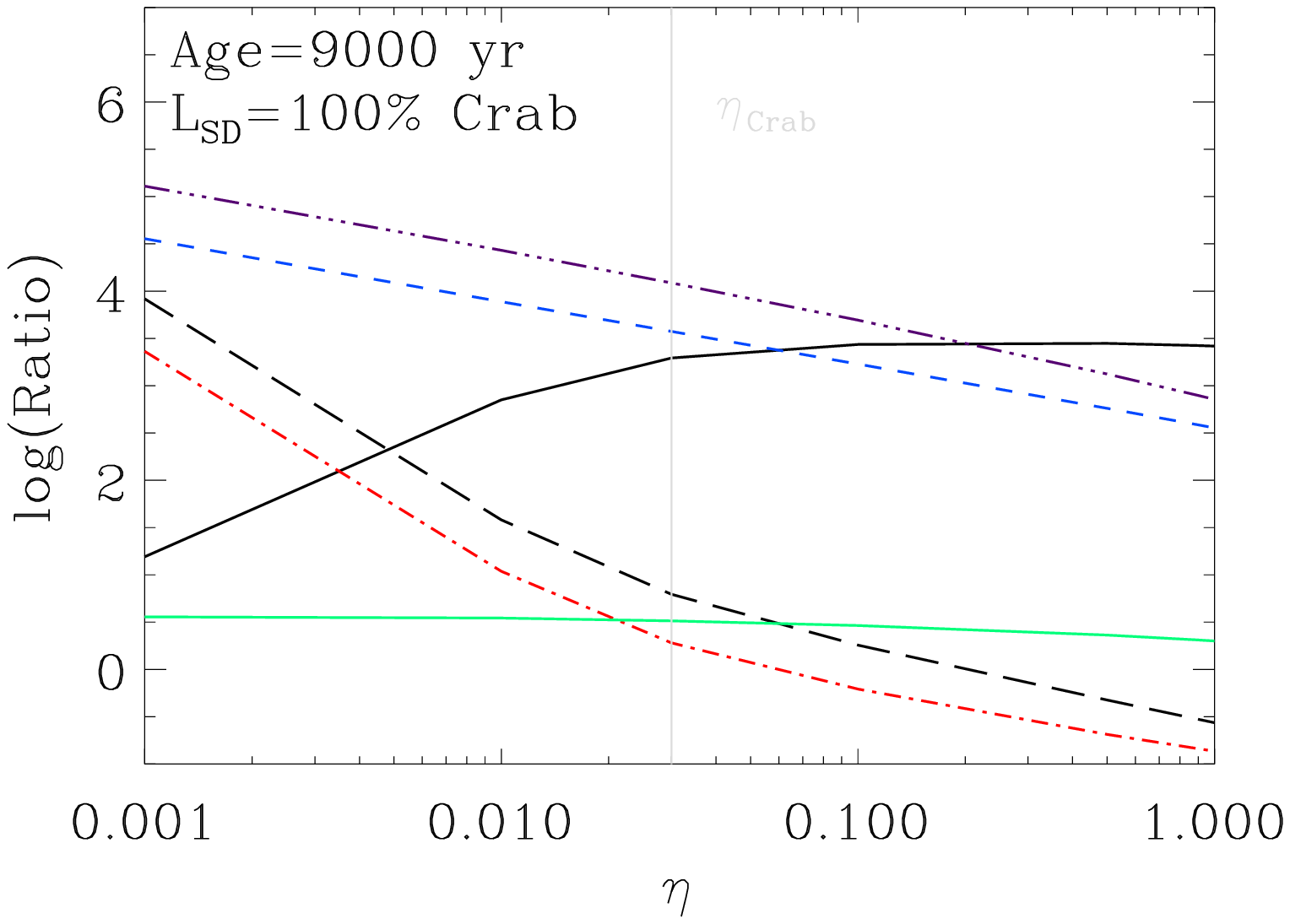}
\end{center}
\caption{Luminosity ratios at 940, 3000, and 9000 years (from top to bottom) as a function of the magnetic fraction. We show the cases for $L_{SD}$=0.1, 1, 10, and 100\% of Crab (from left to right) and depict the following ratios:
X-ray/radio: black solid line,
Gamma-ray/radio: blue dashed line,
VHE/radio: purple triple-dot-dashed line,
Gamma-ray/X-ray: red dot-dashed line,
VHE/X-ray: black dashed line,
VHE/Gamma-ray: green solid line. The bands used for defining the ratios corresponds to the shaded regions in previous figures. For further details, see text.}
\label{lum-ratios}
\end{figure}
\end{landscape}

Fig. \ref{lum-ratios} represents the distance-independent luminosity ratios 
at 940, 3000, and 9000 years (from top to bottom) as a function of the magnetic fraction; for different spin-down powers. 
All the ratios can, of course, be directly measured if such pulsars exist. As an example, the vertical line in the rightmost panels shows the ratios from the spectrum of the Crab Nebula along time (the right-top panel corresponds to the values of the ratios as measured today). They correspond to one and the same magnetic fraction (in the framework of the model assumptions), at $\eta = 0.03$.
Note that the ratios are, mostly, monotonic functions of $\eta$, and thus, a measurement or upper limits on the luminosities of a PWN can be used to estimate a value of the magnetic fraction. 
Exceptions to the monotonic character of the ratios happen. Examples of those are the VHE/X-ray ratio (represented by the black dashed line) of pulsars having 1 to 10\% of the Crab's spin down and ages of 9000 years (the two bottom-middle panels).
If we are to measure a VHE / X-ray ratio only, there could be two magnetic fractions corresponding to it and its value is then degenerate. 
However, not all ratios are, and measurements of other luminosity ratios would break the degeneracy and inform on a plausible value of $\eta$.

One can also consider that the ratios between luminosities can be ideally measured even if we do not know the $P$ and $\dot P$ of the corresponding pulsar, say, after a blind discovery of a PWN in the foreseen CTA Galactic Plane survey. Having several luminosity ratios, if they correlate with a single value of $\eta$ would inform of a plausible value not only of the magnetic fraction, but also of the age and power (always under the assumption of Crab-like injection and environmental variables, which we challenge below).
%
%Fig. \ref{lum-eff} shows 
The efficiencies of the radiative power at each of the bands
play a similar role to Fig. \ref{lum-ratios} when $P$ and $\dot P$ are known quantities.
% and the spin-down power can be computed. 
%This can be particularly useful in cases when there is such knowledge, but a detection of the nebula happens only in one band, e.g. in X-rays. 
%

The caveat  is that for particular PWNe,
neither the injection parameters, nor the densities of the background photons will be exactly as assumed here. Thus, there is no escape from individual modeling;  Fig. \ref{lum-ratios} can only be taken as an approximation if we are to compare with directly measurable quantities. However, we can  imagine having a set of Figs. \ref{lum-ratios}  and/or the efficiencies at each of the bands, spanning
different assumptions for the injection or the environmental parameters. Using such expanded phase space, an automatic procedure of  interpolation 
could inform on plausible values of  $\eta$, age, and luminosity starting only from observational data, like the ratios of luminosities or efficiencies. This is to be considered at CTA times, when 
hundreds of PWN are expected to be discovered blindly.

%%%%%%%%%%%%%%%%%%%%%%%%%%%%%%%%%%%%%%%%%%%%%%%%%%
\section{Discussion}
%%%%%%%%%%%%%%%%%%%%%%%%%%%%%%%%%%%%%%%%%%%%%%%%%%

After considering a phase space of $\sim 100$ Crab-like PWNe of different magnetization, spin-down power, and age we concluded that:

\begin{itemize}

\item  The SSC contribution to the total IC yield is irrelevant for $L_{SD}$=0.1, 1, and 10\% of the Crab power, disregarding the age and the magnetic fraction of the nebulae. It only becomes relevant for highly energetic ($\sim 70\%$ of the Crab) particle dominated nebulae at low ages (of less than a few thousand years). 

\item No young (rotationally powered) PWN would be detectable at TeV energies if the pulsarÕs spin-down power is 0.1\% Crab or lower. For 1\% of the Crab spin-down, only particle dominated nebulae can be detected by H.E.S.S.-like telescopes if young enough (with the detail of the detectability analysis depending on the precise injection and environmental parameters). Above 10\% of the Crab's power, all PWNe are detectable by H.E.S.S.-like telescopes if they are particle dominated, no matter the age.

\item The magnetic fraction is an important order parameter in the TeV observability of nebulae, and induces orders of magnitude variations in the luminosity output for systems that are otherwise the same (same spin-down, $P$, $\dot P$, injection, and environment). 
For Crab-like photon field background and injection parameters, H.E.S.S.-like telescopes would not be sensitive enough to fully explore the $\eta>$0.5 regime, independently from the pulsar age (when lower than $10^4$ years) or spin-down power. 

\end{itemize}

Based on the above results, we pose that if extreme differences between the environmental or injection variables do not occur (in comparison with the Crab Nebula's)  it is 
the magnetic fraction what  decides detectability of TeV nebulae. 
It is thus important to analyze the impact 
that a different injection or environmental parameters have on our conclusions. 
Here we specify on the stability of the results against changes on $\alpha_1$, $\alpha_2$, and $w_{FIR}$.
{
The IR energy density in specific regions of the Galaxy can well exceed the 0.5  eV/ cm$^{-3}$ considered, for instance close to star formation sites.}

\begin{figure*}
\begin{center}
\includegraphics[scale=0.45]{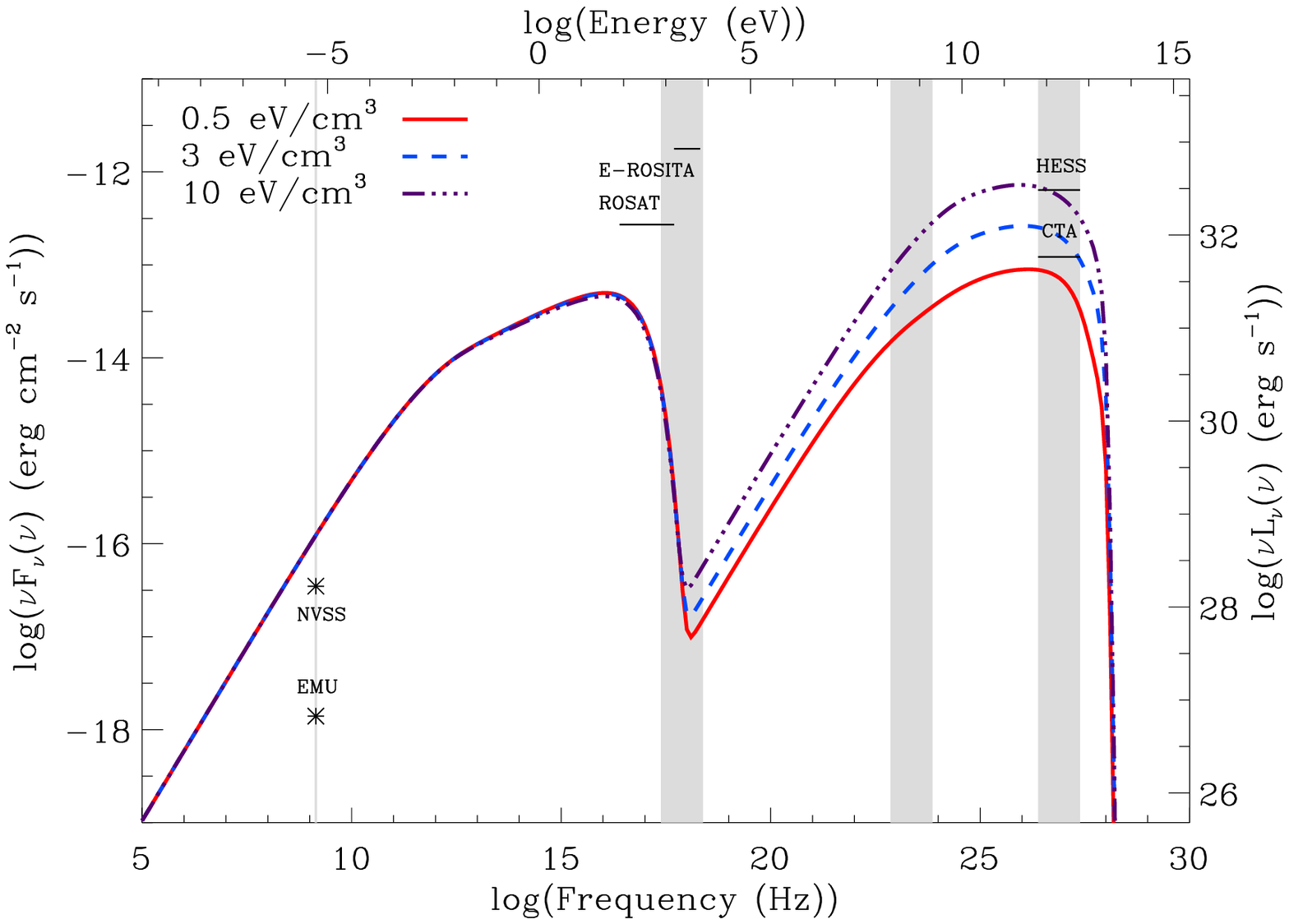}\hspace{0.2cm}
\includegraphics[scale=0.45]{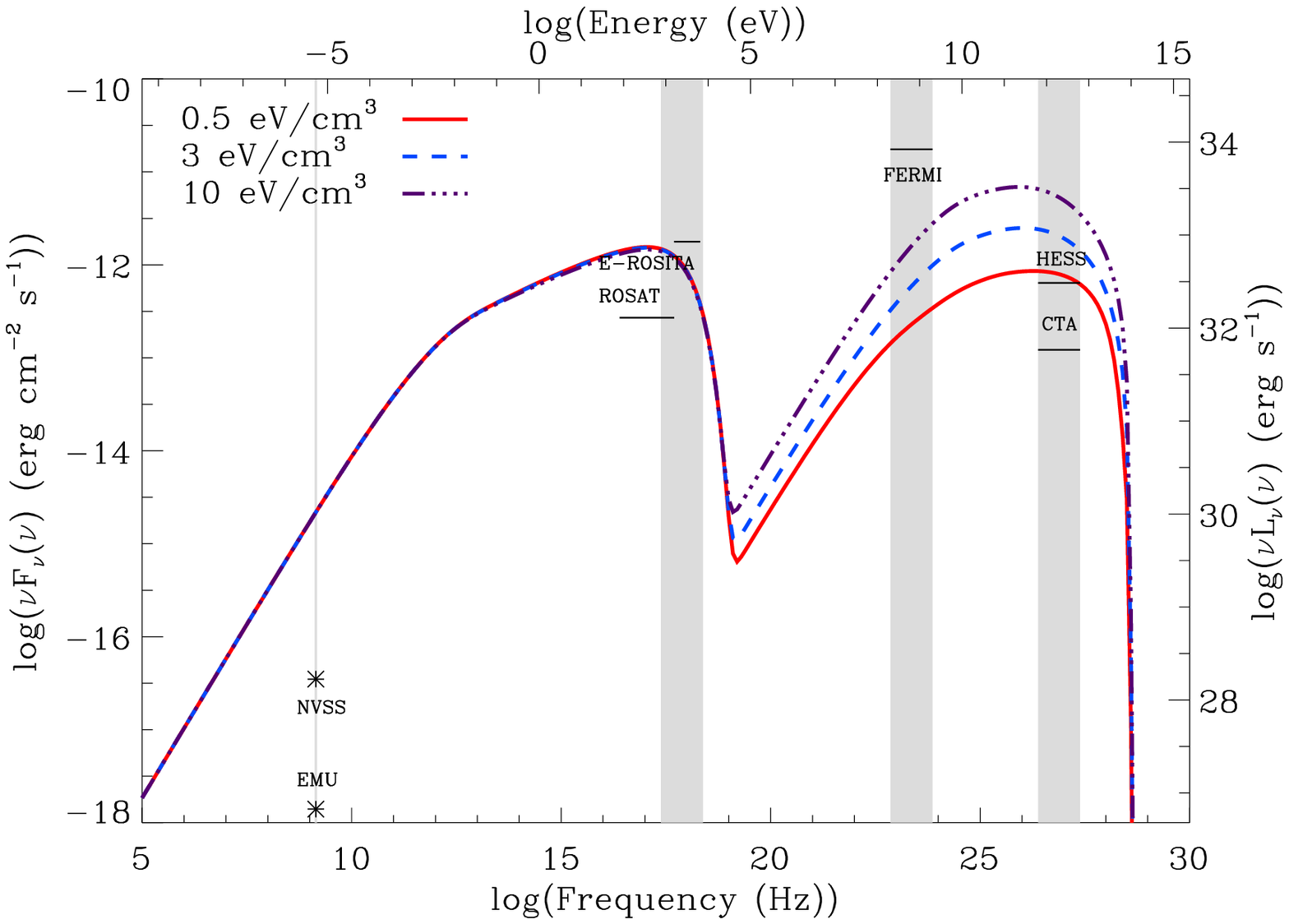} 
\end{center}
\caption{SEDs for Crab's injection parameters and different FIR photon density, from 0.5 to 10 eV cm$^{-3}$. The left panel shows
the results for 0.1\% of Crab's energetics, at an age of 3000 years. The right panel shows 
the same analysis for 1\% of Crab's energetics. A low magnetization of 0.03 is assumed.}
\label{fircomp} 
\end{figure*}

Figure \ref{fircomp} (left) compares 
the results for 0.1\% of Crab's energetic, at an age of 3000 years. It can be seen that even in the extreme case of 10 eV cm$^{-3}$; a factor of 20 in excess of Crab's energy density, a H.E.S.S.-like telescope will not detect this PWN. The conclusion is thus stable for the current generation of telescopes. CTA detectability, instead, will depend on the FIR density. We recall that the distance assumed for the flux-sensitivity comparison is that of Crab, and thus, that for farther PWNe, the ability of the telescope for detection will be diminished further. 
The right panel of Fig. \ref{fircomp} shows 
the same analysis but for 1\% of Crab's energetics.
As stated above, the FIR energy density (and distance, of course) is the decisive parameter concerning the detectability of PWNe
in H.E.S.S.-like telescopes in this case. Note that the extreme case of 10 eV cm$^{-3}$, which at 2 kpc produces about one order of magnitude
in excess of a H.E.S.S.-like telescope sensitivity, would be invisible when the pulsar that drives the nebula is located instead at 5 kpc or beyond. 
Thus, if having the same injection parameters,
most of of the PWNe with 1\% of the Crab's energetics would not be seen by the current generation of instruments.

\begin{figure*}
\begin{center}
\includegraphics[scale=0.45]{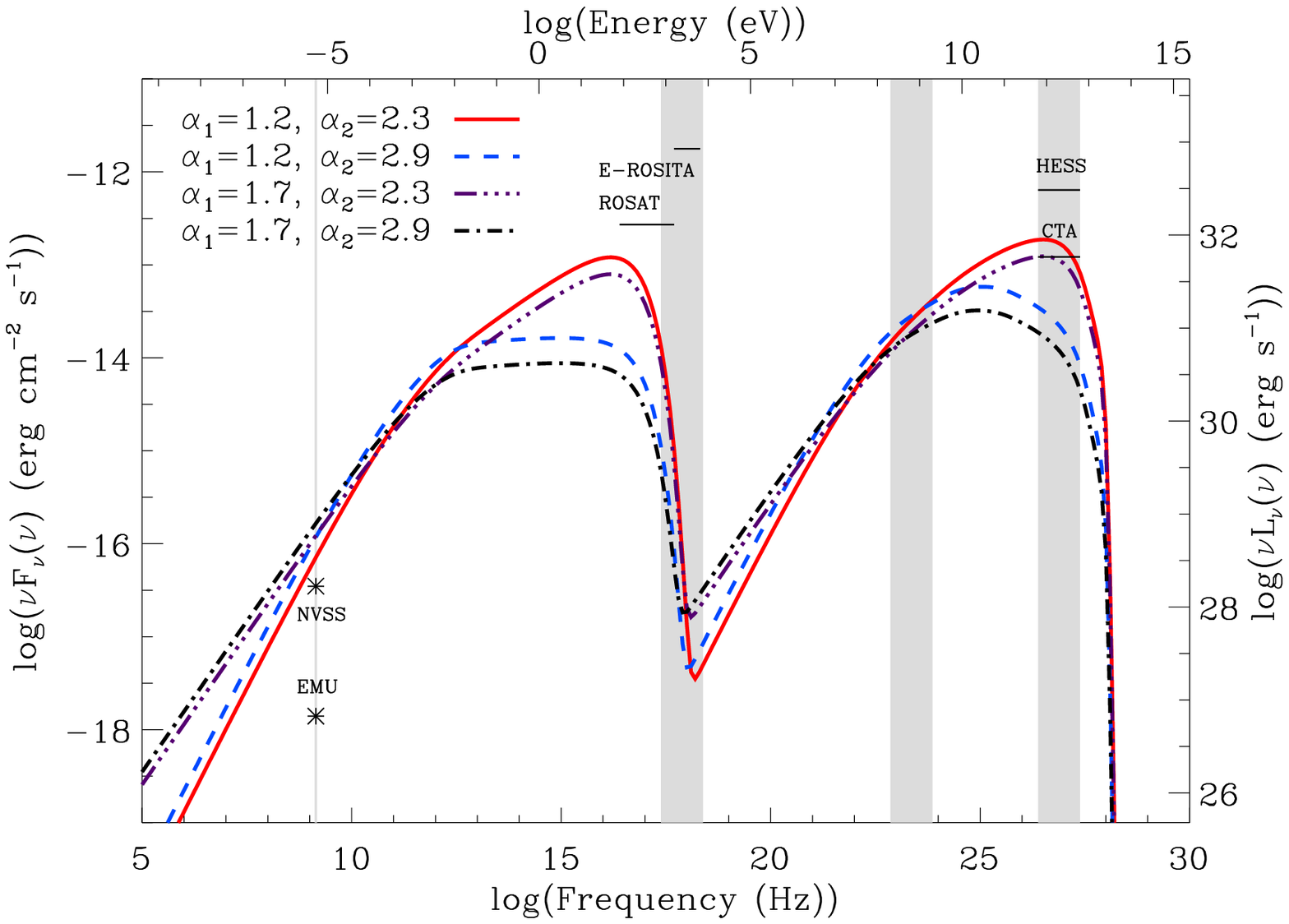}
\includegraphics[scale=0.45]{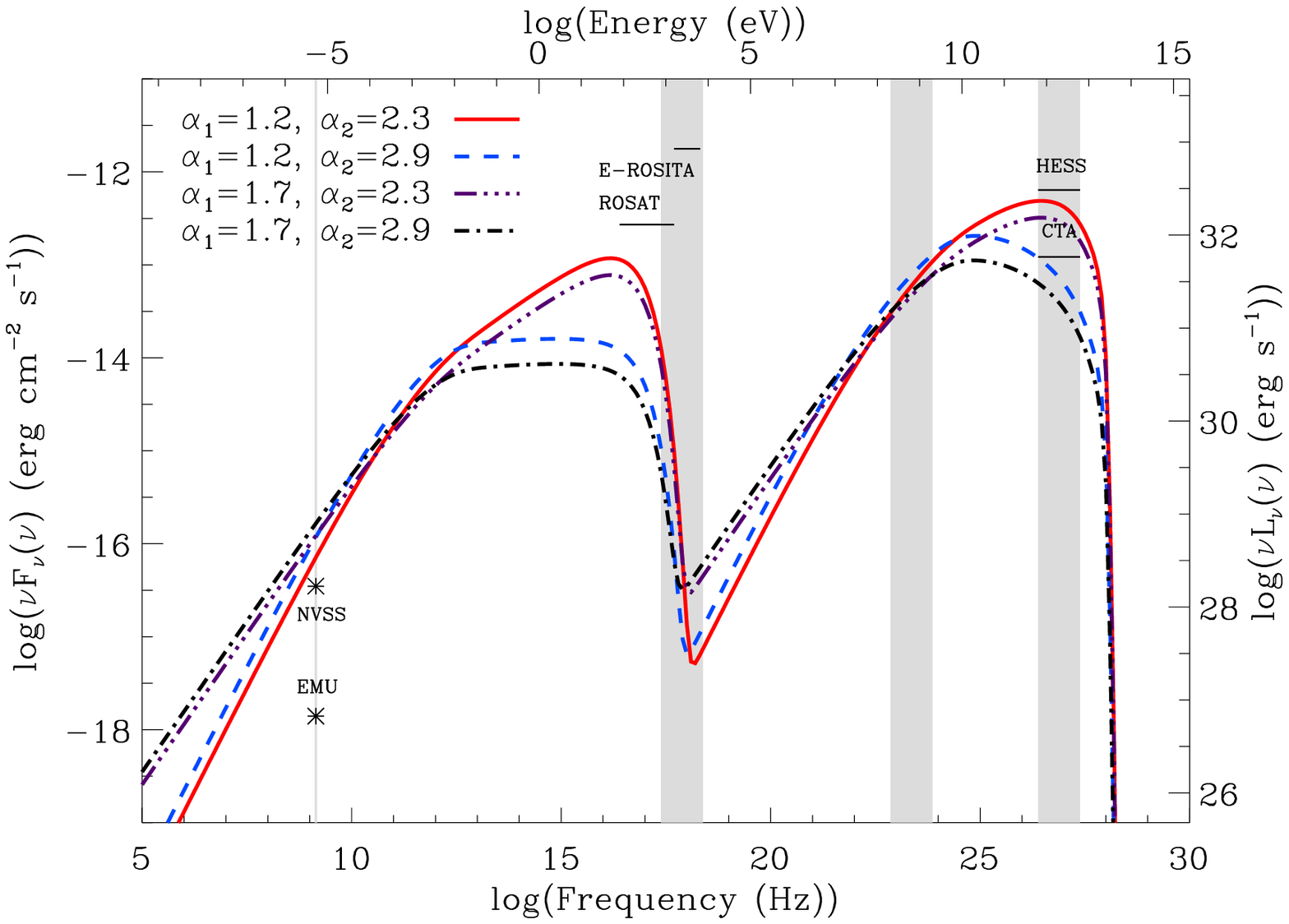}\\
\includegraphics[scale=0.45]{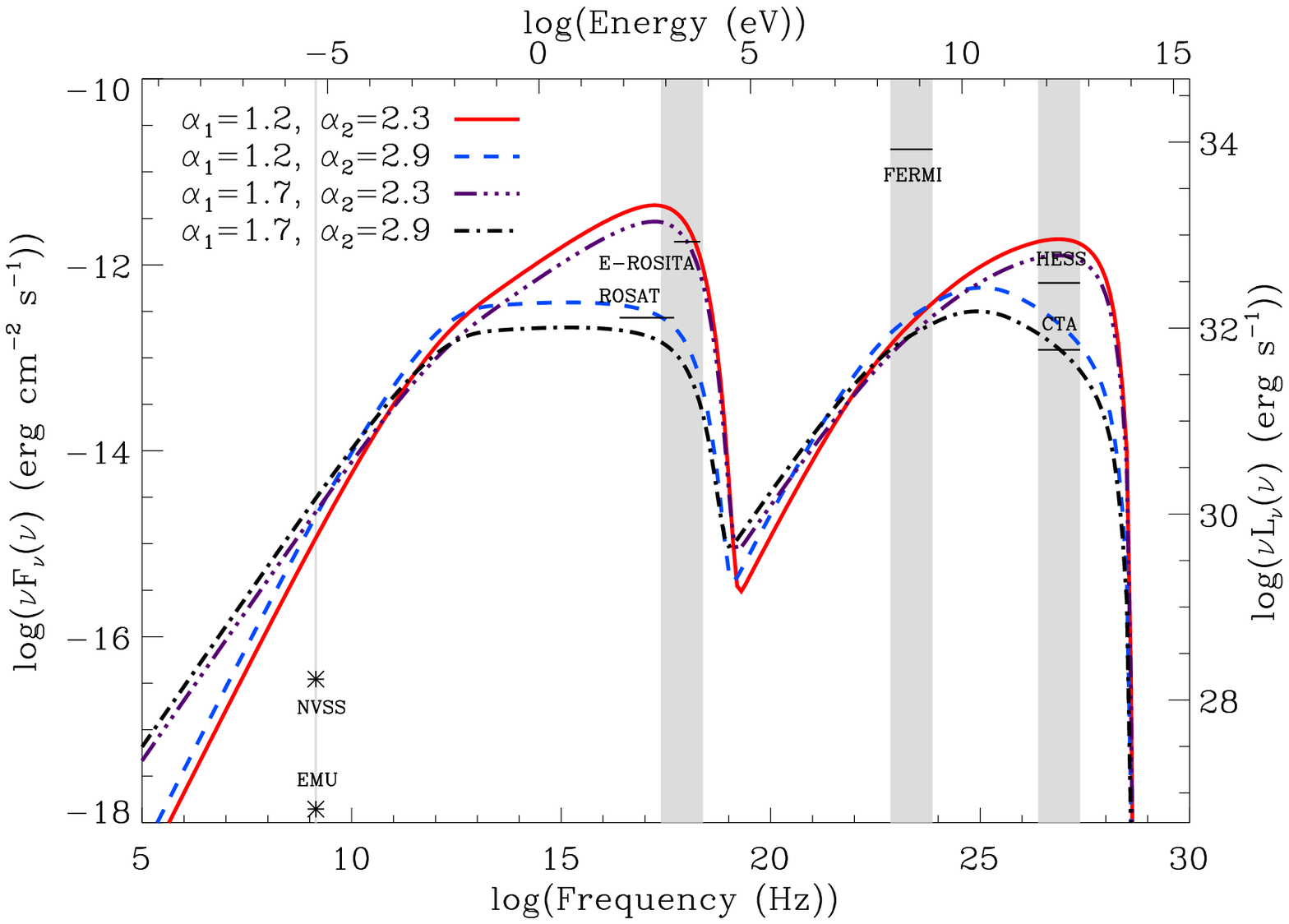}\hspace{0.24cm}
\includegraphics[scale=0.45]{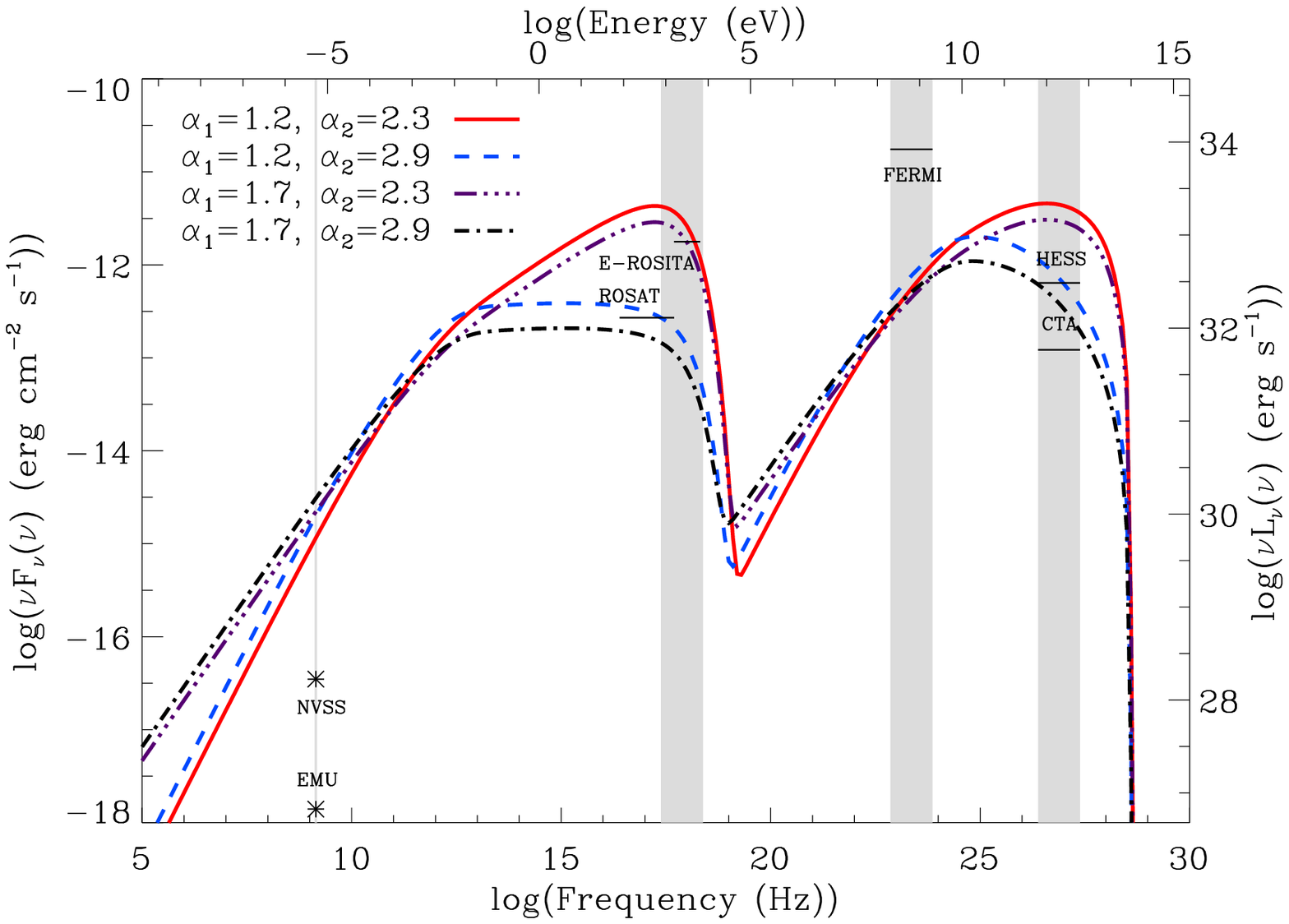} 
\end{center}
\caption{SEDs for different injection parameters (as detailed in the legend) and different FIR 
photon density (0.5 eV cm$^{-3}$ in the left panel,  and 3 eV cm$^{-3}$ in the right one) for a pulsar with 0.1\% (top row)
and 1\% (bottom row) of Crab's energetics. A low magnetization of 0.03 is assumed.}
\label{injtest} 
\end{figure*}

The detectability of PWNe with 0.1\% of Crab's energetics in H.E.S.S.-like telescopes is not affected either by changes in 
the injection parameters. The top row of Fig. \ref{injtest}
shows such changes together with an increased FIR background. Four different pairs of 
injection slopes are assumed and results are shown for an age of 3000 years. The differences produced by the 
the changes in injection are indeed large, as expected, but still, a low FIR background would preclude most of these PWNe having 0.1\% of Crab's energetics
to be detected even by CTA.
The bottom row of Fig. \ref{injtest} shows the same results for the case of 1\% of Crab.
Note that for an average value (0.5 eV cm$^{-3}$ in the left example) or even a significantly increased value (3 eV cm$^{-3}$ in the right example)
of FIR photon density, 
only hard spectra will lead to a clear detection in the current generation of instruments. 
None of these pulsars featuring 1\% Crab's energetics, if located at 5 kpc instead of 2 kpc, would be detected by H.E.S.S.-like instruments.

It is interesting to { mention that 3C58 (PSR J0205+6449) 
%and  PSR J1747-2809/PWN G0.9+0.1. They both 
has a spin-down power of ($\sim 3 \times 10^{37}$ erg cm$^{-3}$), and a characteristic age of 5300 years. 
Compared with Crab at the same age, it is about twice as luminous. } 3C58 has however been observed by both MAGIC and VERITAS and only upper limits were imposed up to now (Anderhub et al. 2010, Konopelko et al. 2007). 3C58 and Crab differ significantly, particularly, the 
X-ray luminosity of 3C58 is 2000 times smaller than that of Crab (Torii et al. 2000). Comparing instead with the X-ray luminosity that Crab will have at 5000 years, 
we would still obtain at least one order of magnitude difference, implying that the injection parameters  from one PWN to another
change significantly. This is what we have found when fitting a PWN model to 3C58 (see Torres et al. 2013), where $\alpha_1=1.05$ and $\alpha_2=2.9$. As shown generically in Fig. \ref{injtest},  the spectrum is flattened, and unless a significant increase of the FIR background is present (which might be indeed 
the case after the results shown by Abdo et al. 2013 in the adjacent GeV band), the PWN would remain invisible at VHE.

\begin{figure*}
\begin{center}
\includegraphics[scale=0.45]{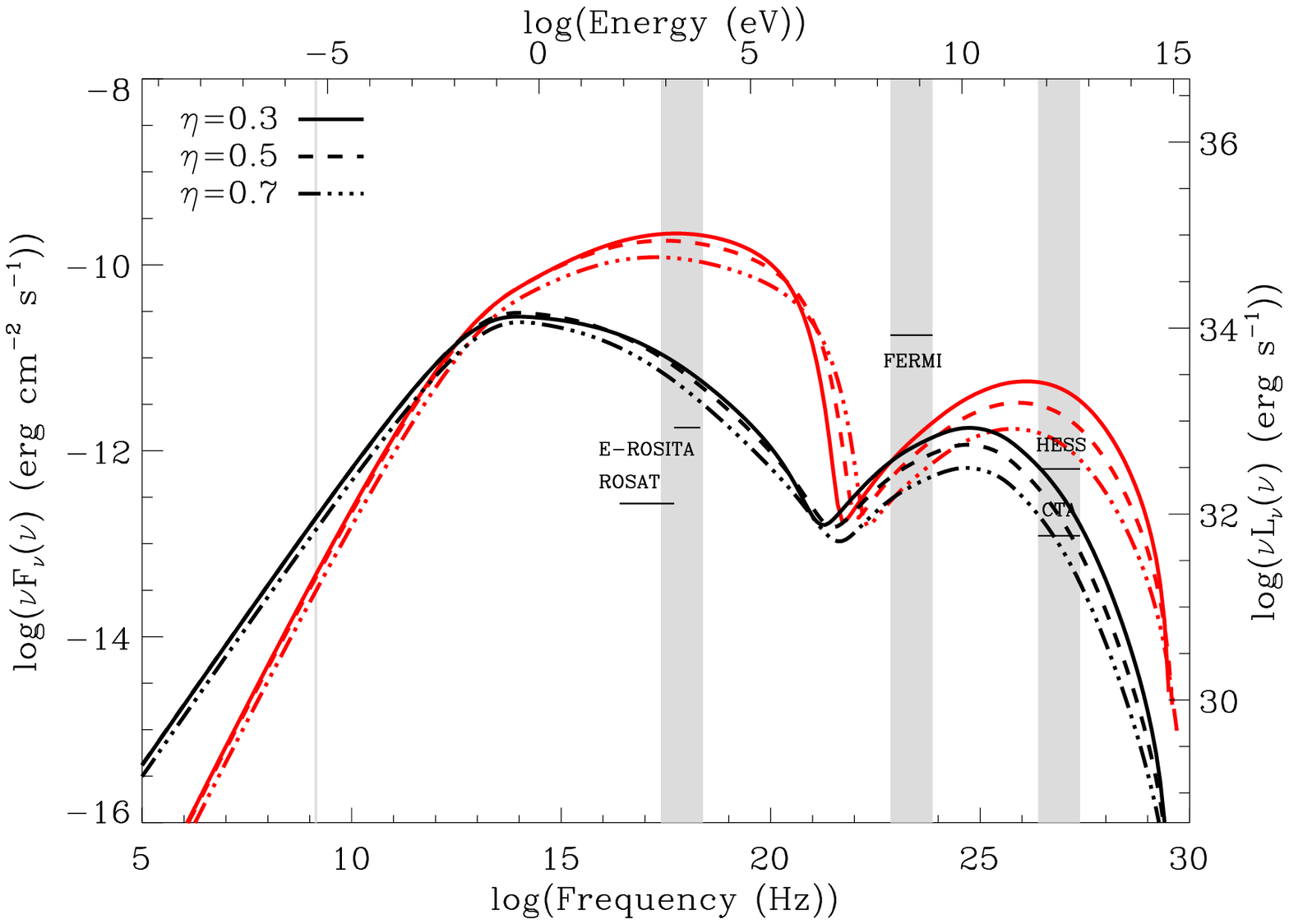}
\includegraphics[scale=0.45]{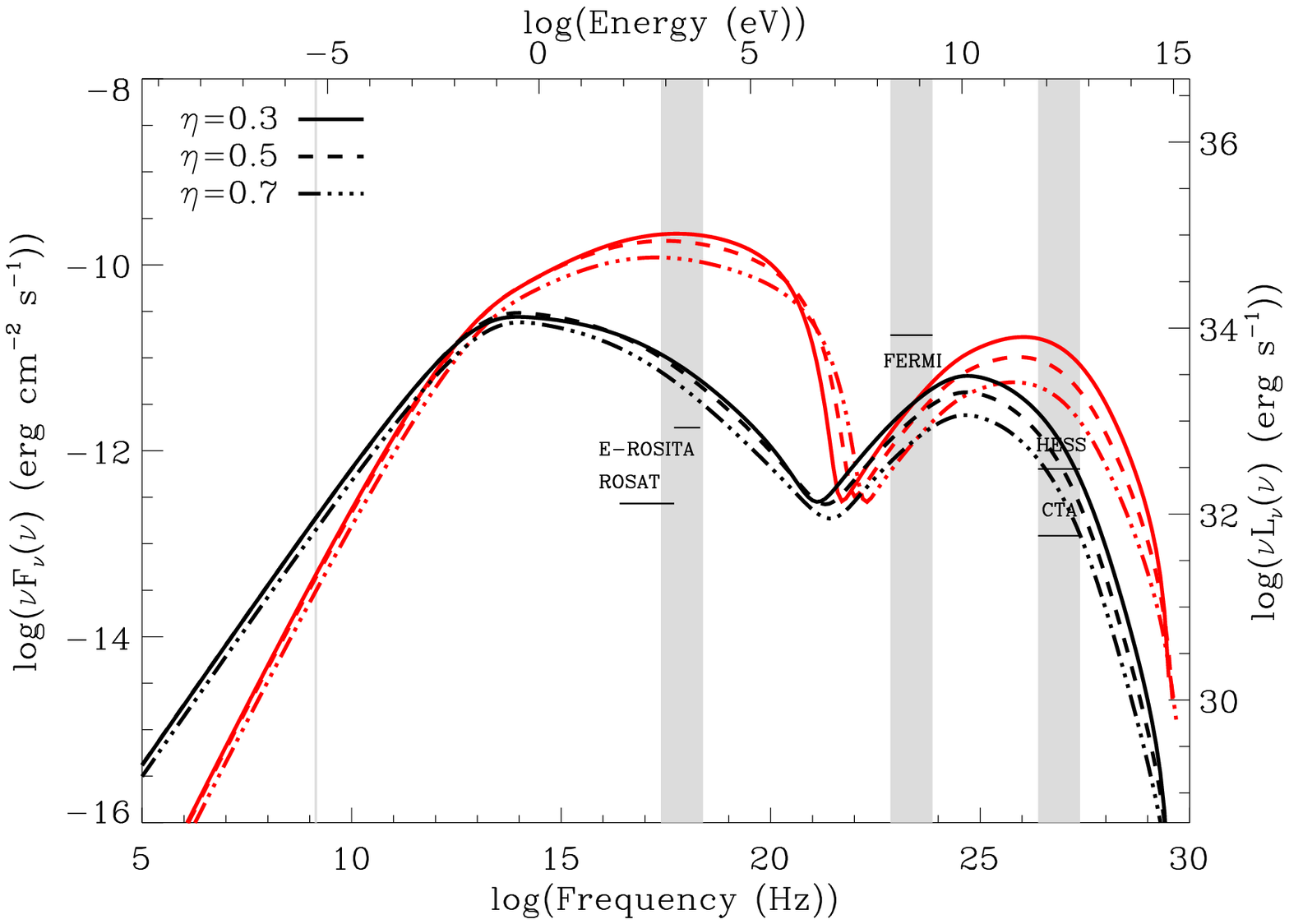}
\end{center}
\caption{SEDs for different magnetic fraction  (as detailed in the legend) and different FIR 
photon density (0.5 eV cm$^{-3}$ in the left panel,  and 3 eV cm$^{-3}$ in the right one) for a pulsar with 10\% of Crab's energetics. 
In red, a hard spectrum of particles with 
$\alpha_1=1.2$, $\alpha_2=2.3$ is assumed, whereas the black curves stand for 
a steep case with
$\alpha_1=1.7$, $\alpha_2=2.9$.
}
\label{eta-fir-inj}
\end{figure*}

We also consider how stable is  the assertion 
that at 10\% of Crab's spin down, a
H.E.S.S.-like instrument 
needs $\eta<0.5$ for detecting PWNe, against variations of
injection or FIR background. To do that we consider a PWN of a pulsar with 10\% of Crab's energetics, subject to two energy densities, 0.5 and 3 eV cm$^{-3}$,
with  electron distributions having  slopes of
$\alpha_1=1.2$, $\alpha_2=2.3$ and $\alpha_1=1.7$, $\alpha_2=2.9$ represented in red and black in Fig. \ref{eta-fir-inj}, respectively.
Obviously, the most  favorable cases for detection are given by the harder slopes of injection and the largest FIR densities. In those particular cases with $w_{FIR}$=3 eV cm$^{-3}$ (0.5 eV cm$^{-3}$)
H.E.S.S.-like instruments could detect the PWNe up to a magnetic fraction of 0.7 if located closer than $\sim$5 kpc ($\sim$3.5 kpc).
Results for pulsars with 10\% of Crab's energetic 
uniformly produce detectable PWNe for magnetization parameters lower than 0.5. The fact that most of the PWNe detected have strong particle dominance
is thus not affected by observational biases when their spin-down exceeds 10\% of the Crab.

\subsection*{Acknowledgments}

Work done in the framework of the grants AYA2012-39303, SGR2009-811, and iLINK2011-0303. DFT was additionally supported by a Friedrich Wilhelm Bessel Award of the Alexander von Humboldt Foundation. ANC acknowledges CONICET (Consejo Nacional de Investigaciones Cient\'ificas y T\'ecnicas) for financial support (PIP CONICET Nr 305/2011).
We acknowledge an anonymous referee for useful comments on the paper.

\section*{Appendix}

%\section{Underlying model components}

Here, we will briefly summarize the components of the underlying models we consider. 
We assume a broken power-law, time-dependent  injection of particles where $\gamma_b$ is the break energy and $\alpha_1$  and $\alpha_2$
are the spectral indices:
\begin{equation}
\label{injection}
Q(\gamma,t)=Q_0(t)\left \{
\begin{array}{ll}
\left(\frac{\gamma}{\gamma_b} \right)^{-\alpha_1}  & \text{for }\gamma \le \gamma_b,\\
 \left(\frac{\gamma}{\gamma_b} \right)^{-\alpha_2} & \text{for }\gamma > \gamma_b.
\end{array}  \right .
\end{equation} 
We consider that particles are subject to synchrotron, inverse Compton, self-synchrotron Compton, adiabatic, and bremsstrahlung processes, and escape via Bohm diffusion. No radiative approximation is made.
Solving the diffusion loss equation we obtain the
lepton population in the nebula as a function of time.

The spin-down power of the pulsar  is: 
\begin{equation}
L(t)=L_0 \left(1+\frac{t}{\tau_0} \right)^{-\frac{n+1}{n-1}},
\label{Ldet}
\end{equation}
where $L_0$ is the initial luminosity, $\tau_0$ is the spin-down timescale of the pulsar, 
\begin{equation}
\label{tau0}
\tau_0=\frac{P_0}{(n-1)\dot{P}_0}=\frac{2\tau_c}{n-1}-t_{age},
\end{equation}
and 
\be n= \ddot P P / \dot P^2
\label{n}
\ee
 is the braking index. 
Here, $P_0$ and $\dot{P}_0$ are the initial period and its first derivative and $\tau_c$ is the characteristic age of the pulsar, 
$
\tau_c={P}/{2\dot{P}}.
$

The normalization of the injection function of Eq. (\ref{injection}) is given by
\begin{equation}
\label{normalization}
(1-\eta)L(t)=\int_0^\infty \gamma m c^2 Q(\gamma,t) \mathrm{d}\gamma,
\end{equation}
where $\eta=L_B(t)/L(t)$ ($L_B(t)$ is the magnetic power) is the magnetic energy fraction. The latter bears a resemblance with
the magnetization parameter 
$
\sigma(t)= {L_B(t)} / {L_p(t)},
$
where $L_p(t)$ is the relativistic particle's fraction of the spin-down power, since 
$\eta = \sigma / (\sigma + 1)$. We will say a nebula is in 
equipartition when $\eta=0.5$. Recall that $B(t)$ is here
the average magnetic field across the nebula.
This magnetic field $B(t)$ is defined by the equation 
\be
\int_0^t{\eta L(t')R_{PWN}(t')dt'}=(4\pi/3)R_{PWN}^4(t)B^2(t)/(8\pi),
\ee
%This equation is equivalent to 
%\begin{equation}
%$(dW_B/dt)=\eta L-W_B(dR_{PWN}/dt)/R_{PWN}$
%\end{equation}
%where 
%$W_B=(4\pi/3)R_{PWN}^3(t)B^2(t)/(8\pi)$,
which includes the adiabatic losses due to nebular expansion (e.g., Pacini \& Salvati 1973).

The maximum Lorentz factor of the particles is limited by confinement, i.e., the 
Larmor radius should be smaller than the termination shock, 
\be 
\gamma_{max}(t)=({\varepsilon e \kappa}/{m_e c^2})\sqrt{\eta {L(t)}/{c}},
\ee
where $e$ is the electron charge and $\varepsilon$ is the fractional size of the radius of the shock. 
The Larmor Radius is
$R_{L}=(\gamma_{max} m_e c^2)/(e B_s)$, where $B_s$ is the post-shock field strength, defined as
$B_s \sim (\kappa (\eta L(t)/c)^{0.5})/R_s ,$
with $R_s$ the termination radius.
%The latter becomes a free parameter, but we have explored its impact and found it is not the most relevant. 
We have fixed $\kappa$, the magnetic compression
ratio, to 3 (as in Venter \& de Jager 2006, 
Holler et al. 2012). 

Finally, we have adopted 
the free expanding expansion phase in the model by van der Swaluw et al. (2001) and 
van der Swaluw et al. (2003), where the radius of the
PWN is 
\be
 R_{PWN}(t) \sim \left({L_0 t}/{E_0} \right)^{1/5} V_{ej} t,
 %R_{PWN}(t) =C \left(\frac{L_0 t}{E_0} \right)^{1/5} V_0 t,
\ee
with 
$V_{ej}=\sqrt{{10 E_0}/{3 M_{ej}}}$ and where 
$E_0$ and $M_{ej}$ are the energy of the supernova explosion and the ejected mass, respectively.

\clearpage

%%%%%%%%%%%%%%%%%%%%%%%%%%%%%%%%%%%%
% FIGURES
%%%%%%%%%%%%%%%%%%%%%%%%%%%%%%%%%%%%

%\begin{landscape}
%\begin{figure}
%\begin{center}
%\includegraphics[scale=0.3245]{newratios_940_0.1.ps}  \includegraphics[scale=0.3245]{newratios_940_1.0.ps} 
%\includegraphics[scale=0.3245]{newratios_940_10.0.ps}  \includegraphics[scale=0.3245]{newratios_940.ps} \\
%\includegraphics[scale=0.3245]{newratios_3000_0.1.ps}  \includegraphics[scale=0.3245]{newratios_3000_1.0.ps} 
%\includegraphics[scale=0.3245]{newratios_3000_10.0.ps}  \includegraphics[scale=0.3245]{newratios_3000.ps} \\
%\includegraphics[scale=0.3245]{newratios_9000_0.1.ps}  \includegraphics[scale=0.3245]{newratios_9000_1.0.ps} 
%\includegraphics[scale=0.3245]{newratios_9000_10.0.ps}  \includegraphics[scale=0.3245]{newratios_9000.ps}
%\end{center}
%\caption{Efficiencies at 940, 3000, and 9000 years (from top to bottom) as a function of the magnetic fraction. We show the cases for $L_{SD}$=0.1, 1, 10, \& 100\% of Crab (from left to right) and depict the following efficiencies: 
%$L_{radio}/L_{sd}$: black solid line,
%$L_X/L_{sd}$: blue dashed line,
%$L_\gamma/L_{sd}$: purple triple-dotted dashed line,
%$L_{VHE}/L_{sd}$: red dot dashed line,
%The bands used for defining the ratios corresponds to the shaded regions in previous figures. For further details, see text.}
%\label{lum-eff}
%\end{figure}
%\end{landscape}

\clearpage

%%%%%%%%%%%%%%%%%%%%%%%%%%%%%%%%%%%%
% TABLES
%%%%%%%%%%%%%%%%%%%%%%%%%%%%%%%%%%%%

\label{lastpage}
\end{document}